%% file: main.tex
\documentclass{aastex631}
\usepackage[utf8]{inputenc}
\usepackage[T1]{fontenc}
\usepackage{color}
\usepackage{amsmath}
\usepackage{natbib}
\usepackage{ctable}
\usepackage{bm}
\usepackage[normalem]{ulem} % Added by MS for \sout -> not required for final version
\usepackage{xspace}
\usepackage{hyperref}
\usepackage{soul}
\usepackage{amsmath}
\let\oldbibliography\thebibliography % killin' me.
\renewcommand{\thebibliography}[1]{%
  \oldbibliography{#1}%
  \setlength{\itemsep}{0pt}%
  \setlength{\parsep}{0pt}%
  \setlength{\parskip}{0pt}%
  \setlength{\bibsep}{0ex}
  \raggedright
}
\newcommand{\gaia}{{\em Gaia}}

\newcommand{\dchi}{\Delta \chi^2}
\newcommand{\svi}{SV1}
\newcommand{\sviii}{One-Percent Survey}

\newcommand{\bitem}{\begin{itemize}}
\newcommand{\eitem}{\end{itemize}}
\definecolor{orange}{rgb}{1,0.5,0}

\begin{document} \sloppy\sloppypar\frenchspacing 

\title{DESI Bright Galaxy Survey: Final Target Selection, Design, and Validation}

%%%%%%%%%%%%%%%%%%%%%%%%%%%%%%%%%%%%%%%%%%%%%%%%%%%%%%%%%%%%%%%%%%%%%%%%%%%%%%%%%%%%
% first tier 
%%%%%%%%%%%%%%%%%%%%%%%%%%%%%%%%%%%%%%%%%%%%%%%%%%%%%%%%%%%%%%%%%%%%%%%%%%%%%%%%%%%%
% e-mails: michael.j.wilson@durham.ac.uk, omar.ruiz.macias@gmail.com, Shaun.Cole@durham.ac.uk, weinberg.21@osu.edu, jmoustakas@siena.edu, akremin@lbl.gov, jlt12@nyu.edu, alexander.smith@cea.fr
\newcounter{affilcounter}
\author[0000-0003-1197-0902]{ChangHoon Hahn}
\altaffiliation{changhoon.hahn@princeton.edu}
\affil{Department of Astrophysical Sciences, Princeton University, Peyton Hall, Princeton NJ 08544, USA}
\affil{Lawrence Berkeley National Laboratory, One Cyclotron Road, Berkeley CA 94720, USA}

\author{Michael J. Wilson}
\affil{Institute for Computational Cosmology, Department of Physics, Durham University, South Road, Durham DH1 3LE, UK}
\affil{Lawrence Berkeley National Laboratory, One Cyclotron Road, Berkeley CA 94720, USA}

\author{Omar Ruiz-Macias}
\affil{Institute for Computational Cosmology, Department of Physics, Durham University, South Road, Durham DH1 3LE, UK}

\author{Shaun Cole} 
\affil{Institute for Computational Cosmology, Department of Physics, Durham University, South Road, Durham DH1 3LE, UK}

\author{David H. Weinberg} 
\affil{Department of Astronomy and the Center for Cosmology and Astroparticle Physics, The Ohio State University, 140 West 18th Avenue, Columbus, OH 43210, USA}

\author{John Moustakas} 
\affil{Department of Physics and Astronomy, Siena College, 515 Loudon Road, Loudonville, NY 12211, USA}

\author{Anthony Kremin}
\affil{Lawrence Berkeley National Laboratory, One Cyclotron Road, Berkeley CA 94720, USA}

\author{Jeremy L. Tinker}
\affil{Center for Cosmology and Particle Physics, Department of Physics, New York University, New York, NY, 10003, USA}

\author{Alex Smith}
\affil{Institute for Computational Cosmology, Department of Physics, Durham University, South Road, Durham DH1 3LE, UK}
\affil{IRFU, CEA, Universit\'e Paris-Saclay, F-91191 Gif-sur-Yvette, France}
\affil{Institute for Astronomy, University of Edinburgh, Royal Observatory, Blackford Hill, Edinburgh EH9 3HJ, UK}

\author{Risa H. Wechsler} 
\affil{Department of Physics, Stanford University, 382 Via Pueblo Mall, Stanford, CA 94305, USA}
\affil{Kavli Institute for Particle Astrophysics and Cosmology, Stanford University, Stanford, CA 94305, USA}
\affil{SLAC National Accelerator Laboratory, Menlo Park, CA 94025, USA}

%%%%%%%%%%%%%%%%%%%%%%%%%%%%%%%%%%%%%%%%%%%%%%%%%%%%%%%%%%%%%%%%%%%%%%%%%%%%%%%%%%%%
% Alphabetical Second tier 
%%%%%%%%%%%%%%%%%%%%%%%%%%%%%%%%%%%%%%%%%%%%%%%%%%%%%%%%%%%%%%%%%%%%%%%%%%%%%%%%%%%%
% e-mails: ahlen@bu.edu, salam@roe.ac.uk, stephenbailey@lbl.gov, david.brooks@ucl.ac.uk apcooper@gapp.nthu.edu.tw, tamarad@physics.uq.edu.au, kdawson@astro.utah.edu,
% arjun.dey@noirlab.edu, biprateep@pitt.edu, seftekharzadeh@usra.edu, deisenstein@cfa.harvard.edu, fanning.59@buckeyemail.osu.edu
% j.e.forero.romero@gmail.com, c.s.frenk@durham.ac.uk, gazta@ice.cat, satyagontcho@lbl.gov, jguy@lbl.gov, kh@physics.osu.edu
% kehoe@physics.smu.edtskisner@lbl.gov, mlandriau@lbl.gov, llg@lpnhe.in2p3.fr, melevi@lbl.gov, christophe.magneville@cea.fr, martini.10@osu.edu, aaron.meisner@noirlab.edu
% amyers14@uwyo.edu, jdnie@bao.ac.cn, peder.norberg@durham.ac.uk, nathalie.palanque-delabrouille@cea.fr, will.percival@uwaterloo.ca, clpoppett@lbl.gov,
% d.sierrap@uniandes.edu.co, fprada@iaa.es, ross.1333@osu.edu, sasha.safonova@yale.edu, csaulder@kasi.re.kr, schlafly1@llnl.gov, djschlegel@lbl.gov, gtarle@umich.edu, benjamin.weaver@noirlab.edu, 
% rwechsler@stanford.edu, pauline.zarrouk@cea.fr, zmzhou@bao.ac.cn, zouhu@nao.cas.cn, 

\author{Steven Ahlen}
\affil{Physics Dept., Boston University, 590 Commonwealth Avenue, Boston, MA 02215, USA}

\author{Shadab Alam}
\affil{Institute for Astronomy, University of Edinburgh, Royal Observatory, Blackford Hill, Edinburgh EH9 3HJ, UK}

\author{Stephen Bailey}
\affil{Lawrence Berkeley National Laboratory, One Cyclotron Road, Berkeley, CA 94720, USA}

\author{David Brooks}
\affil{Department of Physics \& Astronomy, University College London, Gower Street, London, WC1E 6BT, UK}

\author{Andrew P. Cooper}
\affil{Institute of Astronomy and Department of Physics, National Tsing Hua University, 101 Kuang-Fu Rd. Sec. 2, Hsinchu 30013, Taiwan}

\author{Tamara M.\ Davis}
\affil{School of Mathematics and Physics, University of Queensland, 4101, Australia}

\author{Kyle Dawson}
\affil{Department of Physics and Astronomy, The University of Utah, 115 South 1400 East, Salt Lake City, UT 84112, USA}

\author{Arjun Dey}
\affil{NSF’s NOIRLab, 950 N. Cherry Ave., Tucson, AZ 85719, USA}

\author{Biprateep Dey}
\affil{Department of Physics \& Astronomy and Pittsburgh Particle Physics, Astrophysics, and Cosmology Center (PITT PACC), University of Pittsburgh, 3941 O'Hara Street, Pittsburgh, PA 15260, USA}

\author{Sarah Eftekharzadeh}
\affil{Universities Space Research Association, NASA Ames Research Centre}

\author{Daniel J.\ Eisenstein}
\affil{Center for Astrophysics $|$ Harvard \& Smithsonian, 60 Garden Street, Cambridge, MA 02138, USA}

\author{Kevin Fanning}
\affil{Department of Physics, The Ohio State University, 191 West Woodruff Avenue, Columbus, OH 43210, USA}
\affil{Center for Cosmology and AstroParticle Physics, The Ohio State University, 191 West Woodruff Avenue, Columbus, OH 43210, USA}

\author{Jaime E. Forero-Romero} 
\affil{Departamento de F\'{i}sica, Universidad de los Andes, Cra. 1 No. 18A-10, Bogot\'{a}, Colombia}

\author{Carlos S. Frenk} 
\affil{Institute for Computational Cosmology, Department of Physics, Durham University, South Road, Durham DH1 3LE, UK}

\author{Enrique Gazta\~{n}aga}
\affil{Institut de C\`{i}encies de l'Espai, IEEC-CSIC, Campus UAB, Carrer de Can Magrans s/n, 08913 Bellaterra, Barcelona, Spain}

\author{Satya Gontcho A Gontcho}
\affil{Lawrence Berkeley National Laboratory, One Cyclotron Road, Berkeley, CA 94720, USA}
\affil{Department of Physics and Astronomy, University of Rochester, 500 Joseph C. Wilson Boulevard, Rochester, NY 14627, USA}

\author{Julien Guy}
\affil{Lawrence Berkeley National Laboratory, One Cyclotron Road, Berkeley, CA 94720, USA}

\author{Klaus Honscheid}
\affil{Department of Physics, The Ohio State University, 191 West Woodruff Avenue, Columbus, OH 43210, USA}

\author[0000-0002-6024-466X]{Mustapha Ishak}
\affiliation{Department of Physics, The University of Texas at Dallas, Richardson, TX, 75080, USA}

\author{St\'{e}phanie Juneau} 
\affil{NSF's National Optical-Infrared Astronomy Research Laboratory, 950 N. Cherry Avenue, Tucson, AZ, 85719, USA}

\author{Robert Kehoe} 
\affil{Department of Physics, Southern Methodist University, 3215 Daniel Avenue, Dallas, TX 75275, USA}

\author{Theodore Kisner}
\affil{Lawrence Berkeley National Laboratory, One Cyclotron Road, Berkeley, CA 94720, USA}

\author{Ting-Wen Lan}
\affil{Department of Physics, National Taiwan University, Taipei 10617, Taiwan}
\affil{Graduate Institute of Astrophysics, National Taiwan University, Taipei 10617, Taiwan}

\author{Martin Landriau}
\affil{Lawrence Berkeley National Laboratory, One Cyclotron Road, Berkeley, CA 94720, USA}

\author{Laurent Le Guillou}
\affil{Sorbonne Université, Université Paris Diderot, CNRS/IN2P3, Laboratoire de Physique Nucléaire et de Hautes Energies, LPNHE, 4 place Jussieu, F-75252 Paris, France}

\author{Michael E. Levi}
\affil{Lawrence Berkeley National Laboratory, One Cyclotron Road, Berkeley, CA 94720, USA}

\author{Christophe Magneville} 
\affil{CEA Saclay, IRFU F-91191 Gif-sur-Yvette, France}

\author{Paul Martini}
\affil{Center for Cosmology and AstroParticle Physics, The Ohio State University, 191 West Woodruff Avenue, Columbus, OH 43210, USA}
\affil{Department of Astronomy, The Ohio State University, 4055 McPherson Laboratory, 140 W 18th Avenue, Columbus, OH 43210, USA}

\author{Aaron Meisner}
\affil{NSF's National Optical-Infrared Astronomy Research Laboratory, 950 N. Cherry Avenue, Tucson, AZ 85719, USA}

\author{Adam D.\ Myers}
\affil{Department of Physics \& Astronomy, University of Wyoming, 1000 E. University, Dept.~3905, Laramie, WY 82071, USA}

\author{Jundan Nie}
\affil{National Astronomical Observatories, Chinese Academy of Sciences, A20 Datun Rd., Chaoyang District, Beijing, 100101, P.R. China}

\author{Peder Norberg} 
\affil{Institute for Computational Cosmology, Department of Physics, Durham University, South Road, Durham DH1 3LE, UK}
\affil{Center for Extragalactic Astronomy, Department of Physics, Durham University, South Road, Durham DH1 3LE, UK}

\author{Nathalie Palanque-Delabrouille}
\affil{Lawrence Berkeley National Laboratory, One Cyclotron Road, Berkeley, CA 94720, USA}
\affil{CEA Saclay, IRFU F-91191 Gif-sur-Yvette, France}

\author{Will J. Percival}
\affil{Waterloo Centre for Astrophysics, University of Waterloo, 200 University Ave W, Waterloo, ON N2L 3G1, Canada}
\affil{Department of Physics and Astronomy, University of Waterloo, 200 University Ave W, Waterloo, ON N2L 3G1, Canada}
\affil{Perimeter Institute for Theoretical Physics, 31 Caroline St. North, Waterloo, ON N2L 2Y5, Canada}

\author{Claire Poppett}
\affil{Lawrence Berkeley National Laboratory, One Cyclotron Road, Berkeley, CA 94720, USA }

\author{Francisco Prada}
\affil{Instituto de Astrof\'{i}sica de Andaluc\'{i}a (CSIC), Glorieta de la Astronom\'{i}a, s/n, E-18008 Granada, Spain}

\author{Anand Raichoor}
\affil{Lawrence Berkeley National Laboratory, One Cyclotron Road, Berkeley, CA 94720, USA}

\author{Ashley J. Ross}
\affil{Center for Cosmology and AstroParticle Physics, The Ohio State University, 191 West Woodruff Avenue, Columbus, OH 43210, USA}

\author{Sasha Safonova}
\affil{Physics Department, Yale University, P.O. Box 208120, New Haven, CT 06511, USA}

\author{Christoph Saulder}
\affil{Korea Astronomy and Space Science Institute, 776, Daedeokdae-ro, Yuseong-gu, Daejeon 34055, Republic of Korea}

\author{Eddie Schlafly}
\affil{Lawrence Livermore National Laboratory, P.O. Box 808 L-211, Livermore, CA 94551, USA}

\author{David Schlegel}
\affil{Lawrence Berkeley National Laboratory, One Cyclotron Road, Berkeley, CA 94720, USA}

\author{David Sierra-Porta} 
\affil{Facultad de Ciencias Básicas. Universidad Tecnológica de Bolivar. Cartagena 130010, Colombia.}
\affil{Departamento de F\'{i}sica, Universidad de los Andes, Cra. 1 No. 18A-10, Bogot\'{a}, Colombia}

\author{Gregory Tarle}
\affiliation{Department of Physics, University of Michigan, Ann Arbor, MI 48109, USA}

\author{Benjamin A. Weaver}
\affil{NSF's National Optical-Infrared Astronomy Research Laboratory, 950 N. Cherry Avenue, Tucson, AZ 85719, USA}

\author{Christophe Y\`{e}che} 
\affil{CEA Saclay, IRFU F-91191 Gif-sur-Yvette, France}

\author{Pauline Zarrouk} 
\affil{Sorbonne Université, Université Paris Diderot, CNRS/IN2P3, Laboratoire de Physique Nucléaire et de Hautes Energies, LPNHE, 4 place Jussieu, F-75252 Paris, France}

\author{Rongpu Zhou} 
\affil{Lawrence Berkeley National Laboratory, One Cyclotron Road, Berkeley CA 94720, USA}

\author{Zhimin Zhou}
\affil{National Astronomical Observatories, Chinese Academy of Sciences, A20 Datun Rd., Chaoyang District, Beijing, 100101, P.R. China}

\author{Hu Zou}
\affil{National Astronomical Observatories, Chinese Academy of Sciences, A20 Datun Rd., Chaoyang District, Beijing, 100101, P.R. China}

\begin{abstract}
    \noindent Over the next five years, the Dark Energy Spectroscopic Instrument (DESI) 
    will use 10 spectrographs equipped with 5000 fibers on the 4m Mayall
    Telescope at Kitt Peak National Observatory to conduct the first Stage-IV
    dark energy galaxy survey.
    At $z < 0.6$, the DESI Bright Galaxy Survey (BGS) will produce the most detailed map of the 
    Universe during the dark energy dominated epoch with redshifts of >10 million galaxies
    spanning 14,000 deg$^2$.  
    In this work, we present and validate the final BGS target selection and survey design.
    From DR9 of the Legacy Surveys, BGS will target: a $r < 19.5$ magnitude-limited
    sample (BGS Bright); a fainter $19.5 < r < 20.175$ sample, color-selected to have 
    high redshift efficiency (BGS Faint); and a smaller low-$z$ quasar sample (BGS AGN).
    BGS will observe these targets using exposure times that are dynamically scaled to 
    achieve homogeneous completeness and visit each point of the footprint three times on average.  
    We use early spectroscopic observations from the Survey Validation programs conducted 
    prior to the main survey along with realistic simulations to show that BGS 
    can successfully complete this strategy and make optimal use of `bright' time, when the
    moon is above the horizon.  
    Specifically, we demonstrate that BGS targets have stellar contamination below 1\% and
    that their densities do not depend strongly on imaging properties.
    We also confirm that BGS Bright will achieve >80\% fiber assignment efficiency.
    Finally, we show that the BGS Bright and Faint samples will achieve >95\% redshift success 
    rates across a broad range of galaxies, with no significant dependence on observing 
    conditions. 
    Overall, BGS meets the requirements necessary for an extensive range of scientific 
    applications. 
    BGS will yield the most precise Baryon Acoustic Oscillations and Redshift-Space 
    Distortions (RSD) measurements at $z < 0.4$ to date. 
    It also presents unique opportunities to exploit new methods that require highly
    complete and dense galaxy samples (\emph{e.g.} $N$-point statistics, multi-tracer RSD).  
    BGS further provides a powerful tool to study galaxy populations including dwarf 
    galaxies, galaxy groups and clusters, and the relations between  galaxies and dark 
    matter.
    %Accompanying papers detail the higher redshift DESI tracers that will complete the $\simeq{30}$ million DESI galaxy sample.
\end{abstract}

% introduction
\input{intro}

% imaging data  
\input{data}
% target selection 
\input{ts}
% survey design
\input{design}
% survey validation 
\input{sv}
% survey properties 
\input{svda}
% summary
\input{summary}
%acknowledgements
\hfill \break
\input{acknowledgements}

\appendix
\input{sv1}
\input{sky}
\bibliographystyle{mnras}
\bibliography{bgs_sv} 
\end{document}

%% file: intro.tex
\section{Introduction}
\label{sec:intro}
% what is DESI? 
The Dark Energy Spectroscopic Instrument~\citep[DESI\footnote{
We will use DESI to interchangeably refer to the instrument and the cosmological surveys conducted with it.};][]{desicollaboration2016}
is a spectroscopic galaxy survey designed to accurately measure cosmic acceleration and 
determine the nature of dark energy.
On seeing first light on October 22, 2019, DESI made a significant leap in becoming 
the first `Stage-IV'~\citep{albrecht2006} dark energy experiment to be realized.  
Observations are taken with the 4-meter Mayall telescope at Kitt Peak National Observatory with 
a focal plane filled with robotically-actuated fibers that simultaneously direct the light 
of 5000 galaxies to a  set of ten optical spectrographs ($360 < \lambda < 980$ nm with a 
spectral resolution of  $2000 <  \lambda/\Delta \lambda < 5500$). 

Over the next five years of operations, DESI will conduct the Bright Galaxy Survey (BGS) 
of more than 10 million galaxies over the redshift range $0 < z < 0.6$, alongside a 
dark-time redshift survey of 20 million luminous red galaxies (LRGs), emission line 
galaxies (ELGs), and quasars~\citep{desicollaboration2016}. 
From the three-dimensional spatial clustering of these galaxies and quasars, DESI will 
precisely measure the expansion history of the Universe using Baryon Acoustic 
Oscillations (BAO) and the growth of structure using Redshift-Space Distortions
(RSD).
With an expected footprint of 14,000 ${\rm deg}^2$ and longer redshift 
baseline, DESI will achieve a precision 1--2 orders of magnitude better than existing 
surveys~\citep{levi2013}.

% what is BGS? 
While the dark-time survey delivers informative constraints at high redshift,  
BGS will probe the epoch when dark energy becomes dominant and its impact is
expected to be greatest.  
BGS presents a unique opportunity to probe lower redshift regimes, where model 
predictions vary most strongly amongst theories of modified gravity and dark 
energy.
As an added benefit, BGS can proceed in some of the slowest observing conditions 
available, since galaxies are brighter at shorter distances, and optimize the
use of bright time.
The main BGS sample (BGS Bright) constitutes an $r < 19.5$ magnitude-limited sample 
that is an order-of-magnitude larger than that of the Sloan Digital Sky Survey I and II
combined~\citep[SDSS;][]{york2000}.  
A sample of fainter $19.5 < r < 20.175$ galaxies (BGS Faint), color-selected to
achieve high redshift efficiency, ensures BGS will be the highest density DESI
tracer and deliver the highest fidelity measurement of the density field.
Meanwhile, the lower redshift quasar sample~\citep[BGS AGN;][]{bgs_agn} ensures 
the quasar sample obtained by DESI is as complete as possible.  
With these samples, BGS will provide a galaxy sample that will be more than ten times 
the density of the LOWZ  SDSS-III~\citep{eisenstein2011} Baryon Oscillation 
Spectroscopic Survey~\citep[BOSS;][]{dawson2013} and up to two magnitudes fainter than 
the SDSS Main Galaxy Survey~\citep[MGS;][]{strauss2002}, with double the median redshift 
($z \approx 0.2$).  
%As the total BGS target density ($\simeq 1400$ deg$^{-2}$) is insufficient to  fully populate the DESI focal plane, stellar Milky Way Survey (MWS) targets are  assigned  at lower priority than BGS, as described in \todo{Cooper~\etal~(in prep.)}. 

% BGS main science goals 
For dark energy science, BGS will achieve the best measurement of BAO and RSD at
redshifts $z \leq 0.4$ to date. 
The lower redshift of BGS will provide maximum leverage against
high-redshift measurements and constraints from cosmic microwave background
(CMB) observations. 
BGS will also enable new approaches to measuring the growth of structure, 
including the use of galaxy groups and clusters~\citep{Mohammad16}, 
galaxy--galaxy weak lensing~\citep{DES3x2,Heymans21,Miyatake21,Amon22}, higher-order
statistics~\citep{gil-marin2017}, small-scale clustering~\citep{zhai2019}, 
and a range of complementary techniques.  
Its exceptionally high sampling density and wide selection also opens the door
to innovative techniques including: `multi-tracer' methods that exploit galaxy
populations with different clustering
properties~\citep[\emph{e.g.}][]{mcdonald2009, seljak2009, blake2013, wang2020}
and methods that forward model the small-scale density 
field~\citep[\emph{e.g.}][]{Seljak17}.
More rigorous tests of systematic effects based on, for instance, splits 
by galaxy type~\citep{ross2014} will also be possible with BGS. 
Beyond dark energy, BGS will also enable novel tests of modified gravity theories 
using the velocity fields of cluster infall regions~\citep{zu2014} and precise
measurements of the sum of neutrino masses with higher-order 
clustering~\citep{hahn2020b} and other statistics.

% BGS secondary science goals 
The BGS will also provide an extraordinary resource for advancing our
understanding of the physical processes that drive galaxy formation and
 evolution. 
With spectra and photometry for each of its >10 million galaxies, BGS will enable 
measurements of their physical properties~\citep{hahn2022} 
and provide the most precise measurements of galaxy property statistics
at low redshifts: \emph{e.g.} the stellar mass function~\citep{li2009,
moustakas2013}, star-forming sequence~\citep{noeske2007}, and the mass--metallicity
relation~\citep{tremonti2004}. 
It will also enable studies of galaxy groups~\citep{eke2004}, the connection 
between galaxies and dark matter~\citep{tinker2011, zu2015, xu2018},
and the environment-dependent luminosity 
function~\citep{McNaught-Roberts14,Eardley15,Fang19}. 
By extending to faint apparent magnitudes $r < 20.175$, which corresponds
to galaxies of stellar mass ${\sim}10^7M_\odot$ at $z < 0.025$~\citep{hahn2022}, 
BGS will also be an unprecedented sample for studying dwarf galaxies~\citep{mao2021}.

% BGS requirements to achieve above goals 
To achieve the broad range of science goals with BGS, we establish the
following requirements for the survey. 
BGS will 
(a) sample a wide range of galaxy types, 
(b) have a high and well-characterized completeness, and
(c) be at least an order of magnitude larger than the largest comparable
survey that exists today (SDSS MGS; $10^6$ galaxies).
For (a), we require the BGS Bright sample to be selected using a magnitude
limit and to have high target density, above >800 targets/deg$^2$.
For (b) and (c), we require that BGS targets have $<1\%$ stellar
contamination rates, that $>80\%$ of them are assigned to a fiber, and that 
$>95\%$ of those assigned produce successful redshifts. 
This ensures that we successfully assign fibers to and measure redshifts
for the vast majority of galaxies targeted.
Lastly, we require a 20\% margin in the forecasted BGS operations to ensure
that DESI will be able to complete the survey even with unforeseen events.

% what is survey validation phase? 
In this paper, we present the final target selection, design, and validation 
of the BGS. 
We demonstrate that the choices we make for BGS will achieve all of the stated
requirements and can be executed in the bright time available over the 5 years
of DESI operations.
This work presents significant updates and new results from preliminary versions
of the BGS target selection presented by~\cite{ruiz-macias2021}
and~\cite{zarrouk2021}.
These advancements are based on early spectroscopic observations from the
Survey Validation (SV) programs conducted by DESI prior to the main survey. 
SV was conducted between December 2020 and May 2021 with the primary goal of
verifying that each survey meets and exceeds its requirements. 
Complementary papers describe the target selection for the dark-time
DESI tracers (LRGs;~\citealt{lrg},  ELGs;~\citealt{elg}, and
QSOs;~\citealt{qso}) and the Milky Way Survey (MWS;~\citealt{mws}). 
\cite{desitarget} presents how the target selections are  implemented in DESI.
\cite{sv} presents an overview of the DESI spectroscopic 
observations.
\cite{vigal} and \cite{viqso} present the visual inspection for 
the galaxies (BGS, LRG, ELG) and QSO targets, respectively, 
used to construct spectroscopic truth tables.

In Section~\ref{sec:data}, we describe the latest imaging data from the Legacy
Surveys~\citep{dey2019} and external catalogs that are used for the BGS target
selection.
We describe the specific selection criteria for each of the BGS samples
in Section~\ref{sec:ts}. 
In Section~\ref{sec:design}, we describe and explain the details of the
BGS survey design and observing strategy.
We describe the SV programs in detail in Section~\ref{sec:sv_obs} and present the
validation of the target selection (Section~\ref{sec:valid_cuts}), redshift success 
(Section~\ref{sec:zsuccess}), and fiber assignment efficiency
(Section~\ref{sec:fibeff}). 
Finally, we showcase the BGS samples for the first public dataset in Section~\ref{sec:survey} 
and summarize our conclusions in Section~\ref{sec:summary}. 
We assume AB magnitudes and a flat $\Lambda$CDM cosmology described by the final 
{\it Planck} results \citep{planckcosmo}.

%% file: data.tex
\section{Imaging Data} \label{sec:data} 
The target selection for BGS is primarily based on the imaging data from the
Legacy Surveys (LS).
We also make use of observations from a number of external catalogs derived
from \gaia~Data Release 2~\citep{collaboration2016}, Tycho-2~\citep{hog2000},
and the Siena Galaxy Atlas~\citep[SGA;][]{moustakas2021, sga}. 
In the following we briefly describe each of these imaging surveys.
\begin{figure}
\begin{center}
    \includegraphics[width=\textwidth]{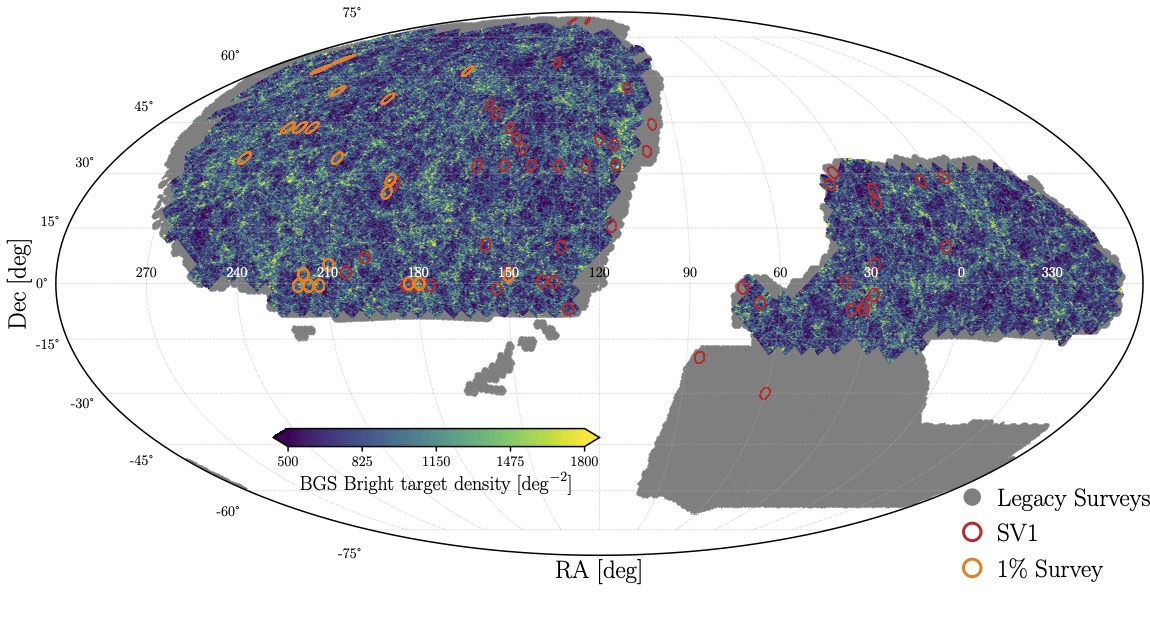}
    \caption{ \label{fig:footprint}
        The 14,000 ${\rm deg}^2$ footprint for the DESI Bright Galaxy Survey
        (color map).  
        Imaging from the Legacy Surveys DR9 allows the selection of BGS targets
        over a larger area, approximately 20,000 deg$^2$ (gray).  
        The color map represents the density of BGS Bright targets (Section~\ref{sec:select}).
        %Based on survey progress, the footprint may expand from the minimum 9000 deg$^2$ required, up to 14,000 deg$^2$.
        During the Survey Validation phases of operations, DESI observed SV1
        and the \sviii~to optimize and validate the BGS target selection and
        survey performance.
        We mark the tiles observed during SV1 and the \sviii~in red and orange.
    }
\end{center}
\end{figure}

%%%%%%%%%%%%%%%%%%%%%%%%%%%%%%%%%%%%%%%%%%%%%%%%%%%%%%%%%%%%
\subsection{Legacy Survey Data Release 9} \label{sec:legacy}
The DESI Legacy Imaging
Surveys\footnote{\url{https://www.legacysurvey.org/dr9/}} (also known as the
Legacy Surveys; LS) comprise the primary photometric catalog that we use for our
target selection~\citep{dey2019}. 
LS covers $\sim$14,000 ${\rm deg}^2$ of the extragalactic sky visible from
the Northern hemisphere split into two contiguous areas by the Galactic plane, 
together with $\sim$6,000 deg$^2$ of the Southern sky.  
It provides photometry with the necessary coverage, depth, and target
density for DESI. 
For BGS we use the ninth data release: LS DR9~\citep{dr9}.
In Figure~\ref{fig:footprint}, we present the footprint of the LS imaging in
gray. 

LS provides photometry in three optical bands, $g$ ($470~\rm{nm}$), $r$
($623~\rm{nm}$), and $z$ ($913~\rm{nm}$), and is observed using three independent
programs: the Beijing-Arizona Sky Survey~\citep[BASS;][]{zou2017}, the Mayall $z$-band Legacy
Survey (MzLS), and the Dark Energy Camera Legacy Survey (DECaLS). 
BASS observed $g$ and $r$-band photometry using the 2.3m Bok telescope over
$\sim$5,100 ${\rm deg}^{2}$ of the Northern Galactic cap (NGC) above Dec >
32.375 deg. 
MzLS observed $z$-band photometry over the same footprint as BASS using the 4m
Mayall telescope. 
Lastly, DECaLS observed $g$, $r$, and $z$-band photometry over the rest of the
LS footprint using the Dark Energy Camera~\citep[DECam;][]{flaugher2015} on 
the 4m Blanco telescope. 
DECaLS expands on the footprint of the Dark Energy Survey~\citep[DES;][]{des2016}, 
for which DECam was initially built, using publicly available DECam time. 
The optical photometry from these surveys are complemented by infrared
Wide-field Infrared Survey Explorer~\citep[WISE;][]{wright2010} $W1$
($3.4~\rm{\mu m}$) and $W2$ ($4.6 ~\rm{\mu m}$) photometry from the custom
``unWISE'' catalog~\citep{meisner2017}. 
The infrared photometry is derived from all WISE imaging through year 6 of the
NEOWISE Reactivation, force-photometered at the locations of the LS optical
sources~\citep{meisner2017a}.

% description of {\sc LegacyPipe} and TRACTOR
All data from LS are first reduced using the National Optical-Infrared
Astronomy Research Laboratory (NOIRLab) DECam Community
Pipline\footnote{\url{https://legacy.noirlab.edu/noao/staff/fvaldes/CPDocPrelim/PL201_3.html}},
which provides instrumental calibration, astrometric calibration, photometric
characterization, and masking of various artifacts.   
Afterwards, the LS source catalog is constructed using the 
{\sc legacypipe}\footnote{\url{https://github.com/legacysurvey/legacypipe}} 
pipeline, which uses the
{\sc Tractor}\footnote{\url{http://thetractor.org/doc/}} software~\citep{lang2016} 
for pixel-level forward modeling of astronomical sources.
{\sc legacypipe} initially detects sources and defines contiguous sets of
pixels associated with each detection.
{\sc Tractor} then fits these pixels with surface brightness models (\emph{e.g.}
point source, de Vaucouleurs galaxy profile) on individual optical images ($g$,
$r$, and $z$ bands), taking into account their different PSF and sensitivity. 
Based on a penalized $\chi^2$, {\sc legacypipe} then determines which surface
brightness model best describes the light profile and whether to keep the
source in the catalog.

The resulting LS source catalog includes source positions, fluxes, $r$ band
fiber flux, and measures of the quality of the source fits.  
Fiber flux represents the predicted flux within a 1.5\arcsec\ diameter fiber
aperture in Gaussian seeing with a Full Width at Half Maximum (FWHM) of
1\arcsec. 
It also provides galactic extinction measurements derived from the
\cite{schlegel1997} (SFD98) maps. 
For a fiducial galaxy target, defined as a source with an exponential profile
with half-light radius of 0.45\arcsec, LS achieves a median $5\sigma$ detection limit
in the $g$, $r$, and $z$ bands of 23.72, 23.27, and 22.22 magnitudes  over the
DECaLS coverage and 23.48, 22.87, and 22.29 magnitudes for BASS/MzLS. 
For more details on the LS imaging data, we refer readers to \cite{zou2017}, 
\cite{dey2019} and \cite{dr9}.

In addition to the small differences in the detection limits between DECaLS
and BASS, there are also slight differences in their measured magnitudes
due to the fact that they were observed using different instruments at
different telescopes. 
\cite{zarrouk2021} quantified this discrepancy in detail using some of the
overlapping region between DECaLS and BASS in the NGC over the range 29 <
Dec < 35 deg. 
Overall, the same objects are slightly brighter in the $r$-band in DECaLS
versus BASS. 
To account for this discrepancy, we impose an $r$-band magnitude offset of
$\Delta r = r_{\rm BASS} - r_{\rm DECaLS} = 0.04$ mag in our target selection. 
We opt for a simple magnitude offset, instead of a color-dependent one, because
it sufficiently accounts for the discrepancy, without imposing a single
color correction on galaxies over the entire broad $r$ range. 
The amplitude of the offset is determined so that the average target density
of BGS Bright targets (Section~\ref{sec:bgs_bright}) in BASS is equal to the
average target density in DECaLS. 
For more details on the magnitude offsets between DECaLS and BASS, we refer
readers to \cite{zarrouk2021}.

\subsection{External Catalogs} \label{sec:secondary}
In addition to the LS imaging, we use additional catalogs in the BGS sample
selection for star-galaxy separation and spatial masking. 
We describe these catalogs below. 

\subsubsection{Gaia Data Release 2} \label{sec:gaia}
\gaia~is a European Space Agency space-based mission launched in
2013 with the goal of observing $\approx 1\%$ of all the stars in
the Milky Way~\citep{collaboration2016}. 
In addition to accurate positions and proper motions of these stars,
\gaia~provides photometry in the $G$ band, which covers the wavelength range
330 - 1050 nm~\citep[hereafter $G_{Gaia}$;][]{carrasco2016}. 
The \gaia~Data Release 2~\citep[DR2;][]{collaboration2018}, which covers 22
months of observations and was released on April 2018, provides observations
for 1.7 billion stars over the entire sky down to $G_{Gaia} = 20.7$. 
%\todo{NECESSARY AT ALL? Perhaps just need to say that sources in the LS are
%    matched to Gaia, or be specific about how Gaia is used: 
%    Data from \gaia~has been extensively used in the LS for \emph{e.g.}
%    astronometric calibration, proper motions and bright star masking. 
%}
Since \gaia~is sufficiently deep to detect stars that contaminate the BGS
sample, we use it in our star-galaxy separation criteria (Section~\ref{sec:select}).

\subsubsection{Tycho-2} \label{sec:tycho}
Bright stars can often impact the measured photometric properties of nearby
sources or cause spurious sources to be detected in imaging catalogs.
We therefore exclude regions around bright stars in our sample selection to
mitigate these effects (Section~\ref{sec:select}). 
For our list of bright stars, in addition to \gaia, we use the Tycho-2 
catalog~\citep{hog2000}, which contains the positions, proper motions, and
photometry for 2.5 million of the brightest stars in the Milky Way. 

\subsubsection{Globular Clusters and Planetary Nebulae} \label{sec:gc}
The bright and extended profiles of globular clusters (GCs) and planetary
nebulae (PNe) can impact our selection of extragalactic sources near them. 
We therefore mask the regions surrounding them in our sample selection 
(see Section~\ref{sec:select} for details). 
For our catalog of GCs and PNe, we select all objects classified as GCs or PNe
in the OpenNGC
catalog\footnote{\href{https://github.com/mattiaverga/OpenNGC}{https://github.com/mattiaverga/OpenNGC}}.
We also include nine additional GCs and compact open clusters from the
literature as well as two Local Group galaxies, Fornax and Sculptor. 
For further details on our list of GCs and PNe, we refer readers to the LS DR9
documentation\footnote{\href{https://www.legacysurvey.org/dr9/external/}{https://www.legacysurvey.org/dr9/external/}}
and~\cite{dr9}. 

\subsubsection{Siena Galaxy Atlas} \label{sec:sga} 
Images of large galaxies can often generate spurious sources in standard 
photometric pipelines.
For instance, HII regions within a galaxy can be mistaken for individual 
sources. 
The outskirts of large galaxies can also be fragmented into spruious sources.  
To mitigate this contamination, we use the Siena Galaxy 
Atlas~\citep[SGA;][]{moustakas2021}\footnote{\href{https://sga.legacysurvey.org/}{https://sga.legacysurvey.org/}}
to select the largest galaxies in the LS. 
Using optical data from the HyperLeda catalog~\citep{makarov2014} and infrared 
data from the ALLWISE catalog~\citep{secrest2015}, the SGA identified large
galaxies with $D(25) > 20\arcsec$, where $D(25)$ is the diameter at the 25 magnitude
arcsec$^{-2}$ surface brightness isophote, a conventional measure of galaxy size.
A separate source extraction was performed in regions around those galaxies in DR9. 
In total, the SGA catalog includes 383,620 galaxies with DR9 $grz$ photometry.

%% file: ts.tex
\section{Target Selection} \label{sec:ts}
Our ultimate goal is to construct BGS to have a highly complete, high
density sample of galaxies with robust redshift measurements that meets the
DESI science requirements.  
This requires a reliable input target catalog derived from the LS and
external catalogs.
In this section, we describe how we construct this input target catalog for
BGS (Section~\ref{sec:select}) and how we select targets from it for the
BGS Bright, Faint, and AGN samples (Sections~\ref{sec:bgs_bright},
~\ref{sec:bgs_faint}, and~\ref{sec:bgs_agn}). 
For additional details on the target selection, we refer readers to
the DESI target selection pipeline paper~\citep{desitarget}.
The target catalog will be publicly available at 
\url{https://data.desi.lbl.gov/public/ets/target/catalogs/}. 

\subsection{Selection Cuts} \label{sec:select}
We design the target catalog to ensure high efficiency and completeness for
the redshift survey and remove any systematic effects that can affect
galaxy clustering analyses. 
To achieve this, we minimize the number of spurious objects and
impose spatial masking, star-galaxy separation, a fiber-magnitude cut, a
bright limit, and quality cuts on objects compiled from the LS and external
catalogs. 
We describe each of these selection cuts below.
\vspace{1mm}

%%%%%%%%%%%%%%%%%%%%%%%%%%%%%%%%%%%%%%%%%%%%%%%%%%%%%%%%%%%%%%%%%%%%%%%%%%%%%%%%
% spatial masking
%%%%%%%%%%%%%%%%%%%%%%%%%%%%%%%%%%%%%%%%%%%%%%%%%%%%%%%%%%%%%%%%%%%%%%%%%%%%%%%%
\noindent\uline{\emph{Spatial Masking}}: 
We want to mask out regions of the sky surrounding bright stars, GCs, and PNe
because these regions are typically contaminated by features such as extended
halos, bleed trails, and diffraction spikes. 
These features in the imaging can not only compromise the photometry of
surrounding objects but also produce spurious objects.  
For the BGS target catalog, we apply spatial masking around bright stars and
globular clusters. 
First, for bright stars, we apply circular masks with a magnitude dependent 
radius, $R_{\rm BS}$.
The masks account for the fact that  {\sc Tractor} underestimates 
galaxy fluxes near bright stars and, thus, reduce the target density 
in these regions. 
The masks are compiled using 773,673 \gaia~DR2  (Section~\ref{sec:gaia})  
objects with $G_{Gaia} < 13$ and  3,349 Tycho-2 (Section~\ref{sec:tycho}) 
objects with visual magnitude brighter than $\mathtt{MAG\_VT}$ < 13. 
The masks have radii
%\todo{check 39.3 number below}
\begin{equation}
    R_{\rm BS}(m) = 815 \times 1.396^{-m}~{\rm arcsec}
    %\begin{cases}
    %    39.3 \times 2.5^{(11-m)/3} & m > 2.9 \\ 
    %    471.6\arcsec & m < 2.9
    %\end{cases}
\end{equation}
where $m$ is either $G_{Gaia}$ or Tycho-2 $\mathtt{MAG\_VT}$ magnitude. 
We use $G_{Gaia}$ when both are available. 
We do not apply spatial masking around stars fainter than 13$^{\rm th}$
magnitude.
Next, for GCs, we apply a circular mask with radius defined by the major axis of 
the object.
We apply this masking around all of the GCs and PNe in the list compiled from 
the OpenNGC catalog (Section~\ref{sec:gc}). 
In total, our bright star and GC masks exclude 0.87 and 0.01\% of the initial area, 
respectively.

We note that we do not apply spatial masking around large galaxies of the
SGA~(Section~\ref{sec:sga}), unlike in the preliminary version of the target 
selection~\citep{ruiz-macias2021}. 
Custom source fitting around the SGA galaxies in DR9 led to a significant
improvement. 
Fits no longer automatically assume that the sources within $D(25)$ are best-fit by
a PSF model and, thus, more accurately measure their photometry.  
Since $\sim 40-50\%$ of sources within the large galaxy mask are galaxies
based on visual inspection and GAMA spectroscopy of overlapping galaxies~\citep{zarrouk2021},
we opt to reject spurious objects at a later stage. 
\vspace{1mm}
%%%%%%%%%%%%%%
\begin{figure}
\begin{center}
    \includegraphics[width=0.95\textwidth]{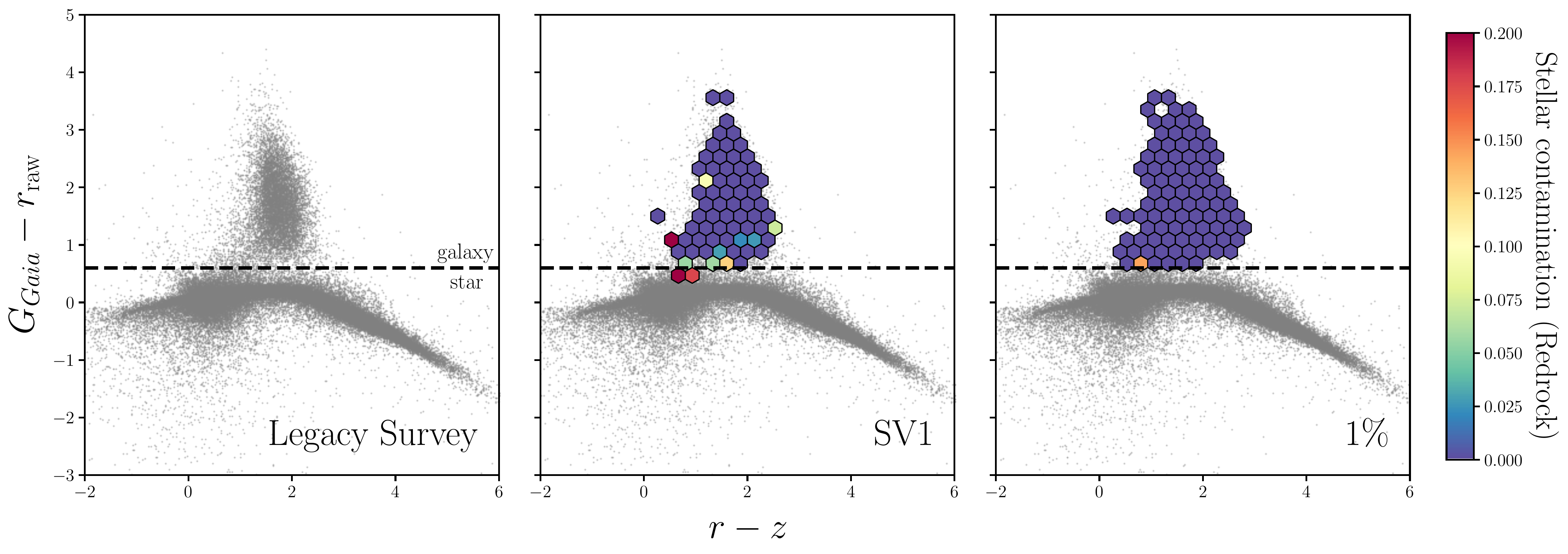}
    \caption{ \label{fig:stargalaxy}
    Star-galaxy separation in BGS is performed using a $G_{Gaia} - r_{\rm raw}$ cut. 
    This criterion exploits the fact that the \gaia~magnitude is measured from
    space with a diffraction-limited PSF while the LS $r$ magnitude captures
    the light from the entire source. 
    LS objects (grey) with $G_{Gaia} - r_{\rm raw} > 0.6$ (black dashed) or
    objects not in \gaia~are classified as galaxies. 
    $G_{Gaia}$ is the $G$-band magnitude from \gaia~DR2; $r_{\rm raw}$ is the
    LS $r$-band magnitude without Galactic extinction correction. 
    In the center and right panels, we present in the color maps the stellar
    contamination fraction for \svi~and the \sviii~respectively.
    The contamination fraction is estimated based on {\sc Redrock} spectral type
    classification and a minimum redshift cut, $z > 300$ km/s. 
    We only include hexbins with at least 10 BGS targets. 
    \emph{Stellar contamination is negligible ($<1\%$) above our star-galaxy 
    separation threshold.}
    }
\end{center}
\end{figure}

%%%%%%%%%%%%%%%%%%%%%%%%%%%%%%%%%%%%%%%%%%%%%%%%%%%%%%%%%%%%%%%%%%%%%%%%%%%%%%%%%%%%%%%%%%% 
% star - galaxy separation 
%%%%%%%%%%%%%%%%%%%%%%%%%%%%%%%%%%%%%%%%%%%%%%%%%%%%%%%%%%%%%%%%%%%%%%%%%%%%%%%%%%%%%%%%%%% 
\noindent\uline{\emph{Star-galaxy separation}}: 
For the BGS target catalog, we only want to include galaxies and exclude stars.  
To classify LS objects as either stars or galaxies, we use a combination of the photometric
data from LS and \gaia~DR2 (Section~\ref{sec:gaia}).
An object is considered a BGS target if either of the following conditions are met:
\begin{enumerate}
    \item object is not in the \gaia~catalog;
    \item object is in \gaia~and has $(G_{Gaia} - r_{\rm raw}) > 0.6$.
\end{enumerate}
$G_{Gaia}$ is the $G$ band magnitude from \gaia \  and $r_{\rm raw}$ is the LS $r$ band 
magnitude that is \emph{not} corrected for galactic extinction. 
Our $(G_{Gaia} - r_{\rm raw})$ criterion takes advantages of the fact that $G_{Gaia}$ 
is measured assuming that the object is a point source for a (narrow) 
diffraction-limited point spread function (PSF) measured from space.
Hence, $G_{Gaia}$ will be systematically fainter than the {\sc Tractor} magnitudes for 
galaxies, which capture light from the entire galaxy.
We use $r_{\rm raw}$ magnitudes because $G_{Gaia}$ is also uncorrected for galactic
extinction.
Our star-galaxy separation has a significantly lower stellar contamination rate than the 
{\sc Tractor} model classifications~\citep{ruiz-macias2021}.
% shaun.cole: At face value this sounds contradictory, i.e. if we agree well with the TRACTOR classifications how do we get lower contamination., but it is all in the numbers,  From the numbers in Fig 5 of the Ruiz paper for the objects in GAIA, TRACTOR and the GAIA classification only disagree 0.12% of the time but this still reduces the stellar contamination (in this magnitude range) by 1.64 objs/sq deg which is 1.7% as there are only 95.1 gals per sq deg at this magnitude limit.

In the left panel of Figure~\ref{fig:stargalaxy}, we present the $(r-z)$ versus 
$(G_{Gaia} - r_{\rm raw})$ distribution for LS objects (grey) and highlight our 
star-galaxy separation criterion (black dashed). 
Figure~\ref{fig:stargalaxy} clearly reveals the locus of galaxies in the color
distribution above our $(G_{Gaia} - r_{\rm raw}) > 0.6$ star-galaxy separation 
criterion.
\vspace{1mm}

\begin{figure}
\begin{center}
    \includegraphics[width=0.8\textwidth]{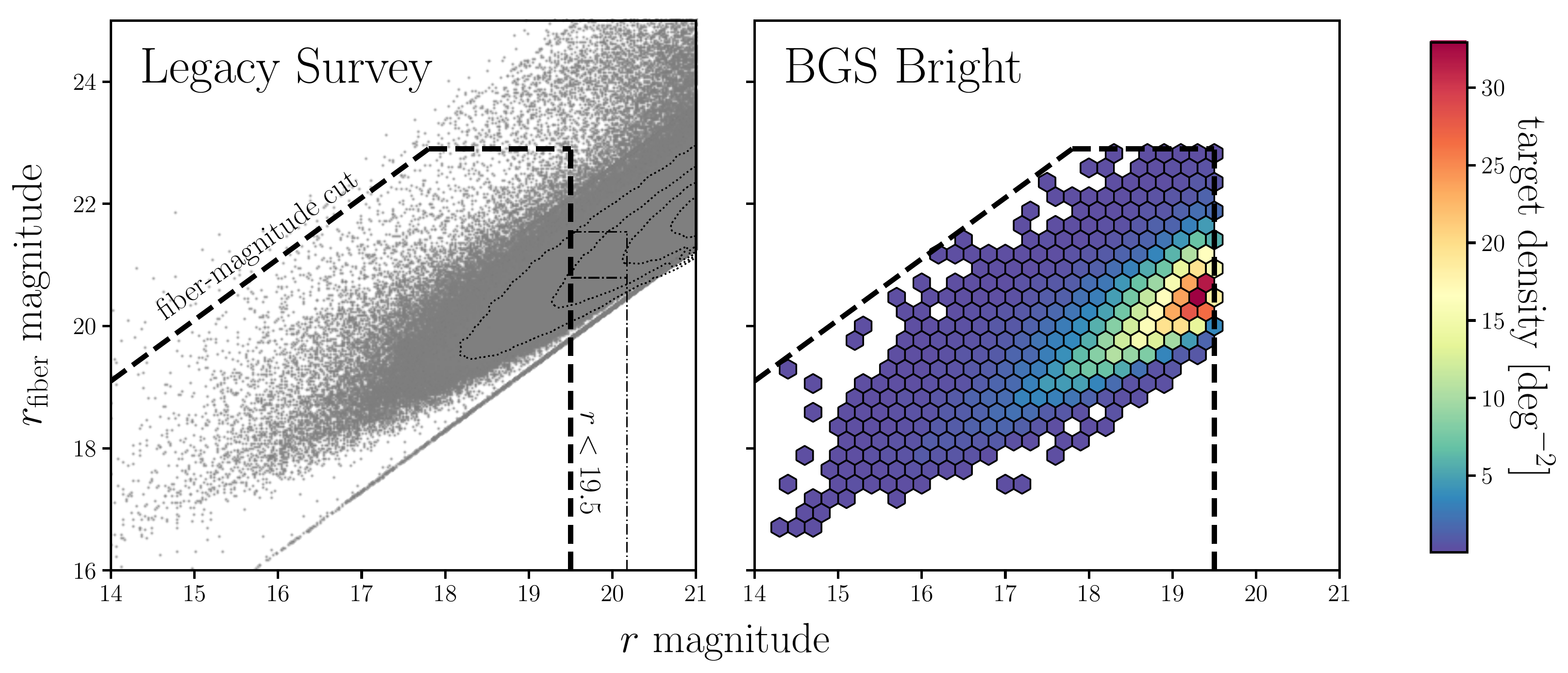}
    \caption{ \label{fig:bgsbright}
    Targets for the BGS Bright sample are chosen based on the selection cuts 
    described in Section~\ref{sec:select} and a $r < 19.5$ magnitude cut.
    In the left panel, we show these cuts (based on fiber magnitude) and the 
    $r < 19.5$ cut (black dashed) on the distribution of $r$ versus $r_{\rm
    fiber}$ magnitude for LS objects that pass our star-galaxy selection (grey). 
    The contours mark the 11.7, 39.3, 67.5, and 86.4 percentiles of the distribution (dotted). 
    We also include the $r$ and $r_{\rm fiber}$ cuts for the BGS Faint sample (dot dashed). 
    We impose selection cuts on BGS targets in order to  minimize the number of spurious
    objects and mitigate any systematic effects that can affect galaxy clustering 
    analyses.
    In the right panel, we present the target density of the BGS Bright targets (color 
    map).
    In total, we have 864 targets/${\rm deg}^{2}$ for the BGS Bright sample.}
\end{center}
\end{figure}

%%%%%%%%%%%%%%%%%%%%%%%%%%%%%%%%%%%%%%%%%%%%%%%%%%%%%%%%%%%%%%%%%%%%%%%%%%%%%%%%%%%%%%%%%%% 
% fiber magnitude cut
%%%%%%%%%%%%%%%%%%%%%%%%%%%%%%%%%%%%%%%%%%%%%%%%%%%%%%%%%%%%%%%%%%%%%%%%%%%%%%%%%%%%%%%%%%% 
\noindent\uline{\emph{Fiber magnitude cut}}: 
Some of the objects in the LS are imaging artifacts or fragments of `shredded' galaxies. 
In order to remove these spurious objects from our BGS target catalog, we apply the fiber 
magnitude cut (FMC) below:  
\begin{equation} \label{eq:fmc}
    r_{\rm fiber} < 
    \begin{cases}
        22.9 + (r - 17.8) & {\rm for}~r < 17.8 \\
        22.9 & {\rm for}~17.8 < r < 20.
    \end{cases}
\end{equation}
$r_{\rm fiber}$ is the $r$-band fiber magnitude derived from the predicted $r$-band flux of the 
object within a 1.5\arcsec diameter fiber and $r$ is the total $r$-band magnitude. 
In the left panel of Figure~\ref{fig:bgsbright}, we present our fiber magnitude cut (black dashed)
in the $r$ versus $r_{\rm fiber}$ distribution of LS objects.  
The bright end of the FMC is set so that it does not remove any
spectroscopically confirmed GAMA galaxies.
Meanwhile, the faint end of the FMC is determined through visual
inspection of LS objects.
The $r_{\rm fiber} < 22.9$ limit rejects spurious objects and retains
genuine galaxies.  
9.2 and 19.3 objects/deg$^{2}$ are removed by the FMC in DECALS and BASS/M$z$LS, respectively.
\vspace{1mm}

%%%%%%%%%%%%%%%%%%%%%%%%%%%%%%%%%%%%%%%%%%%%%%%%%%%%%%%%%%%%%%%%%%%%%%%%%%%%%%%%%%%%%%%%%%% 
% bright limit
%%%%%%%%%%%%%%%%%%%%%%%%%%%%%%%%%%%%%%%%%%%%%%%%%%%%%%%%%%%%%%%%%%%%%%%%%%%%%%%%%%%%%%%%%%% 
\noindent\uline{\emph{Quality cuts}}: 
We want BGS Bright to be complete in all three optical bands of its imaging: $g$, $r$, and $z$. 
We therefore require that there is at least one photometric observations in each of the bands: 
\begin{equation} \label{eq:nobs}
    {\rm nobs}_i  > 0 \quad{\rm for}~i = g, r, z. 
\end{equation}
${\rm nobs}_i$ represents the number of observations (images) at the central
pixel of the source in each band. 
This requirement removes 0.41\% of the imaging footprint. 
    
We also exclude spurious objects (\emph{e.g.}~imaging artifacts or stars) with
extreme colors by requiring: 
\begin{align} \label{eq:quality_color}
    -1 &< (g - r) < 4\\
    -1 &< (r - z) < 4.
\end{align}
The color cuts remove 3.44 and 25.7 objects/deg$^{2}$ for DECALS and BASS/M$z$LS 
respectively. 
A large fraction of these objects are the same as those removed by the FMC.
We do not use any cuts based on {\sc Tractor} photometric quality
flags (unlike the preliminary \citealt{ruiz-macias2021} selection), 
as they are not necessary for DR9.  
\vspace{1mm}

%%%%%%%%%%%%%%%%%%%%%%%%%%%%%%%%%%%%%%%%%%%%%%%%%%%%%%%%%%%%%%%%%%%%%%%%%%%%%%%%
% bright limit
%%%%%%%%%%%%%%%%%%%%%%%%%%%%%%%%%%%%%%%%%%%%%%%%%%%%%%%%%%%%%%%%%%%%%%%%%%%%%%%%
\noindent\uline{\emph{Bright Limit}}: 
Lastly, flux from very bright objects on neighboring fibers can contaminate
the traces of faint objects on the spectrograph CCD, and pollute their
observed fluxes. 
This is particularly problematic at $>9000$\AA, where 10\% of the flux of
the contaminating source is scattered into the wings of its PSF.
Since the faintest fiber magnitudes in BGS have $r_{\rm fiber} \simeq 21.5$, we
remove all objects that meet 
\begin{equation} \label{eq:bright_limit}
    (r > 12)~\&~(r_{\rm fibertot} < 15) 
\end{equation}
from the BGS target catalog.  
Here, $r_{\rm fibertot}$ is the total fiber magnitude derived from the
predicted $r$-band flux within a 1.5\arcsec diameter fiber from all 
sources\footnote{This ensures that if a bright object is modeled by multiple 
sources (shredding), then the intrinsic source will still be masked.}.
Most of the bright objects rejected by this threshold are stars and saturated
point-like sources according to visual inspection. 
Based on similar concerns, MWS also applies a cut to the Milky Way sources
sharing the focal plane~\citep{mws}.  

%%%%%%%%%%%%%%%%%%%%%%%%%%%%%%
\subsection{BGS Bright Sample} \label{sec:bgs_bright}
The highest priority target class in BGS is a magnitude-limited `Bright' sample. 
From the sources that satisfy the selection cuts above, we impose an $r < 19.5$
magnitude limit to select the BGS Bright targets.  
With this simple selection, BGS Bright can achieve a range of scientific goals 
that require a dense sampling of galaxies and a wide range of galaxy properties. 
In the right panel of Figure~\ref{fig:bgsbright}, we present the target density
of the BGS Bright sample as a function of $r$ and $r_{\rm fiber}$ magnitudes.
We represent the target density in each $(r, r_{\rm fiber})$ hex bin with the 
color map and mark the $r < 19.5$ magnitude limit (black dashed).
We note the locus of targets at low $r_{\rm fiber}$ with tightly correlated 
$r$ and $r_{\rm fiber}$  are a small fraction of targets that are likely 
stellar contaminants. 
In total, we have 864 targets/${\rm deg}^{2}$ for the BGS Bright sample. 
These targets can be accessed in DESI catalogs as described in
\cite{desitarget} under the $\mathtt{BGS\_BRIGHT}$ bit-name.

\begin{figure}
\begin{center}
    \includegraphics[width=0.8\textwidth]{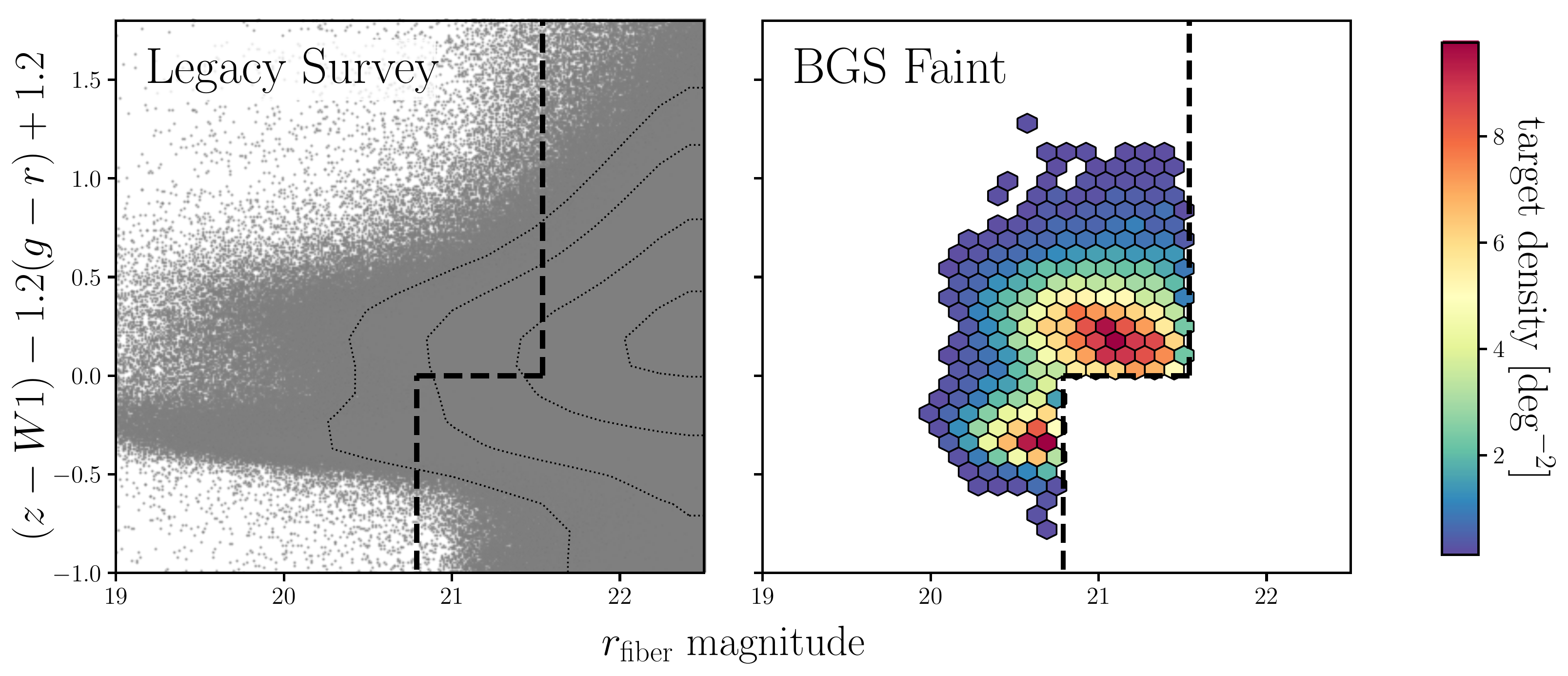}
    \caption{\label{fig:bgsfaint}
    BGS Faint targets include objects fainter than BGS Bright, 
    $19.5 < r < 20.175$, that are within custom $r_{\rm fiber}$ - color
    selection cuts (Eq.~\ref{eq:rfib_color_cut}).
    In the left panel, we show the $r_{\rm fiber}$ - color cut (black dashed)
    on the $r_{\rm fiber}$ versus $(z - W1) - 1.2 (g - r) + 1.2$ distribution of LS
    objects that pass our star-galaxy separation (grey). 
    The contours mark the 11.7, 39.3, 67.5, and 86.4 percentiles of the distribution (dotted). 
    $(z - W1) - 1.2 (g - r) + 1.2$ is a proxy for the strength of H$\alpha$ and 
    H$\beta$, so BGS Faint targets either have bright fiber magnitudes or
    strong emission lines.  
    We impose the $r_{\rm fiber}$ - color cut in order to maintain high redshift 
    efficiency for BGS Faint. 
    In the right panel, we present the target density of the BGS Faint targets (color 
    map).
    In total, we have 533 targets/${\rm deg}^{2}$ for the BGS Faint sample.}
\end{center}
\end{figure}

%%%%%%%%%%%%%%%%%%%%%%%%%%%%%%%%%%%%%%%%%%%%%%%%%%%
\subsection{BGS Faint Sample} \label{sec:bgs_faint}
In addition to BGS Bright, BGS includes galaxies with magnitudes 
fainter than $r > 19.5$.  
This fainter sample will substantially increase the overall BGS target density 
and, thus, enable small-scale clustering measurements with higher signal-to-noise. 
The BGS Faint sample will also enable multi-tracer analyses using multiple
populations of tracers with very different biases, which are forecasted to produce
the tightest constraints on RSD~\citep{seljak2009, mcdonald2009, wang2020}.
It will also include many faint emission line galaxies that will be 
valuable for studies of galaxy evolution and the cosmic star formation history.

The preliminary BGS target selection described by \cite{ruiz-macias2021} only
included an $r$-band magnitude cut. 
This selection, however, included many faint galaxies with low fiber
fluxes, which significantly reduce the redshift success rate of the sample. 
In order to maintain high redshift efficiency and completeness, the final BGS
Faint selection includes the following $r_{\rm fiber}$-color cut
\begin{equation} \label{eq:rfib_color_cut}
    r_{\rm fiber} < 
    \begin{cases}
        20.75   & \text{if}~(z - W1) - 1.2 (g - r) + 1.2 < 0\\ 
        21.5    & \text{if}~(z - W1) - 1.2 (g - r) + 1.2 \ge 0.
    \end{cases}
\end{equation}
Here, $W1$ is the magnitude in the WISE $W1$ band, 3.4$\mu m$ at 6.1\arcsec
angular resolution.  
We also include a $19.5 < r < 20.175$ magnitude limit in order to satisfy a 
${\sim}$1,400 targets/${\rm deg}^{2}$ constraint on the total target density 
imposed by the survey fiber budget allocated to BGS.   
The corresponding fiber budget for the MWS stellar targets is 
$\sim$800 targets/deg$^{2}$~\citep{mws}. 

In a sample of LS-matched emission line galaxies from the AGN and Galaxy Evolution 
Survey~\citep[AGES;][]{moustakas2011, kochanek2012}, 
we identified that H$\alpha$ and H$\beta$ emitting star-forming galaxies 
predominantly lie above a $(z - W1) - 1.2 (g - r) + 1.2$ locus. 
This selection identifies galaxies that either have brighter fiber
magnitudes or have emission lines in their spectra.  
As we discuss in Section~\ref{sec:sv}, the high redshift success rate of this
sample validates this selection. 
In the left panel of Figure~\ref{fig:bgsfaint}, we present the $r_{\rm fiber}$
versus $(z - W1) - 1.2 (g - r) + 1.2$ color distribution of LS objects that pass
our star-galaxy selection (gray) and highlight the BGS Faint $r_{\rm fiber}$-color
cut (black dashed).
In the right panel of Figure~\ref{fig:bgsfaint}, we present the number density
of BGS Faint sample targets.  
We also mark the Eq.~\ref{eq:rfib_color_cut} $r_{\rm fiber}$ and the $19.5 < r < 20.175$ 
limits in the $r$ versus $r_{\rm fiber}$ distribution of LS objects 
in Figure~\ref{fig:bgsbright} (dot dashed). 
In total, we have 533 targets/${\rm deg}^{2}$ for the BGS Faint sample.
These targets can be accessed in DESI catalogs as described in \cite{desitarget}
under the $\mathtt{BGS\_FAINT}$ bit-name.

%%%%%%%%%%%%%%%%%%%%%%%%%%%%%%%%%%%%%%%%%%%%%%%
\subsection{BGS AGN Sample} \label{sec:bgs_agn}
In addition to the Bright and Faint samples, BGS includes a supplementary
selection to recover Active Galactic Nuclei (AGN) host galaxies that are
rejected by the $(G_{Gaia} - r_{\rm raw}) > 0.6$ star-galaxy separation cut, but
would otherwise pass BGS selection criteria.  
This BGS AGN sample is designed to increase the completeness of the DESI dark 
time quasar targets.
It does not overlap with the BGS Bright or Faint samples, by design, and has 
minimal overlap with the dark time quasar targets because we remove $r > 17.5$
targets with a PSF morphological type. 

To select targets with AGN, we exploit optical and infrared colors that
trace the signatures of hot, AGN-heated dust in the spectral energy
distribution. 
The primary AGN selection criteria are: 
\begin{align} \label{eq:agn}
        (z - W2) - (g - r) &> -0.5   \\
        (z - W1) - (g - r) &> -0.7   \\
        (W1 - W2) &> -0.2 \\
        (G_{Gaia} - r) &< 0.6
\end{align}
We also require a $S/N>10$ detection in both $W1$ and $W2$ bands to ensure a robust 
constraint in the infrared regime, and apply quality and magnitude cuts. 
The full sample selection is described in a dedicated BGS AGN paper~\citep{bgs_agn}.
The resulting target density of BGS AGN is $\sim3-4$ targets/${\rm deg}^{2}$, 
distributed uniformly over the DESI footprint. 
These targets can be accessed in DESI catalogs as described in
\cite{desitarget} under the $\mathtt{BGS\_WISE}$ bit-name.

%% file: design.tex
\section{Survey Design} \label{sec:design}
In this section, we present the final design of the BGS. 
We describe how we determine exposure times to achieve uniform $>95\%$ redshift 
efficiency for BGS exposures spanning a broad range of observing conditions 
(Section~\ref{sec:texp}). 
We then present the observing strategy that BGS will employ to observe a
footprint of 14,000 ${\rm deg}^2$ with >80\% fiber assignment
completeness and 20\% operational margins (Section~\ref{sec:strat}). 
Lastly, we present the strategy for assigning fibers to BGS targets
(Section~\ref{sec:fibassign}). 

\begin{figure}
\begin{center}
    \includegraphics[width=0.45\textwidth]{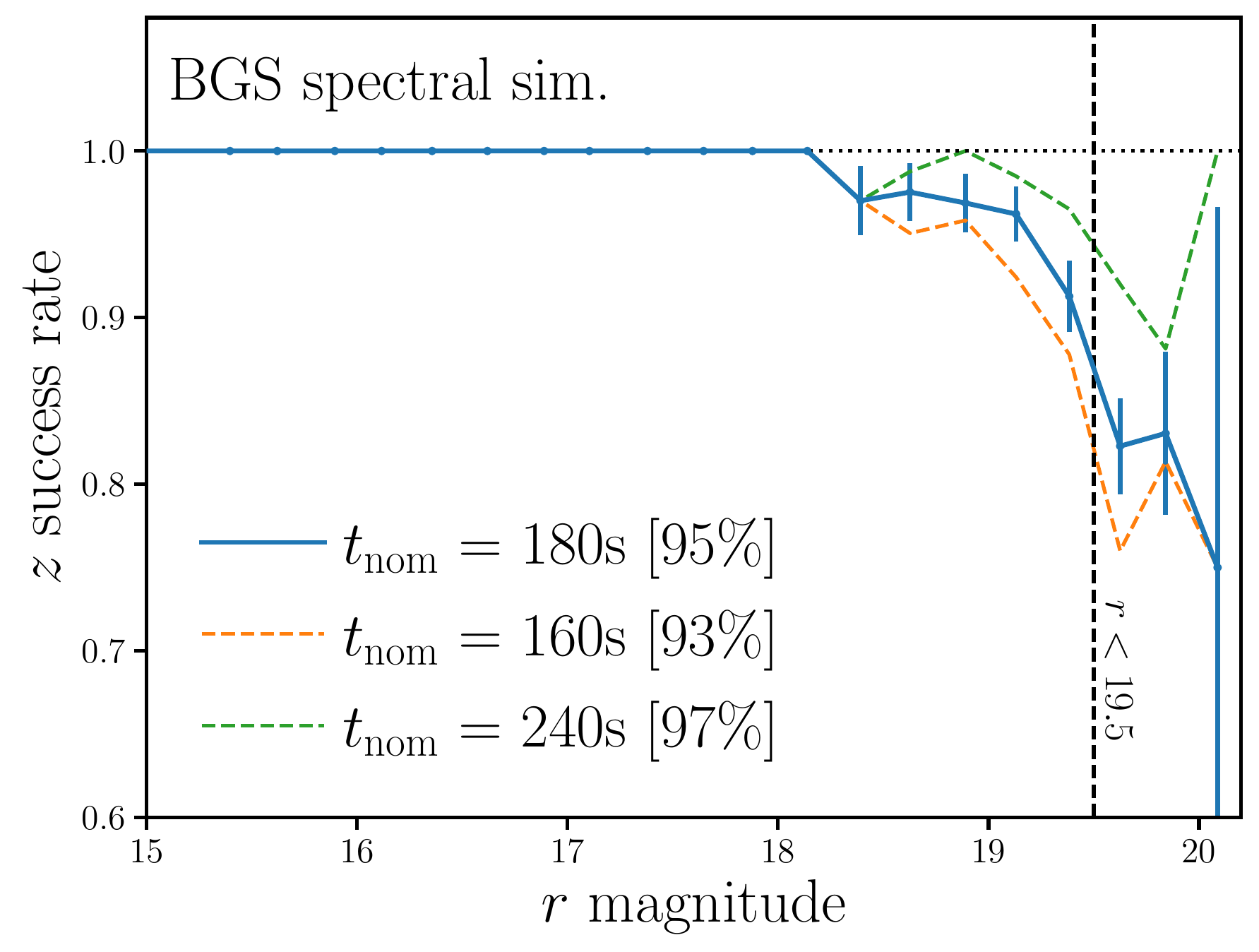}
    \caption{
    Redshift success rate of spectral simulations run using a nominal exposure 
    time of $t_{\rm nom} = 180$s (blue) as a function of $r$-band magnitude. 
    We include $z$ success rates for spectral simulations run using 
    $t_{\rm nom} = 160$s (orange dashed) and $240$s (green dashed) 
    for comparison.
    The overall $z$ success rates of all $r < 19.5$ galaxies are presented 
    in the legend.  
    These simulations assume spectra based on realistic continuum templates 
    derived from AGES, matched to $g$, $r$, $z$ LS photometry, and GAMA 
    emission line fluxes.
    They incorporate realistic noise and throughput for BGS observations.
    \emph{With $t_{\rm nom} = 180{\rm s}$, we predict that the BGS Bright sample can
    achieve an overall redshift success rate of $95\%$}.
    } \label{fig:specsim}
\end{center}
\end{figure}

%%%%%%%%%%%%%%%%%%%%%%%%%%%%%%%%%%%%%%%%%%%%%%%%%%
\subsection{Nominal Exposure Time} \label{sec:texp}
We want BGS to achieve near-homogeneous completeness in the final survey. 
This requires BGS exposures, which are taken over a broad range of
observing conditions, to have uniform redshift efficiencies. 
To achieve this uniformity, exposure times are dynamically scaled by the
Exposure Time Calculator (ETC), depending on the measured sky background,
seeing and transparency. 
For instance, an exposure taken close to the moon with bright sky background will
require a longer exposure time than an exposure taken at larger moon separation.
We define the anchoring \emph{nominal} exposure time as the exposure time required 
to achieve ${>}95\%$ redshift efficiency for the BGS Bright sample during 
{\em nominal dark conditions}, which roughly corresponds to the median expected 
conditions during dark 
time\footnote{The nominal dark condition is defined at zenith ($X=1$), with no 
extinction ($E(B-V)=0.0$), a seeing FWHM of 1.1\arcsec, and a sky background
of $r=21.07$ mags/arcsec$^2$}. 

We use spectral simulations to determine an initial estimate of 
the nominal exposure time.
We construct multiple sets of simulated spectra using a range of exposure times and then
select the exposure time that achieves ${>}95\%$ redshift efficiency.  
To construct these spectra, we first compile a catalog of GAMA galaxies that
would be selected as BGS targets.
We match each galaxy to a continuum template, constructed from galaxy spectra of AGES based on their redshift,
$r$-band absolute magnitude, and $(g-r)$ color.
Next, we add emission lines to the continuum using emission line flux and width 
measurements from GAMA. 
We then normalize the simulated spectra to match the fiber aperture flux from LS,
because DESI spectra only include light within the fiber aperture. 
We construct $1000$ simulated spectra in total that cover the full range of BGS 
galaxy spectral types.

From the noiseless galaxy spectra, we construct realistic DESI-like spectra using
the DESI $\mathtt{specsim}$ package\footnote{\href{https://specsim.readthedocs.io}{https://specsim.readthedocs.io}},
which simulates the source profile, atmosphere, and the DESI instrument characteristics.
The atmosphere model accounts for the variance added to the 
source spectrum (\emph{i.e.} our noiseless galaxy spectra) from the sky emission 
spectrum and the attenuation of the spectrum by its passage through the atmosphere.
In our case, we use the dark sky spectrum derived from UVES~\citep{hanuschik2003}
and the nominal airmass for the attenuation. 
The total photon count entering the fiber is then modeled using the spectrum and 
specified exposure time. 
Afterwards, the DESI instrument model simulates the resolution 
effects and throughput for each of the three spectrograph cameras.
Finally, the sensor electronics response, characterized by gain, dark current, and 
readout noise is simulated to produce realistic DESI-like spectra.  

Next, we measure redshifts of the simulated BGS spectra using 
{\sc Redrock}\footnote{\href{https://redrock.readthedocs.io}{https://redrock.readthedocs.io}}, 
the redshift fitter for DESI~\citep{redrock}. 
{\sc Redrock} finds the best-fit redshift using a linear combination of
Principal Component Analysis (PCA) basis templates derived from galaxy models.
Fitting is performed over a specified redshift range for three template classes:
``stellar'', ``galaxy'', and ``quasar''. 
The redshift and spectral class that give the lowest $\chi^2$ value is 
considered the best description of the spectrum. 
{\sc Redrock} also provides a redshift confidence, $\Delta\chi^2$, based on the 
difference between the $\chi^2$ values of the best-fit {\sc Redrock} model and
the next best-fit model that differs in redshift by >1000 km/s. 
We consider a BGS redshift as successful if: 
(1) no warning flags, \emph{e.g.} for poor fits or bad data, are raised, 
(2) $\Delta\chi^2 > 40$ (Section~\ref{sec:zsuccess}), 
and, in this case, 
(3) the measured redshift, $z'$, closely reproduces the true redshift $z_{\rm true}$ 
of the simulation:
$|z' - z_{\rm true}|/(1+z_{\rm true}) < 0.0033 \simeq 1000$ km/s. 

In Figure~\ref{fig:specsim}, we present the redshift success rate as a function of 
$r$ magnitude for spectral simulations run using exposures times of $t_{\rm exp} = 180$ 
(blue), 160 (orange), and 240 seconds (green). 
We include Poisson uncertainties of the redshift success rate, for reference.
We mark the $r< 19.5$ magnitude limit of the BGS Bright sample (black dashed) 
and  the overall redshift success rate of all $r < 19.5$ galaxies in the legend.
Based on the spectral simulations, we predict that we can achieve an overall 
redshift success rate of ${\sim}95\%$ for a sample limited to $r < 19.5$ with
the nominal BGS exposure time of $t_{\rm nom}$ = 180s.
Every BGS exposure time will be scaled based on this 180s nominal exposure
time and its individual observing conditions to achieve uniform redshift 
efficiency. 
For details on how BGS exposure times are set during observations, we
refer reader to \cite{ops}.

%%%%%%%%%%%%%%%%%%%%%%%%%%%%%%%%%%
\subsection{Observing Strategy} \label{sec:strat}
During bright conditions, DESI will observe the bright time programs: BGS and
MWS.  
The decision on whether to observe the dark or bright time programs is
determined by a threshold on `survey speed', which is a diagnostic based on observing
conditions, such as seeing, transparency, airmass, and sky brightness.
BGS will aim to cover 14,000 deg$^2$ with a footprint that closely 
matches the dark program. 
The footprint will be covered by 5675 `tiles', which are planned DESI
pointings, that are arranged according to a non-overlapping `best packing' scheme. 
The scheme requires tiles in a single pass to have a minimum separation of 3.411 deg.
BGS will be observed with four passes such that each point in the footprint
will be visited three times on average, to satisfy the requirement that 80\%
of targets are observed spectroscopically. 

To assess the feasibility of the BGS strategy above, we use survey simulations
to forecast the progress of the DESI survey.
The survey simulations, as described in detail in~\cite{ops},
simulate the nightly operations of DESI as a sequence of tile
exposures\footnote{\href{https://surveysim.readthedocs.io/}{https://surveysim.readthedocs.io/}}. 
They account for the expected configuration and dead time of both the telescope
and instrument and assume historical weather and environmental factors
appropriate for Kitt Peak.
Lunar conditions, which are defined by tabulated ephemerides, play a key role 
in the simulation. 
The simulated exposure time is scaled according to the predicted sky
brightness relative to nominal conditions at the time of observation for
the tile, which will depend on the lunar phase and position. 
We use an empirical sky background model derived from SV exposures and the 
nominal exposure time defined in Section~\ref{sec:texp}.
We describe the sky model in detail in Appendix~\ref{sec:sky}.

\begin{figure}
\begin{center}
    \includegraphics[width=0.9\textwidth]{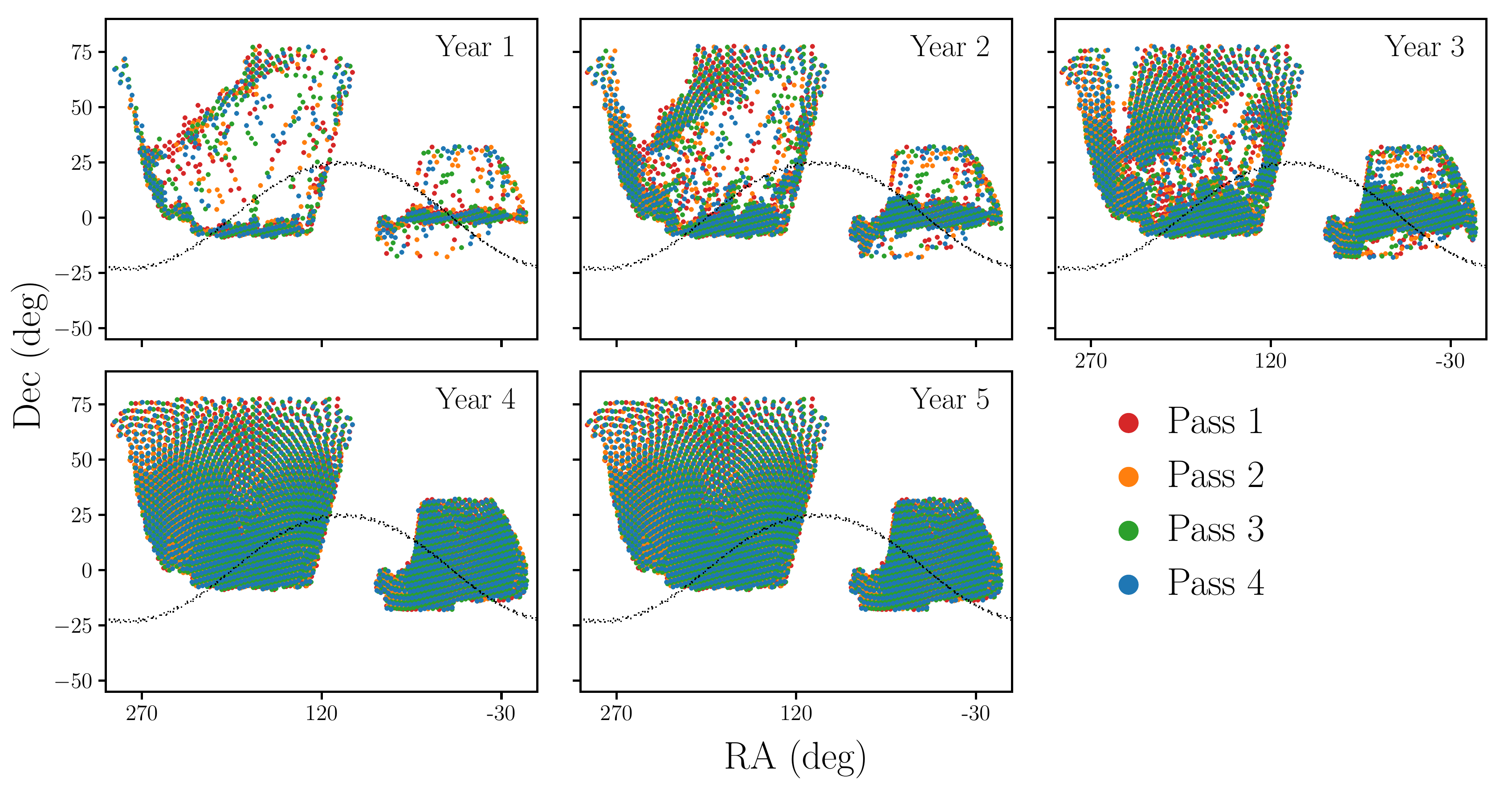}
    \caption{
    BGS tiles observed by DESI over its five years operation according 
    to survey simulations. 
    In each panel, we present BGS tiles that will be observed up to and 
    including that year. 
    BGS will observe a footprint of 14,000 deg$^2$, which will match
    the footprint of the dark program as closely as possible, with a 
    four-pass strategy. 
    Each point on the footprint will be visited $3$ times on average. 
    For reference, we mark the ecliptic in each panel (black dotted). 
    According to survey simulations, we achieve the 14,000 deg$^2$ 
    footprint with 4 passes after 4 years and meet the required 20\% operational
    survey margins.
    }
    \label{fig:stratsim}
\end{center}
\end{figure}
In Figure~\ref{fig:stratsim}, we present the survey footprint observed over the 
5 year operation of DESI, as predicted by survey simulations. 
Each panel shows the BGS tiles that are completed up to and including that year. 
We plot the ecliptic for reference (black dotted).
We assume a lunar exclusion zone of radius 50 deg, which is apparent in 
the reduced coverage close to the ecliptic in early years. 
Based on the survey simulation, \emph{the effective three visit coverage 
is achieved for the 14,000 deg$^2$ footprint after 4 years, which leaves
20\% of the total DESI operation time as margin.
}

%%%%%%%%%%%%%%%%%%%%%%%%%%%%%%%%%%%%%%%%%
\subsection{Fiber Assignment Strategy} \label{sec:fibassign}
The focal plane of DESI contains 5,000 fibers arranged in 10 wedge-shaped
petals. 
Each fiber is controlled by a robotic fiber positioner, which can rotate on
two arms and be positioned within a circular patrol region of radius
1.48 arcmin~\citep{schubnell2016, desicollaboration2016, martini2022, silber2022}.
The patrol regions of adjacent positioners slightly overlap; however, there are
gaps in the regions between the petals. 
For each tile we dedicate a minimum of 40 `sky' fibers per petal to
measure the sky background for accurate sky subtraction. 
An additional 10 fibers per petal are assigned to standard stars for flux 
calibration~\citep{spec2022}. 
The rest of the `science' fibers are assigned to BGS targets on each tile 
according to the following fiber assignment strategy~\citep{fba}.

First, we assign a primary priority to BGS targets.
Our first goal is to obtain a magnitude-limited BGS Bright sample that is as
complete as possible to simplify clustering analyses. 
So we assign the highest priority to BGS Bright targets. 
Next, we assign lower priorities to 80\% of the BGS Faint 
targets\footnote{The numerical values of these priorities are 2100 for the
highest priority and 2000 for the lower priority}. 
If we were to assign lower priorities to all BGS Faint targets, measurements of
clustering in the BGS Faint sample would suffer significantly from uncertain 
fiber  assignment incompleteness corrections. 
For instance, in regions with a high density of BGS Bright targets, BGS Faint targets
would not be assigned to fibers~\citep{smith2019}. 
This would lead to certain galaxy pairs having zero probability of assignment,
which would be impossible to correct in later clustering 
analyses~\citep[\emph{e.g.}][]{hahn2017, bianchi2018}.
To reduce this effect and to facilitate corrections for fiber assignment
incompleteness, we randomly promote 20\% of BGS Faint targets to the same 
priority as BGS Bright\footnote{
    In the DESI catalogs these BGS Faint targets with higher priority are
    labeled under the $\mathtt{BGS\_FAINT\_HIP}$ bitname.
}. 
Lastly, the BGS AGN targets are assigned at the same priority as BGS Faint.  
BGS shares the focal plane with the MWS, whose targets enter at a lower priority
than both BGS Faint and BGS AGN targets. 

After the primary priorities are assigned, a uniform random sub-priority in 
the range (0, 1) is generated for each object.
The total priority is the sum of the primary and sub-priority values.
Fibers are assigned to targets in their patrol region in rank order of total priority.
With this strategy, fiber assignments to targets with the same primary priority is
randomized but a higher primary priority target will always be assigned a fiber
in preference to a lower primary priority target.
We note that targets with higher priority than BGS Bright are rare and are 
typically standard stars or MWS white dwarfs.
Also, the lowest priority targets are occasionally `bumped' on a tile by tile
basis to satisfy the sky fiber requirement.
Occasionally, targets whose redshifts were unsuccessfully measured are
re-observed but reassigned lower priority (see~\citealt{fba} for details).
In Section~\ref{sec:fibeff}, we present the fiber efficiencies that result from 
this strategy for each of the BGS target types.

%% file: sv.tex
%%%%%%%%%%%%%%%%%%%%%%%%%%%%%%%%%%%%%%%%%%
\section{Survey Validation} \label{sec:sv}
Before beginning its five years of operations, DESI conducted the Survey Validation
(SV) campaign with the primary goal of verifying that the main survey will meet its 
requirements.
In this section, we use SV observations to validate the BGS selection cuts 
(Section~\ref{sec:ts}) and demonstrate that BGS will meet the requirements we set 
to ensure that it will achieve its broad range of science goals. 
The requirements include stellar contamination rates less than 1\%, redshift 
efficiency above 95\%, and fiber assignment efficiency above 80\%.
In Section~\ref{sec:sv_obs}, we describe the SV observations in further detail. 
We then use these observations to validate the selection cuts (Section~\ref{sec:valid_cuts}), 
redshift  efficiency (Section~\ref{sec:zsuccess}), and fiber assignment efficiencies
(Section~\ref{sec:fibeff}). 

\subsection{SV Observations} \label{sec:sv_obs}
The SV campaign was divided into two main phases: 
the first, \svi, observed fields spanning the expected footprint and aimed to characterize
performance for different observing conditions and optimize sample selection.  
The second program, the \sviii, aimed to observe a dataset that can be used for
representative clustering measurements and deliver a `truth' sample with
high completeness ($\gtrsim$ 99\% for BGS Bright) 
over an area at least 1\% of the expected main survey footprint. 
We use the \sviii~data to predict the redshift efficiency for BGS targets 
in the main survey. 
In the following, we describe the characteristics of the \svi~and \sviii~programs 
in further detail and explain the relevance for the tests performed in this work. 

% details on SV1 exposures from 
% https://data.desi.lbl.gov/desi/survey/observations/SV1/sv1-html/index.html#obsconds-cumul
\svi~observed on 76 nights over ${\sim}5$ months from December 2020 to April 2021. 
During this time, we observed $562$ bright time exposures that cover 50 
unique BGS tiles with an effective area 310 deg$^2$.  
In Figure~\ref{fig:footprint}, we mark the BGS tiles observed during \svi~in red. 
The tiles span the expected DESI footprint and cover both the NGC and SGC.
They also cover imaging from both DECaLS and BASS+MzLS.
Some targeted regions where the imaging surveys overlap to investigate
differences in target selection due to the differing imaging quality 
and photometric systems.  
Additional tiles were chosen to overlap with external surveys, \emph{e.g.}
the GAMA G02 and G12 fields, in order to compare redshifts from them to
those from DESI.
Tiles were also chosen to include regions with challenging imaging conditions, 
including strong dust extinction and high stellar density near the Galactic plane.  
Furthermore, \svi~targets were chosen with a broader selection to determine the selection 
that performs best in those conditions.
We describe the full \svi~selection in Appendix~\ref{sec:sv1}. 

% Sub-select those that meet quality cuts first. 
In addition to their positions, \svi~exposures were observed at different times
to span a broad range of observing conditions (airmass, Galactic  extinction, 
transparency, and seeing). 
Moreover, the exposures were observed during times with widely varying sky
brightness: 17 - 24 mags based on both the Guide Focus Array 
cameras~\citep[GFA;][]{desicollaboration2016a}  and sky spectra. 
Exposures were taken during nearly all lunar conditions spanning different
combinations of moon illumination, altitude, and separation.
Each \svi~field was required to have one dark exposure. 
The remaining exposures were taken with $\sim$300s exposure times on different
nights such that they were observed over a range of conditions and hour angles. 
We compile `cumulative' coadds of all exposures of a given tile to date
to serve as deeper `truth tables'.
%This strategy enabled custom coadds of varying effective exposure time to be produced  for each field. For instance, we compiled `cumulative' coadds of all exposures of a given tile to date to serve as deeper `truth tables'.

% details on SV3 exposures from 
% https://data.desi.lbl.gov/desi/survey/observations/SV3/sv3-html/
Shortly after the \svi~ program, the SV2 program dedicated a short amount of available 
time to developing main survey operations. 
The \sviii~(or SV3) then observed on 38 nights from April 2021 to the end of May 2021. 
During this time, we observed $288$ bright time exposures that cover 214 BGS tiles. 
The \sviii~was designed to operate similarly to the main DESI survey but at much 
higher completeness. 
\sviii~pointings, therefore, targeted sets of 11 overlapping tiles with centers
arranged around a 0.12 deg circle, forming a `rosette' completeness pattern.
In total, the \sviii~observed 20 rosettes that cover an area of 180 deg$^2$, of 
which  140 deg$^2$ was of the desired completeness for BGS Bright.  
The rosettes spanned the NGC footprint and several were chosen to overlap with 
external surveys, including GAMA, DEEP2, AGES and HSC. 
In Figure~\ref{fig:footprint}, we mark the \sviii~tiles on the DESI footprint in orange.

Although the \sviii~exposures were not observed in widely varying observing
conditions, they are nevertheless representative of conditions expected during
bright time. 
Unlike the \svi~exposures, $t_{\rm exp}$ were set by the ETC, as in the main 
survey. 
The ETC dynamically scaled the $t_{\rm nom}$ = 180s nominal exposure time 
according to observing conditions (Section~\ref{sec:texp}).
For long observation sequences, the ETC also decided how the observation
is split into multiple exposures. 
To achieve a very high spectroscopic completeness, exposure times were 20\% 
longer than that expected of the main survey. 
Targets with failed redshifts  were reobserved at lower priority. 
A target could be reobserved up to nine times if we continued to fail in measuring
its redshift (see \citealt{desitarget} for details).  
The longer exposure time and reassignment of targets allowed for the construction
of accurate redshift `truth tables' at greater depth for the vast majority of 
\sviii~targets.  
The ETC was also continuously updated and calibrated throughout \sviii. 
Hence, the exposure times are more variable than what we expect for the main 
survey. 
We refer readers to \cite{sv} for further details on the SV programs.

After the SV programs, the main survey began on May 14, 2021 and proceeded until 
the seasonal Arizona  monsoon in July prevented observations for a number of months.  
The 343 main survey tiles observed over the first $\sim$2 months delivered 
532,796 BGS redshifts. 
Some of the later tests we present in this work include observations from this 
period to further ensure that we meet the stated requirements for the final 
survey design choices.  
% The pre-monsoon main survey was observed using a `wide' strategy, which  prioritized footprint coverage over depth, to yield more  time to ensure  tiles had been successfully completed. 
% For BGS, this means that observation from this period has a relatively high fraction of BGS Bright targets as only first pass data is acquired. 
%For more details on the SV observations and the SVDA, we refer readers to \todo{Dawson~\etal~(in prep.)}. 

\begin{figure}
\begin{center}
    \includegraphics[width=0.8\textwidth]{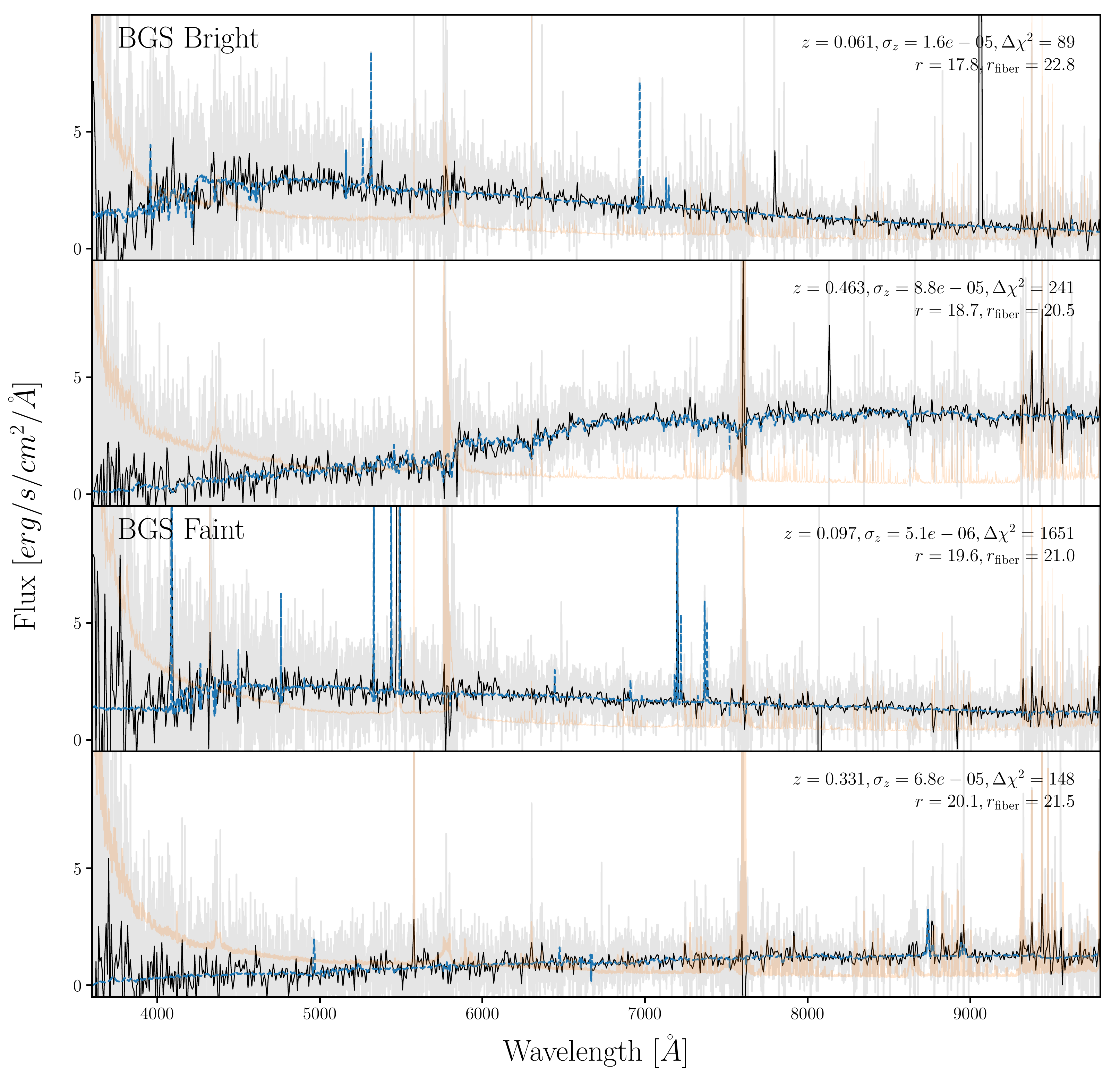}
    \caption{\label{fig:bgs_spec}
    BGS galaxy spectra from the \sviii~(gray). 
    We present spectra of a blue and a red galaxy from the BGS Bright sample
    and a blue and a red galaxy from the BGS Faint samples (top to bottom
    panels). 
    In each panel, we also plot the spectrum rebinned to a coarser wavelength 
    grid (black), the measured uncertainties (orange), and include the best-fit 
    {\sc Redrock} template used to measure the redshift (blue). 
    The redshift measurement, uncertainty, and $\Delta \chi^2$ from 
    {\sc Redrock} are included in the top right along with the $r$ band magnitude
    and fiber magnitude of the galaxy.
    } 
\end{center}
\end{figure}

All BGS targets observed during \svi~and the \sviii~are reduced using the 
spectroscopic data reduction pipeline.
Briefly, spectra are first extracted from the spectrograph CCDs using the 
\emph{Spectro-Perfectionsim} algorithm~\citep{bolton2010}.
Then, fiber-to-fiber variations are corrected by flat-fielding and a sky model, 
derived from the (at least 400) sky fibers, is subtracted from each spectrum. 
Afterwards, fluxes are calibrated using stellar model fits to standard stars.
The calibrated spectra are then co-added across exposures of the
same tile to produce the final processed spectra with an effective exposure
time equal to $t_{\rm nom}$.  
The full spectroscopic data reduction is described in \cite{spec2022}. 
We present a few examples of BGS spectra selected from the \sviii~observations in 
Figure~\ref{fig:bgs_spec}. 

%%%%%%%%%%%%%%%%%%%%%%%%%%%%%%%%%%%%%%
\subsection{Validating Selection Cuts} \label{sec:valid_cuts}
To construct BGS Bright and Faint samples appropriate for science, we apply 
spatial masking, star-galaxy separation, fiber magnitude cut, quality 
cut, and bright limit to potential targets (Section~\ref{sec:select}).
We validate each of these selections using SV data in the following.  

First, we validate the spatial masking around bright stars by examining the
number density of BGS targets inside and outside the masks. 
We compare the density of BGS objects with the mean density as a function of
angular separation from the bright star as in~\cite{ruiz-macias2021}. 
Within, the spatial masking radius, $R_{\rm BS}$, the target density is 
significantly lower than the mean  density. 
This is because {\sc Tractor} fits objects near bright stars with a PSF
model, so they are often excluded from the target catalog by the BGS
selection cuts.
On the other hand, outside $R_{\rm BS}$, we find that the BGS target 
density is in good agreement with the mean density. 
This confirms that excluding the regions within our bright star masks 
sufficiently accounts for the impact of bright stars. 

Next we validate our star-galaxy separation criteria using \svi~observations,
where we imposed a more relaxed star-galaxy separation
(Appendix~\ref{sec:sv1}). 
In \svi, in addition to objects that pass our star-galaxy separation, we also
select  objects below the $(G_{Gaia} - r_{\rm raw}) = 0.6$ threshold that are 
not classified as PSF by {\sc Tractor}. 
In the center panel of Figure~\ref{fig:stargalaxy}, we present the stellar
contamination fraction of BGS objects in \svi.  
Stellar contamination is determined using DESI spectra based on spectral
classification by {\sc Redrock} and a $z < 300$ km/s redshift limit. 
We represent the contamination rate  in hex bins using the color mapping.
We only include bins with more than 10 galaxies to ensure accurate estimates. 
Below the $(G_{Gaia} - r_{\rm raw}) = 0.6$ threshold, we find significant
stellar contamination; some of the bins have >20\% stellar contamination. 
In contrast, we find low stellar contamination rates above the threshold. 
The low stellar contamination rate is further confirmed by the right panel of
Figure~\ref{fig:stargalaxy}, where we present the stellar contamination
fraction of BGS objects in the \sviii. 
The \sviii~only includes objects that pass our star-galaxy separation.
We find <1\% stellar contamination throughout the entire sample. 
We also find no significant difference in the stellar contamination 
rate near the galactic plane. 

For the fiber-magnitude cuts, the bounds of the cuts are already determined by
visual inspection and by comparison to GAMA redshifts of matched galaxies. 
We further confirm, spectroscopically, that the BGS targets in \svi~observations 
that fail these cuts are not galaxies and have low redshift efficiencies. 
Next, to test the quality cuts proposed in \cite{ruiz-macias2021}, we use
2000 BGS targets spread across a 420 ${\rm deg}^2$ area of DECaLS that were
visually inspected with the Legacy Surveys Visual Inspection
(LSVI\footnote{\href{https://lsvi-webtool.herokuapp.com/}{https://lsvi-webtool.herokuapp.com/}})
interactive web interface. 
By visual inspection, we confirm that {\sc Tractor} based quality cuts discussed 
in \cite{ruiz-macias2021} removes a significant number of real galaxies in DR9. 
Removing these quality cuts increases the BGS target completeness without
introducing a significant number of spurious objects~\cite[see also][]{zarrouk2021}. 
We similarly validate the bright limit using visual inspection and confirm that 
most of the sources excluded by the cut are not galaxies --- nearly all are saturated stars. 

\begin{figure}
\begin{center}
    \includegraphics[width=0.95\textwidth]{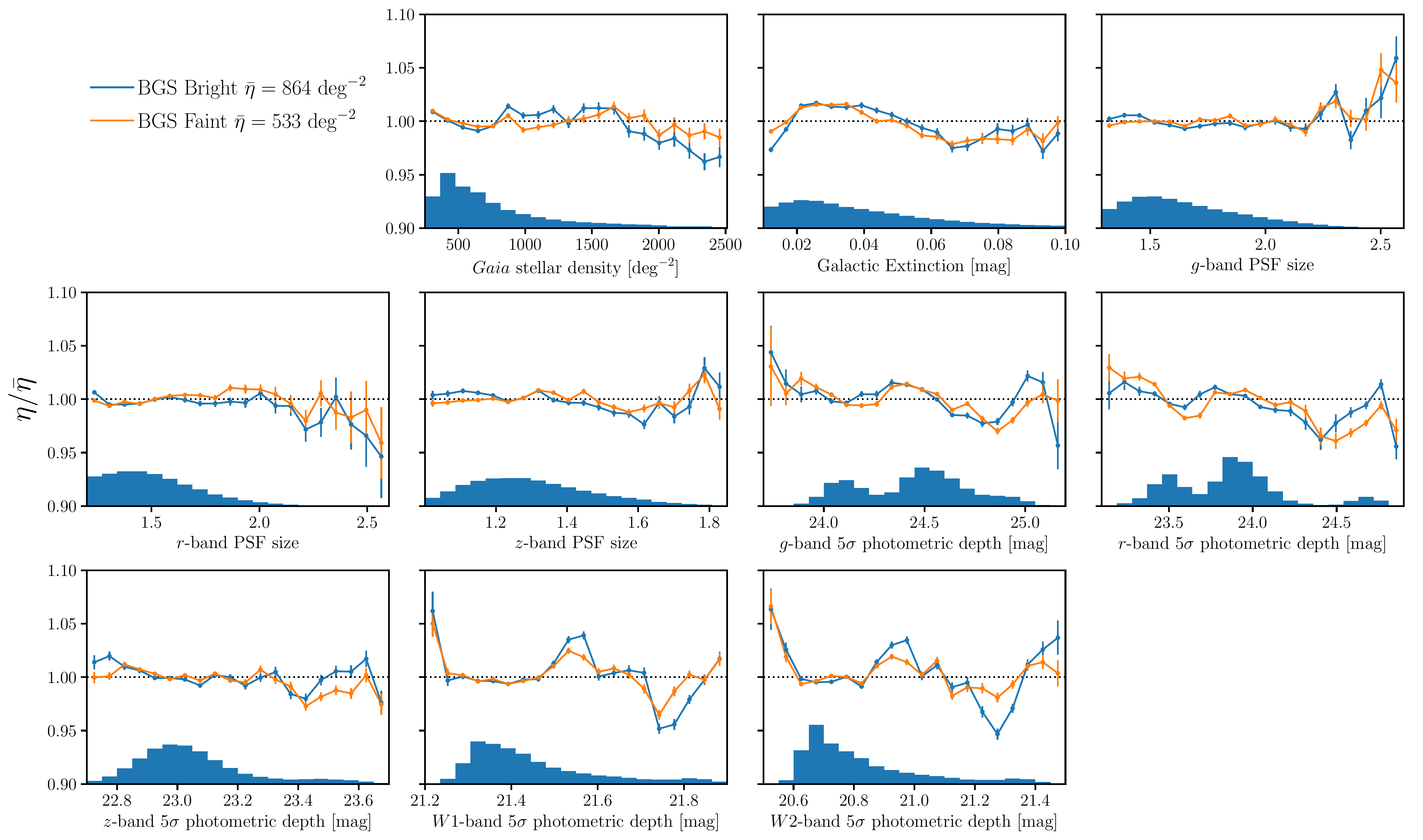}
    \caption{
        The dependence of BGS Bright (blue) and Faint (orange) target densities
        on different imaging properties: stellar density, galactic extinction,
        PSF size (in $g$, $r$, and $z$ bands), and photometric depth 
        (in $g$, $r$, $z$, $W1$, and $W2$ bands). 
        The target densities, $\eta$, are divided by the average target
        density of the samples over the 14,000 deg$^2$ DESI footprint 
        ($\bar{\eta}$; included in the legend) to highlight variations. 
        The error bars represent the uncertainties on the mean.  
        The histogram in each panel represents the normalized distribution of
        each imaging property. 
        We find the consistency of Bright and Faint target densities especially
        encouraging because we use more complex cuts in the latter.
        Overall, \emph{we find <5\% variation in the target densities of BGS 
        Bright and Faint samples and no strong systematic dependence on imaging
        properties.}
    } \label{fig:image_sys}
\end{center}
\end{figure}

In addition to validating the selection cuts that we impose, we
further examine whether any imaging property systematically impacts the 
BGS target density.
In Figure~\ref{fig:image_sys}, we examine whether stellar density, galactic 
extinction, PSF size in $g$, $r$, $z$ bands, or photometric depth in $g$, 
$r$, $z$, $W1$, $W2$ bands impact the target densities, $\eta$, of the BGS 
Bright (blue) and Faint (orange) samples.  
The target densities are measured in 
{\sc HEALPix}\footnote{\url{http://healpix.sourceforge.net}}
pixels with resolution of $N_{\rm side}=256$ (equivalent to an area of 0.05 deg$^2$).
We divide $\eta$ by the average target density of the samples over the 14,000 deg$^2$
DESI footprint, $\bar{\eta}$, to highlight any variations or dependencies.  
$\bar{\eta} = 864$ and 533 targets/${\rm deg}^{2}$ for the BGS Bright and Faint 
samples, respectively. 
In each panel, we also represent the distribution of the imaging property
with a normalized histogram. 
Stellar density is measured using \gaia~DR2 stars with $12 < G_{Gaia} < 17$. 
Galactic extinction is measured using SFD98 dust maps
(Section~\ref{sec:legacy}). 
PSF size denotes the FWHM in arcseconds. 
Photometric depth is characterized by the $5\sigma$ AB magnitude detection
limit for a 0.45\arcsec round exponential galaxy profile. 

Overall, we find <5\% variation in the target densities of both BGS Bright and
Faint samples. 
These variations are significantly lower than for the preliminary BGS target selection 
described by \cite{ruiz-macias2021}. 
We find no evidence for significant stellar contamination, which
would increase the target density in high stellar density regions.  
The target density is slightly lower for regions with stellar density >2000
deg$^{-2}$; however, the relative area is small.
In principle, this effect could be due to photometry being impacted for sources near bright
stars outside our bright star spatial masking.
However, we rule out this possibility because we find little change in
$(\eta/\bar{\eta})$ even when the spatial masking is extended significantly. 
We also find slightly lower target densities in regions with high dust
extinction, which spatially correlates with stellar density. 
A more detailed investigation on the correlation of BGS target densities with 
stellar density and dust extinction is necessary. 
Clustering analyses will likely require systematic weights derived using
linear regression or machine learning techniques to mitigate the effect 
of any spatial correlations~\citep[\emph{e.g.}][]{rezaie2020, ruiz-macias2021}. 
We find no strong dependence in the target densities for the PSF size or
photometric depth in any of the photometric bands, and we find the agreement 
of BGS Bright and Faint to be encouraging, given the more complex selection of
the latter.
Overall, the variations in target densities seem consistent with random
fluctuations from large scale structure; however, further investigation 
with simulations and observation is necessary. 

%%%%%%%%%%%%%%%%%%%%%%%%%%%%%%%%%%%%%%%%%%%%%%%%%%%%%
\subsection{Redshift Efficiency} \label{sec:zsuccess}
One of the main goals of SV is to validate the BGS redshift efficiency
for the main survey. 
We seek to verify that we can achieve our desired 95\% redshift efficiency with
the 180s nominal exposure time (Section~\ref{sec:design}) under BGS
conditions.   
Out of the 562 \svi~and 288 \sviii~BGS exposures, we focus  on spectra collected 
from exposures with effective exposure times close to 180s, the effective exposure
time expected in the main survey.
In practice, we use exposures with 160s $< \mathtt{BGS\_EFFTIME\_BRIGHT} <$ 200s,
where $\mathtt{BGS\_EFFTIME\_BRIGHT}$ is our best spectroscopically derived estimate 
of the effective exposure time achieved for BGS targets.  
It takes into account the transparency measured from standard stars together with
ETC-derived fiber losses and sky background. 

% paragraph on how we classify redshift success 
We measure redshifts for all the spectra from the selected SV exposures using 
{\sc Redrock}, as we did for the spectral simulations in Section~\ref{sec:design}.  
However, unlike the spectral simulations, we do not know the true redshifts. 
Instead, we use the `deep' coadds constructed from the \svi~and \sviii~exposures
as our redshift `truth table` (Section~\ref{sec:sv_obs}). 
We exclude spectra from SV exposures that do not have deep redshifts, because
we do not know their `true' redshifts.
We further exclude spectra that have corresponding deep coadds with individual exposure times
<2000s for \svi~and $\mathtt{BGS\_EFFTIME\_BRIGHT} <$ 100s for the \sviii. 
This is to ensure that our `true' redshifts are derived from spectra with sufficient 
depth. 
We also exclude spectra where there were any known issues with the fiber
assignment.
Furthermore, for the \sviii, we exclude spectra that required more than one
observation to determine a valid redshift. 
These are spectra of targets that have failed initial redshifts, so they may
bias our redshift efficiency estimates.
For some of the spectra, {\sc Redrock} classifies their corresponding deep spectra as 
stellar or measures a redshift outside the $0 < z < 0.6$ BGS range. 
We consider these cases as failures in targeting and, thus, exclude them when 
estimating redshift efficiency.  

From the remaining spectra (125,472 in \svi~and 176,688 in~\sviii), we classify a 
{\sc Redrock} redshift as a success if the following criteria are met: 
\begin{enumerate}
    \item No {\sc Redrock} warning flags are raised
    \item The best-fit {\sc Redrock} $\mathtt{SPECTYPE}$ is `galaxy'
    \item The reported redshift error is small relative to the measured
        redshift, $\mathtt{ZERR} < 0.0005 (1 + z)$
    \item The redshift confidence --- as judged by the difference in $\chi^2$ 
    between the two best-fitting models --- is significant: $\Delta \chi^2 > 40$
    \item $z_{\rm deep}$ is reliable --- \emph{i.e.} the deep spectrum
        meets criteria 1, 3, and 4.
    \item $z$ is consistent with the corresponding $z_{\rm deep}$ at  1000 km/s: 
    $|z_{\rm deep} - z| / (1 + z_{\rm deep})< 0.0033$.
\end{enumerate}
The specific limits in criteria 3 and 6 are BGS requirements that we set 
on the statistical error (< 150 km/s) and catastrophic redshift failures. 

\begin{figure}
\begin{center}
    \includegraphics[width=0.8\textwidth]{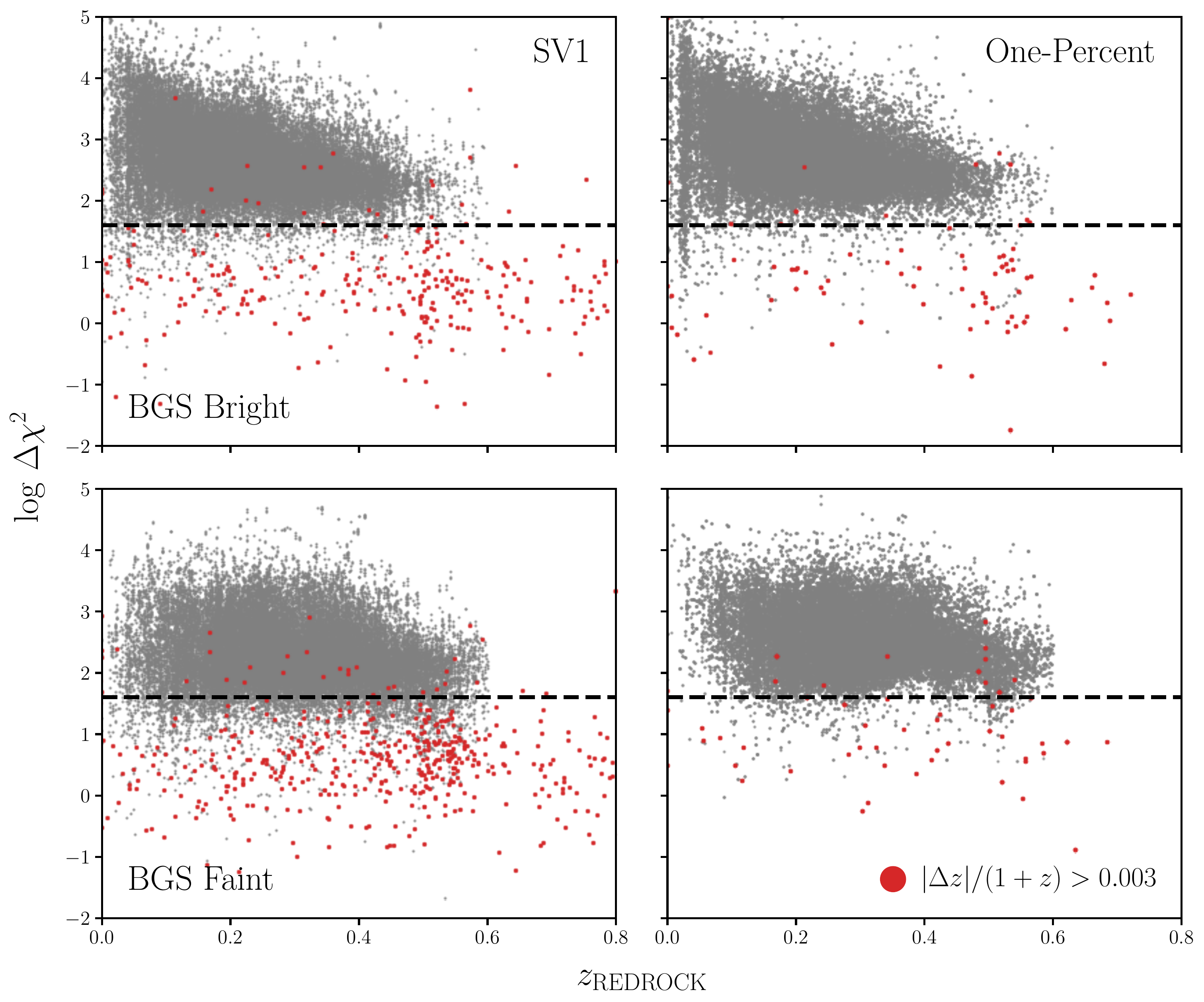}
    \caption{
    {\sc Redrock} $\Delta \chi^2$ as a function of best-fit redshift, 
    $z_{\rm Redrock}$, for BGS Bright (top) and Faint (bottom) galaxies 
    from the \svi~(left) and \sviii.
    $\Delta \chi^2$ and $z_{\rm Redrock}$ are measured using spectra from
    single BGS-like exposures. 
    We highlight the spectra that have catastrophic redshift discrepancies 
    with the redshift measured from corresponding deep exposures, 
    $|\Delta z|/(1+z) > 0.0033$, in red. 
    Our $\Delta \chi^2 > 40$ criterion for redshift success (black dashed) 
    excludes the majority of catastrophic redshift failures.
    Conversely, below the $\Delta \chi^2$ threshold, ${\sim}30$ and 
    ${\sim}15\%$ of BGS Bright and Faint galaxies have catastrophic redshift 
    failures.
    } \label{fig:dchi2_z}
\end{center}
\end{figure}

\begin{figure}
\begin{center}
    \includegraphics[width=0.45\textwidth]{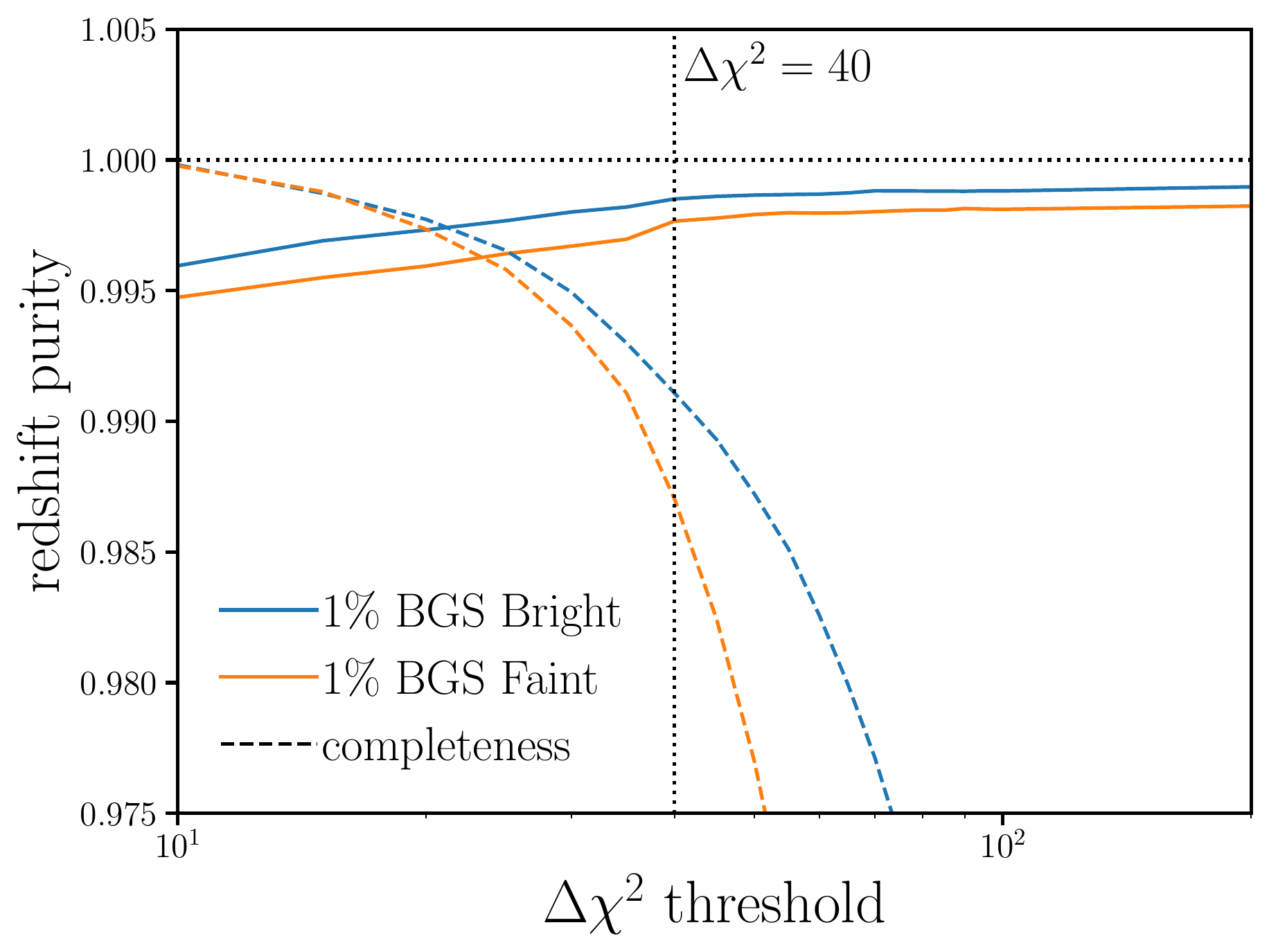}
    \caption{Redshift purity (solid) as a function of the {\sc Redrock} 
    $\Delta \chi^2$ threshold for the redshift success definition of the
    BGS Bright (blue) and Faint (orange) samples. 
    We also present the redshift completeness, the fraction of successful
    redshifts included for a given $\Delta \chi^2$ threshold. 
    We assess redshift success according to the criteria listed in 
    Section~\ref{sec:zsuccess} based on redshifts measured from
    corresponding deep exposures. 
    \emph{The $\Delta \chi^2 > 40$ threshold we impose to determine redshift
    success results in a BGS Bright sample with >99.5\% purity and >99\%
    completeness.}
        } \label{fig:dchi2_purity}
\end{center}
\end{figure}

We choose the $\dchi > 40$ threshold in criterion 4 to ensure a high redshift
purity is delivered by our redshift success criteria.
In Figure~\ref{fig:dchi2_z}, we present $\dchi$ as a function of measured
redshift, $z_{\rm Redrock}$ for the BGS Bright (top) and Faint (bottom) samples
in \svi~(left) and the \sviii~(right). 
We mark spectra that fail criterion 6 (catastrophic redshift failures) in red.
The $\dchi > 40$ threshold (black dashed) removes the majority of spectra with catastrophic
failures for all of the samples.  
Conversely, below the threshold, $\sim$40\% of BGS Bright galaxies and 
$\sim$25\% of BGS Faint galaxies are redshift failures, making this regime
unreliable for statistical or clustering studies. 
Furthermore, in Figure~\ref{fig:dchi2_purity}, we present overall redshift
purity as a function of the $\dchi$ threshold limit for the BGS Bright (blue)
and Faint (orange) samples. 
Redshift purity corresponds to the fraction of grey points that lie above the 
dashed line in Figure~\ref{fig:dchi2_z}. 
For reference, we also include the redshift completeness as a function of the 
$\dchi$ threshold. 
The completeness here is the fraction of the total `accurate' redshifts included
in the $\Delta\chi^2$ threshold. 
We consider a {\sc Redrock} redshift as `accurate' if it passes the criteria
1, 2, 3, 5, and 6 above. 
A similar calculation of purity and completeness using select galaxies with 
visually inspected redshifts produce consistent values (\citealt{vigal}, Figure 7). 
Given the choice of $t_{\rm nom}$= 180s, our $\dchi > 40$ threshold 
provides a BGS Bright sample with >99.5\% purity and >99\% completeness as 
well as a BGS Faint sample  with >99.5\% purity and > 98.5\% completeness. 

\begin{figure}
\begin{center}
    \includegraphics[width=0.45\textwidth]{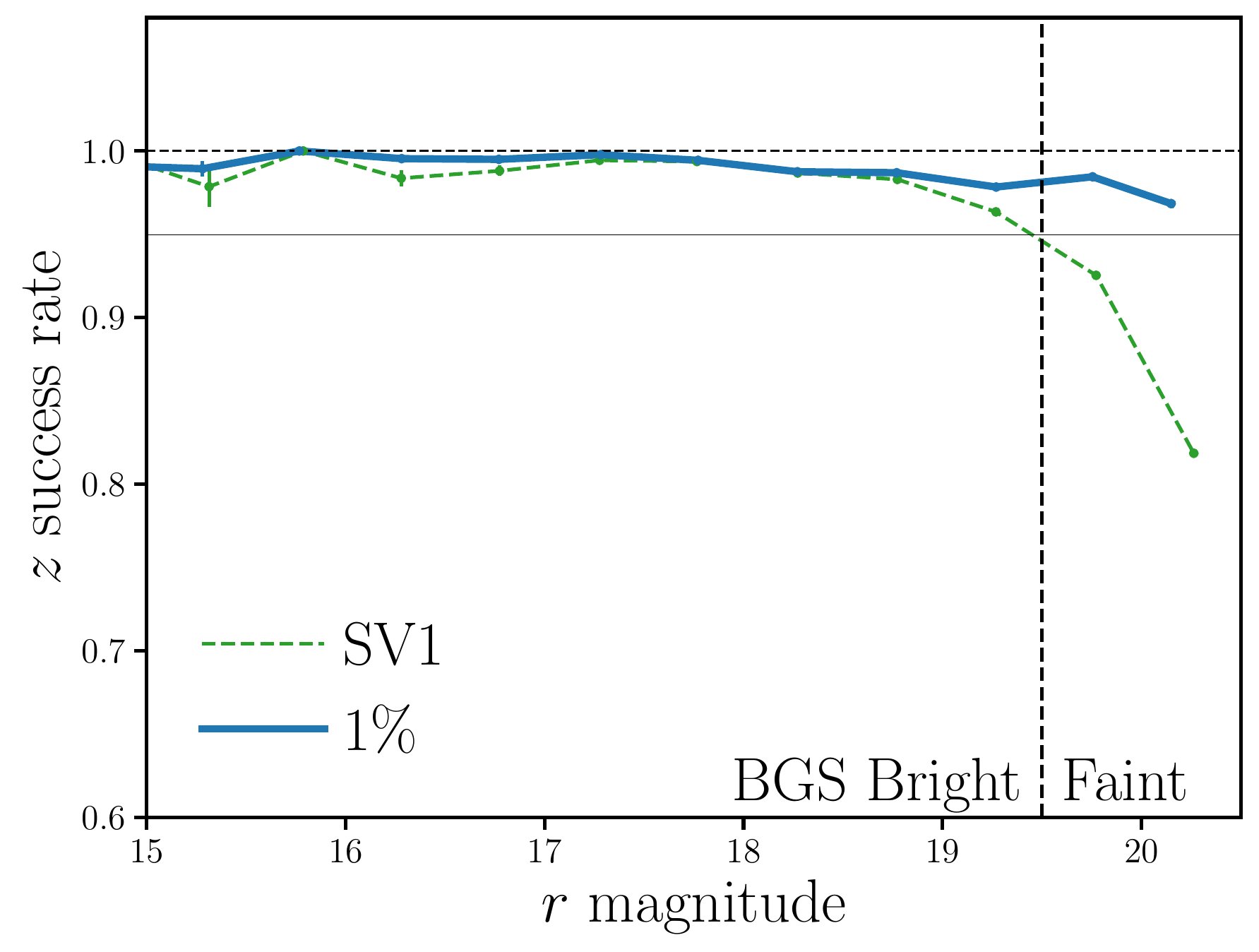}
    \caption{
    Redshift success rate as a function of $r$-band magnitude for galaxies in
    the \sviii~observations (blue). 
    The $z$ success rates are estimated using the criteria listed in
    Section~\ref{sec:zsuccess} and only include spectra from exposures that
    correspond to $t_{\rm nom} \sim 180s$. 
    The error bars represent the Poisson uncertainties for each magnitude bin. 
    We mark the $r = 19.5$ magnitude cut that separates the BGS Bright and
    Faint samples (black dashed).
    We include the $z$ success rate for \svi~galaxies, excluding the \emph{Low
    Quality} class, for reference.
    BGS galaxies in \svi~are more broadly selected than in the final target
    selection.
    \emph{We achieve $z$ success rate ${>}95\%$ for the entire BGS sample
    throughout its full $r$ magnitude range. 
    We achieve overall $z$ success rates of 98.4 and 97.9\% for BGS  Bright 
    and Faint in the \sviii.}
    } \label{fig:zsuccess}
\end{center}
\end{figure}

In Figure~\ref{fig:zsuccess}, we present the redshift success rate as a
function of $r$-band magnitude for BGS galaxies from the 
\sviii~observations (blue). 
We also include the redshift success rate of BGS galaxies from \svi~observations
(green dashed), excluding those in the {\em Low Quality} class
(Appendix~\ref{sec:sv1}). 
We remind readers that the BGS galaxies, especially the faint $r > 19.5$
galaxies, in the \svi~observations are more broadly selected than in the 
\sviii~(see Section~\ref{sec:sv_obs} for details).
We represent the Poisson uncertainties of each magnitude bin with the error
bars. 
Focusing first on the $r < 19.5$ BGS Bright sample, Figure~\ref{fig:zsuccess}
demonstrates that we achieve >95\% redshift success rate for $r < 19.5$. 
Furthermore, we find little magnitude dependence on the redshift success rate
throughout the $r < 19.5$ magnitude range. 
\emph{Hence, the SV observations clearly demonstrate that BGS Bright sample can
achieve the >95\% redshift efficiency requirement with the $180s$ nominal
exposure time.}

\begin{figure}
\begin{center}
    \includegraphics[width=0.8\textwidth]{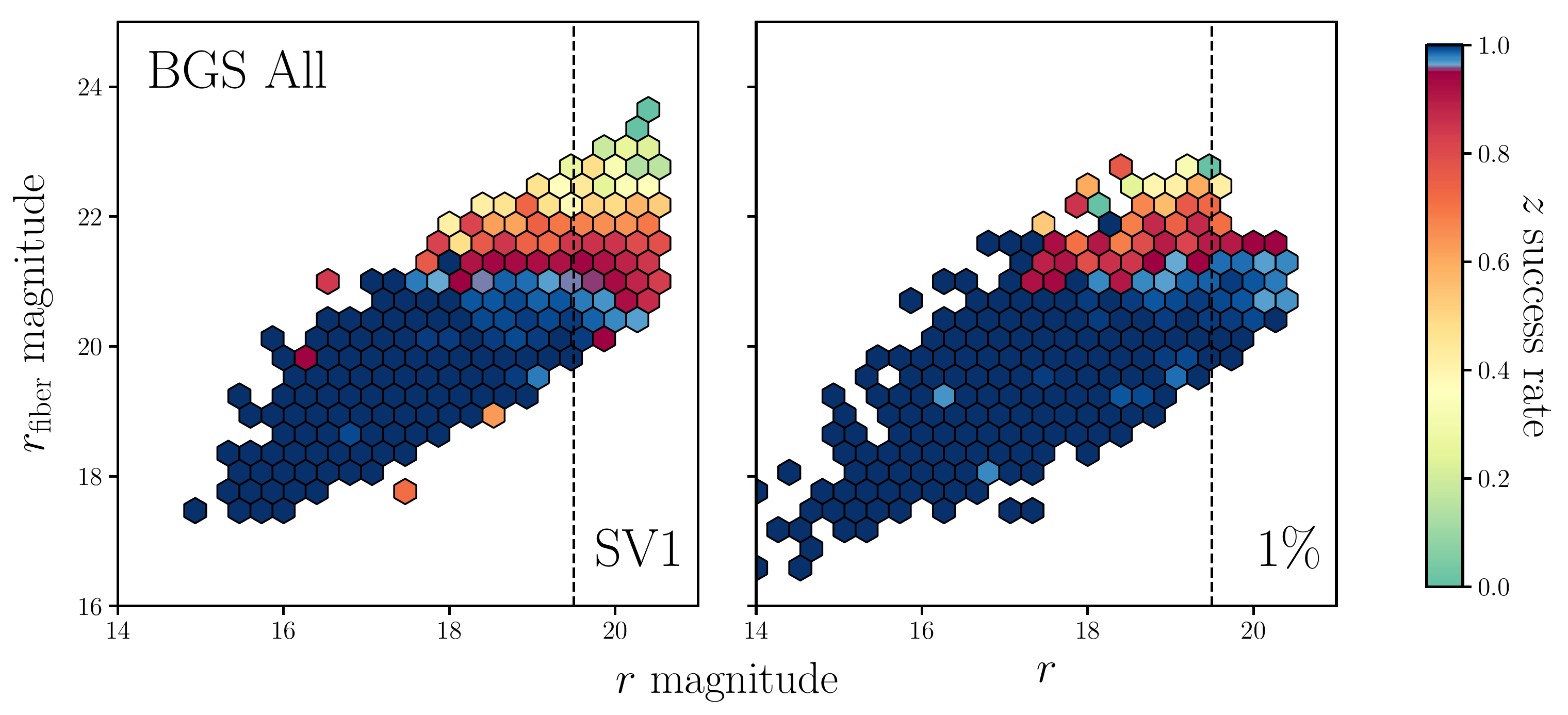}
    \caption{Redshift success rate of BGS galaxies as a function of $r$ and
    $r_{\rm fiber}$ magnitudes. 
    We present BGS spectra from \svi~(left) and the \sviii~(right).
    The color map represents the $z$ success rate; hex bins with >95\% redshift 
    success rate are marked with a blue color mapping.
    For the BGS Bright sample, we find overall >95\% redshift success rates. 
    Galaxies with fainter $r_{\rm fiber} > 21.5$ have lower redshift success 
    rates, as expected. 
    However, even for the small fraction of galaxies with $r_{\rm fiber} > 21.5$,
    BGS Bright maintains a $\gtrsim90\%$ redshift success rate with the final 
    selection~(right). 
    For the BGS Faint sample, the $r_{\rm fiber}$--~color cut in the
    \sviii~selection excludes galaxies with the lowest redshift succcess 
    rates in \svi.
    As a result, \emph{the BGS Faint sample has high redshift success rates
    (>90\%) throughout its entire $r$-$r_{\rm fiber}$ range}.
    }\label{fig:zsuccess_rmag_rfib}
\end{center}
\end{figure}

% all r-rfib   
We further examine the redshift success rate in Figure~\ref{fig:zsuccess_rmag_rfib},
as a function of $r$ and $r_{\rm fiber}$ magnitudes for the \svi~(left) and \sviii~(right). 
The color map represents the redshift success rate. 
For \svi, we again exclude the spectra in the {\em Low Quality} class.
To highlight the desired 95\% threshold, we present the bins with >95\%
redshift success rates with a blue color mapping.
The remainder are shown with an independent color bar of greater range. 
We only include bins with more than 10 galaxies to ensure accurate estimates. 
Consistent with Figure~\ref{fig:zsuccess}, we find mostly >95\% redshift success
rate for the BGS Bright sample (left of the black dashed line). 
We note that some of the bins with $r_{\rm fiber} > 21.5$ do not meet the 95\%
threshold.
This is more apparent for the \svi~sample, which has a more relaxed separation 
cut (Appendix~\ref{sec:sv1}).
For the final selection, even at $r_{\rm fiber} > 21.5$, the redshift success 
rate remains >90\%.

\begin{figure}
\begin{center}
    \includegraphics[width=0.75\textwidth]{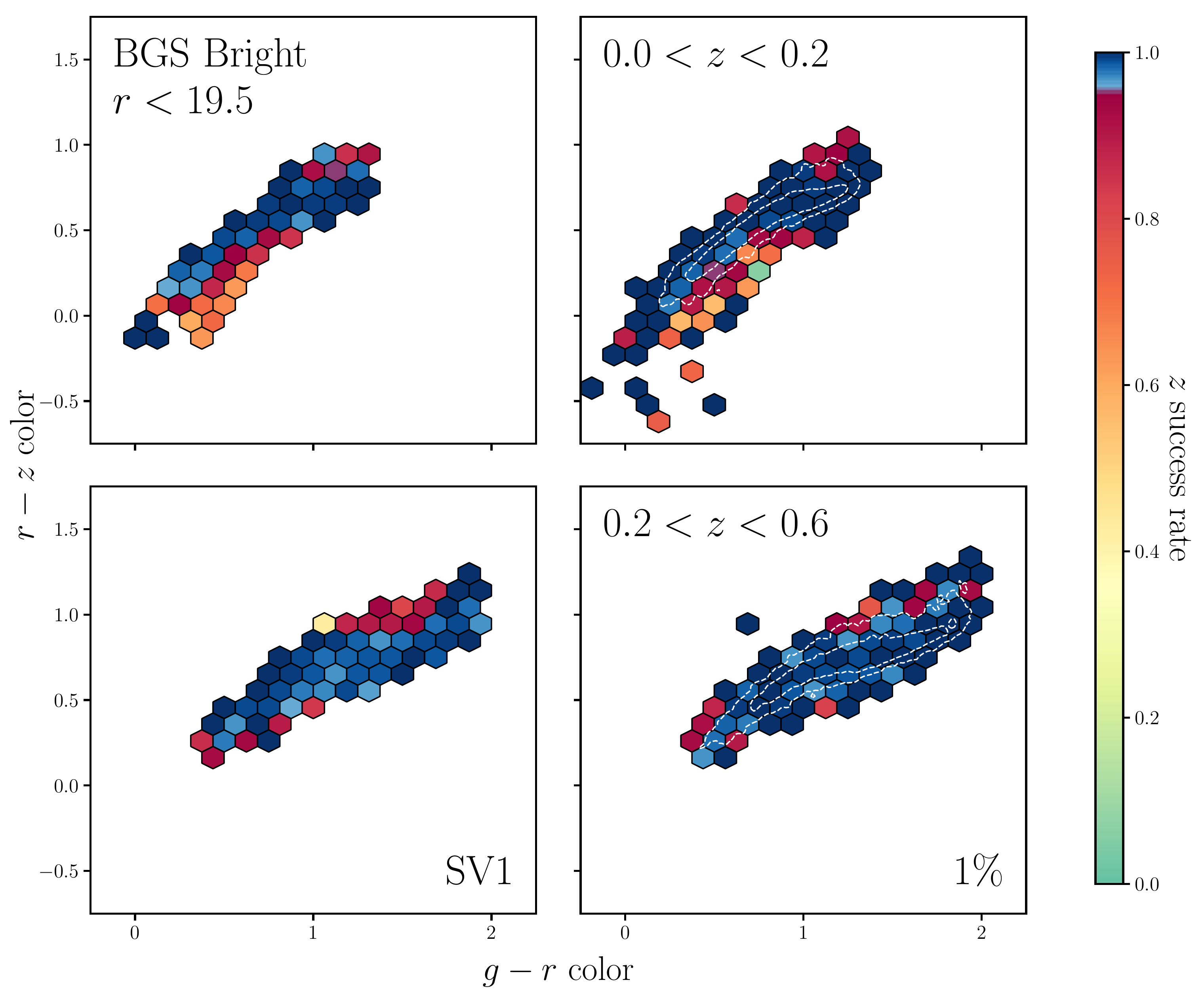}
    \caption{
    Redshift success rate of BGS Bright galaxies from \svi~(left) and the 
    \sviii~(right) as a function of $g-r$ and $r-z$ colors. 
    The top panels include galaxies with $0.0 < z < 0.2$; 
    the bottom panels include galaxies with $0.2 < z < 0.6$. 
    For reference, we mark the 68 and 95 percentiles of the color distribution 
    of \sviii~BGS Bright galaxies (white contours).
    The color map represents the $z$ success rate, where redshift success rates
    above >95\% are marked with a blue color mapping.
    We find slightly lower redshift success rates in the bluest galaxies ($g -
    r< 0.4$) because they are systematically fainter.
    \emph{Overall, we find >90\% redshift success rate throughout the
    color-space and no strong color dependence for the 1\% Survey.}
    } \label{fig:zsucc_color}
\end{center}
%\mw{This would make a lot more sense if it were intrinsic color?  John gives us the continua fits and rest-frame magnitudes so it'd be possible.}
\end{figure}

% bright color-color  
We also examine the redshift success rate of the BGS Bright sample as a
function of $g-r$ and $r-z$ color in Figure~\ref{fig:zsucc_color}. 
Again, we present the success rates for the \svi~and \sviii~in the left 
and right panels, respectively. 
We split the sample by the median redshift of the BGS Bright sample: 
$0.0 < z < 0.2$ galaxies (top panels) and $0.2 < z < 0.6$ galaxies 
(bottom panels).
The color map represents the redshift success rate and we use a blue color map
for the bins with >95\% redshift success rate.  
We exclude bins with less than 10 galaxies. 
We mark the 68 and 95 percentile contours of the color distribution of the
\sviii~BGS Bright galaxies in the right panels for reference (white dashed). 
Galaxies in the BGS Bright sample form a tight locus in color-space, well within
the color cuts we impose in the quality selection cuts
(Eq.~\ref{eq:quality_color}). 
Moreover, we find no strong color dependence in the redshift success rate. 
The slightly lower redshift success rate in the bluest galaxies with 
$g - r< 0.4$ is primarily driven by their overall fainter magnitudes. 
Only a small fraction of BGS Bright targets lie in regions of lower redshift 
success rate (top right).  
Overall, we find >90\% redshift success rate throughout the galaxy color-space
for the \sviii. 

\begin{figure}
\begin{center}
    \includegraphics[width=0.8\textwidth]{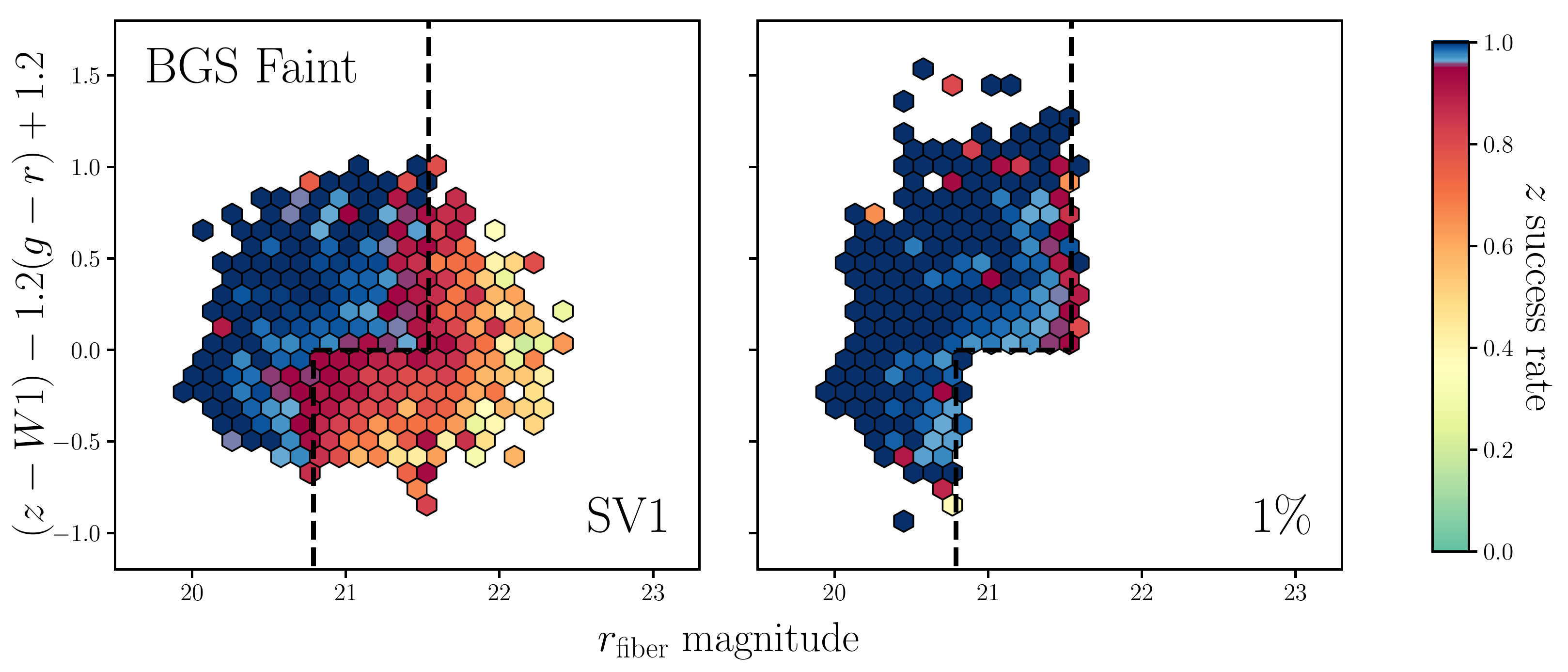}
    \caption{Redshift success rate of the BGS Faint sample as a function of 
    $r_{\rm fiber}$ and $(z - W1) - 1.2 (g - r) + 1.2$ color from the \svi~(left)
    and the \sviii~(right) observations. 
    We mark the $r_{\rm fiber}$--~color BGS Faint selection in black-dashed and 
    represent the redshift success rate with the color map. 
    The \svi~BGS Faint sample (left) is selected using only a $19.5 < r < 20.0$ 
    magnitude cut instead of the $r_{\rm fiber}$--~color selection. 
    The redshift success rate of \svi~galaxies outside the $r_{\rm fiber}$--~color
    cut illustrate that \emph{the BGS Faint selection is effective at excluding
    objects with low redshift success and produces a sample with high redshift
    efficiency.}
    } \label{fig:zsucc_faint}
\end{center}
\end{figure}

Next, we focus on the BGS Faint sample. 
In Figure~\ref{fig:zsuccess}, the \sviii~observations demonstrate that we meet the
>95\% redshift success rate for the BGS Faint sample over its entire $r$
magnitude range. 
We also note that the redshift success rate of the BGS Faint sample in the 
\sviii~is significantly higher than in the \svi~observations. 
This is driven by the $r_{\rm fiber}$--~color cut used in the \sviii, which is not used in
the \svi~BGS Faint selection.
As Figure~\ref{fig:zsuccess_rmag_rfib} demonstrates, $r_{\rm fiber}$--~color cut
excludes faint $r_{\rm fiber} > 21.5$ galaxies with low redshift success rate and
significantly increases the redshift success rate for the BGS Faint sample. 
With the $r_{\rm fiber}$--~color cut, even at the $r_{\rm fiber}\sim 21.5$ limit, we
maintain >87\% redshift success rate.

We further illustrate the effectiveness of the $r_{\rm fiber}$--~color BGS Faint
selection in Figure~\ref{fig:zsucc_faint}. 
We present the redshift success rate of the BGS Faint sample as a function of 
$r_{\rm fiber}$ and $(z - W1) - 1.2 (g - r) + 1.2$ color. 
The color map represents the redshift success rates and we mark the $r_{\rm
fib}$--~color cut (black dashed).
\svi~BGS Faint objects outside the $r_{\rm fiber}$--~color cut have significantly
lower redshift success rates. 
Meanwhile, \svi~BGS Faint objects within the $r_{\rm fiber}$--~color cut and 
the \sviii~BGS Faint objects have >90\% redshift success rates.
The $(z - W1) - 1.2 (g - r) + 1.2$ color is a proxy for emission line strengths 
(\emph{e.g.} H$\alpha$ and H$\beta$).
\emph{Therefore, the $r_{\rm fiber}$--~color cut successfully removes objects with
low redshift efficiency and by selecting on this cut we achieve >95\% redshift
success rate for the BGS Faint sample.}

\begin{figure}
\begin{center}
    \includegraphics[width=0.45\textwidth]{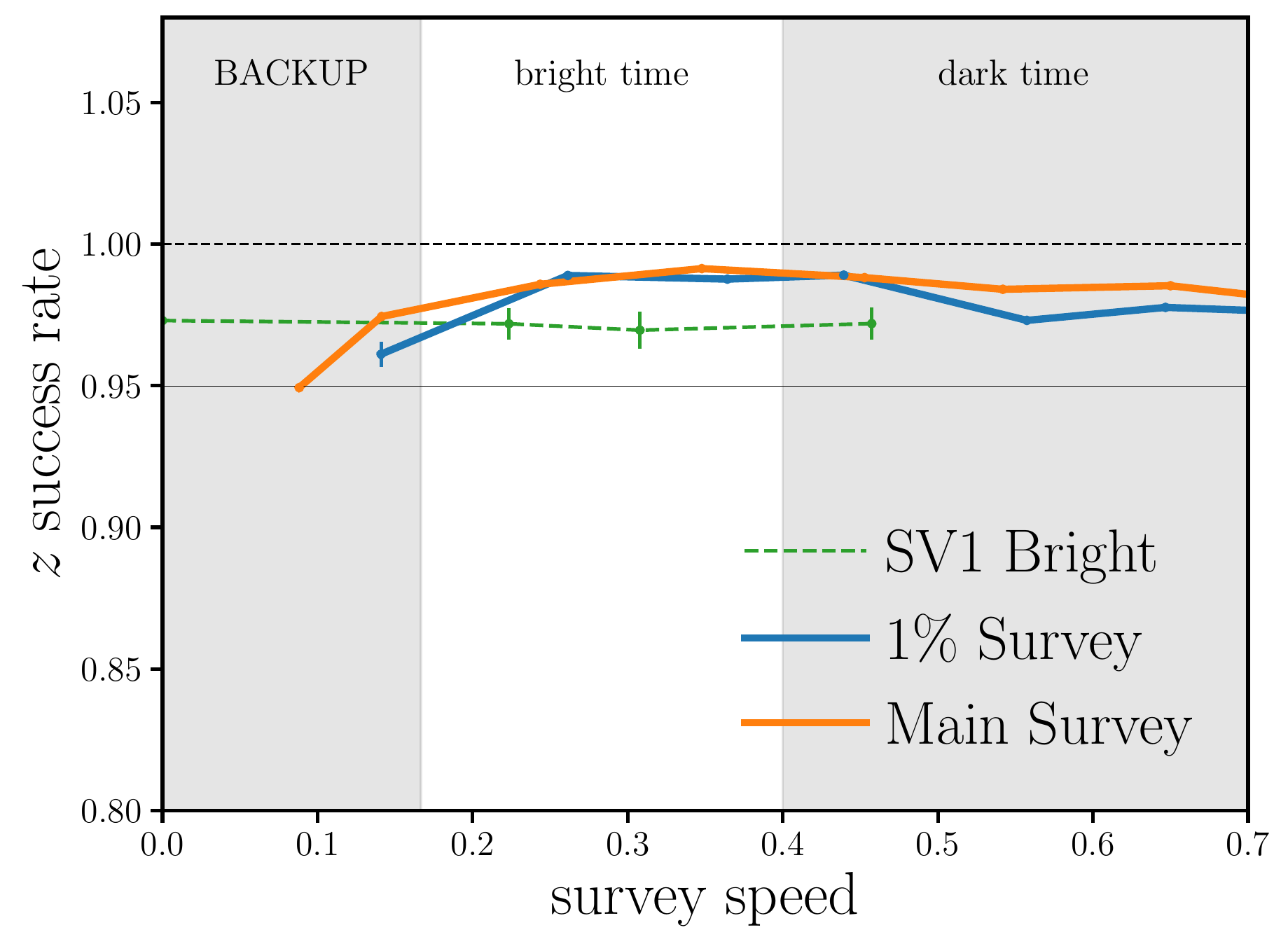}
    \caption{
        Redshift success rate as a function of `survey speed' for all BGS
        galaxies in the \sviii~(blue) and main survey (orange).  
        We also include the redshift success rate for the BGS Bright sample
        from \svi~(green dashed).
        Survey speed is used to determine whether to observe the bright time,
        dark time, or BACKUP program and is estimated by the Exposure Time
        Calculator. 
        It is derived from seeing, transparency, airmass, and sky brightness 
        of the exposure and serves as a metric for the observing conditions.
        The BGS survey speed boundaries are currently defined to be $[\frac{1}{6}, 0.4]$.
        \emph{BGS achieves a redshift success rate above the required >95\%
        threshold throughout its entire survey speed range.}
    } \label{fig:zsucc_speed}
\end{center}
\end{figure}

BGS will be observed under bright time conditions. 
In practice, observing conditions are classified as bright or dark time based on
the survey speed metric~\citep{ops}.
Survey speed assesses the effective seconds that would be accumulated per
second if the observations were taken at zenith and with zero extinction. 
In clouded-out conditions, survey speed is 0; in the best conditions, survey
speed is $\sim 2.5$. 
When survey speed is within a nominal range of $[\frac{1}{6}, 0.4]$, DESI will observe the
bright time programs (BGS and MWS). 
Above this range, DESI observes the dark time programs, and below it a `BACKUP'
program of particularly bright stars. 
Survey speed boundaries may be adjusted during the main survey to ensure the dark 
and bright surveys proceed at the planned rate. 
In Figure~\ref{fig:zsucc_speed}, we examine whether the redshift success rate
of BGS galaxies has any significant dependence on survey speed. 
We present redshift success rate of all BGS galaxies in the \sviii~(blue) and main
survey (orange) as well as BGS Bright galaxies in \svi~(green).  
Throughout the BGS survey speed range, we find little dependence on survey
speed. 
Moreover, \emph{we achieve >95\% redshift success rate throughout the bright
time observing conditions.}

\begin{figure}
\begin{center}
    \includegraphics[width=0.55\textwidth]{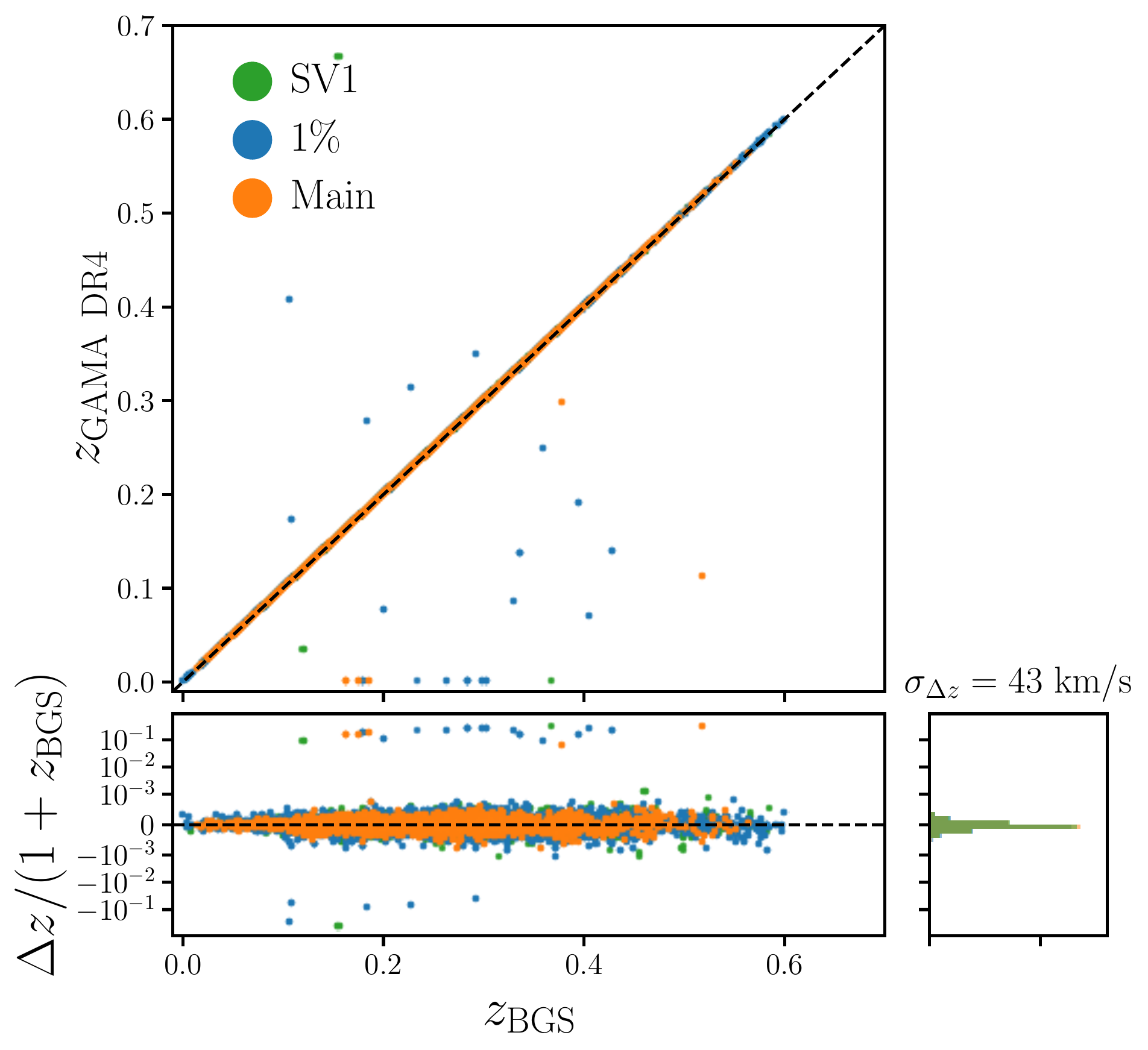}
    \caption{
    Comparison of BGS redshifts to GAMA redshifts for BGS galaxies from the 
    \svi~(green), \sviii~(blue), and main survey (orange) that are also in GAMA.
    In the top panel, we plot GAMA redshift, $z_{\rm GAMA}$,
    versus BGS redshift $z_{\rm BGS}$.
    In the bottom panels, we plot the fractional redshift residual, 
    $\Delta z/(1+z_{\rm BGS})$  as a function of $z_{\rm BGS}$ (left) and 
    the normalized histogram of $\Delta z/(1+z_{\rm BGS})$ (right). 
    The comparison only includes high quality GAMA redshifts with quality 
    flag ${\rm NQ} > 2$. 
    Of the ${\sim}$25,000 overlapping BGS galaxies,  99.7\% have
    $|\Delta z/(1+z_{\rm BGS})| < 0.001$.
    The scatter in the redshift difference between BGS and GAMA is small:
    $\sigma_{\Delta z} = 43$ km/s. 
    \emph{BGS redshifts show excellent overall agreement with GAMA redshifts.}
    } \label{fig:gama_z}
\end{center}
\end{figure}

In addition to the internal assessments, we can also compare BGS redshift
measurements to previous spectroscopic surveys because SV exposures were
observed in multiple overlapping regions (Section~\ref{sec:sv_obs}). 
The GAMA survey DR4~\citep{driver2022}, with its comparable magnitude limited selection, 
provides an ideal sample for assessing BGS redshifts.  
The GAMA sample extends to $r_{\rm SDSS} < 19.8$, where $r_{\rm SDSS}$ 
is the SDSS $r$ band Petrosian magnitude. 
In total, ${\sim}25,000$ BGS galaxies in the \svi, \sviii, and main survey were 
observed by GAMA and  have high quality redshifts (${\rm NQ} > 2$).
In Figure~\ref{fig:gama_z}, we compare the BGS redshifts, $z_{\rm BGS}$, to 
GAMA redshifts, $z_{\rm GAMA}$, for these overlapping galaxies from the 
\svi~(green) \sviii~(blue), and main survey (orange).
The top panel shows $z_{\rm GAMA}$ as a function of $z_{\rm BGS}$, with 
$z_{\rm GAMA} = z_{\rm BGS}$ included for reference (black-dashed).
The bottom left panel presents the fractional redshift residual 
$|\Delta z / (1 + z_{\rm BGS})|$ as a function of $z_{\rm BGS}$
and the bottom right panel presents the normalized $|\Delta z / (1 + z_{\rm BGS})|$
histogram.
Overall, BGS redshifts are in excellent agreement with GAMA redshifts.
We find over 99.7\% of the overlapping BGS galaxies have 
$|\Delta z/(1+z_{\rm BGS})| < 0.001$. 
\cite{vigal} presents a similar assessment of BGS redshifts, 
but using  redshifts from visual inspection in place of GAMA.

% summary of everything 
An overall $>95\%$ redshift success rate for the BGS Bright sample is necessary
to achieve the BGS science requirements, based on cosmological forecasts of
galaxy clustering analyses. 
We demonstrate above that we \emph{exceed} this requirement. 
In fact, we achieve an overall $>95\%$ redshift success rate for the BGS Faint
sample as well, with the updated selection criteria
(Figures~\ref{fig:zsuccess_rmag_rfib} and~\ref{fig:zsucc_faint}).
Furthermore, we demonstrate that the BGS Bright sample has a weak $r_{\rm fiber}$
and optical color dependence (Figures~\ref{fig:zsuccess_rmag_rfib}
and~\ref{fig:zsucc_color}). 
Lastly, we demonstrate that the BGS redshift efficiency does not depend
significantly on observing conditions (Figure~\ref{fig:zsucc_speed}). 

%%%%%%%%%%%%%%%%%%%%%%%%%%%%%%%%%%%%%%%%
\subsection{Fiber Assignment Efficiency} \label{sec:fibeff}
As described in Section~\ref{sec:design}, BGS will observe its footprint
with 4 passes (effective 3 visits). 
Targets will be assigned fibers using the $\mathtt{fiberassign}$
code\footnote{\url{https://fiberassign.readthedocs.io}},  where
the highest priority for bright time will be given to the BGS Bright targets
and 20\% of the BGS Faint targets. 
The other 80\% of BGS Faint targets are assigned at a lower priority. 
To achieve the scientific objectives of BGS, we require a fiber 
assignment efficiency of >80\% for BGS Bright --- \emph{i.e.} >80\% of
BGS Bright targets must be assigned fibers over the course of the survey.

\begin{figure}
\begin{center}
    \includegraphics[width=0.45\textwidth]{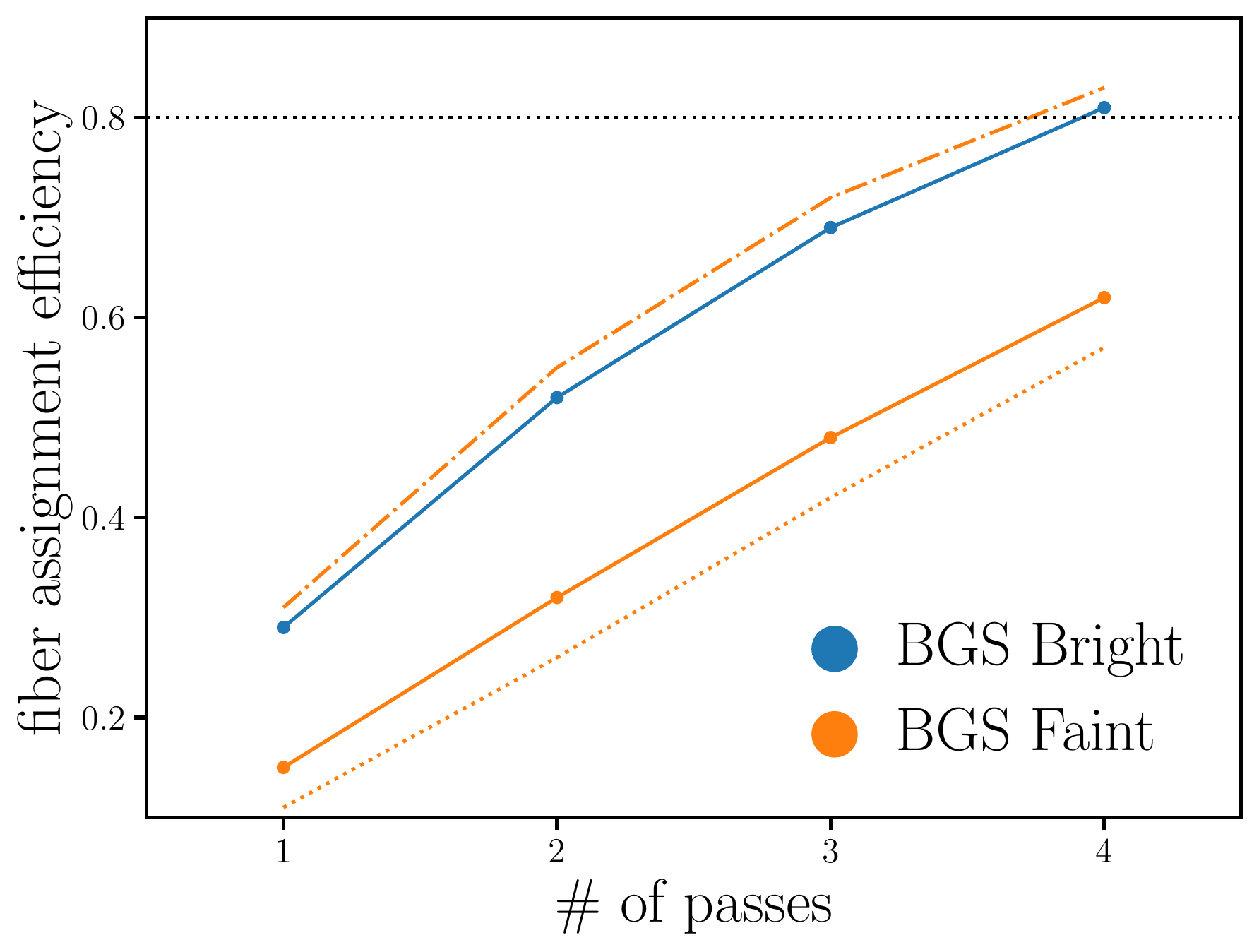}
    \caption{\label{fig:fa_eff}
    Fiber assignment efficiency of BGS Bright (blue) and Faint
    (orange) as a function of the number of passes.
    BGS will be observed with four passes; each position in the
    footprint will be visited on average 3 times to ensure the required 
    completeness. 
    The BGS Bright targets and 20\% of the BGS Faint targets (orange 
    dot-dashed) are assigned fibers with the highest priority during 
    bright time. 
    The rest of the BGS Faint targets are assigned lower priorities 
    (orange dotted).
    We assign higher priorities to a fraction of BGS Faint targets
    to facilitate corrections for fiber assignment incompleteness
    in future clustering analyses. 
    The fiber assignment efficiencies are estimated for the 
    focal plane condition as of October 2021, where 4,174 of 5,000 
    fiber positioners are available for science targets.
    Some of the remaining malfunctioning positioners are used 
    to meet the sky fiber budget.  
    Even if the number of functional positioners does not increase in the
    future, \emph{with four passes we achieve the 80\% fiber assignment 
    efficiency requirement for BGS Bright}.
    } 
\end{center}
\end{figure}

At the beginning of SV operations, the state and knowledge of the focal 
plane condition evolved rapidly, including the discovery of poor 
positioning and faulty electronics. 
As of October 2021, 4,174 of the 5000 positioners in the focal plane are
fully functional and currently being used for science targets.  
The remaining malfunctioning positioners are currently being used to satisfy
the sky fiber budget.  
The reduction in functional fiber positioners significantly reduces the overall
fiber assignment efficiency for DESI.
In Figure~\ref{fig:fa_eff}, we present the fiber assignment efficiency of
the BGS Bright (blue) and Faint (orange solid) samples as a function of 
the number of passes for the focal plane status as of October 2021. 
We also include the fiber assignment efficiency for the higher (orange 
dot-dashed) and lower priority (orange dotted) BGS Faint targets. 
Even with the reduced focal plane capabilities, \emph{we confirm that with 4 
passes we achieve the >80\% fiber assignment efficiency requirement for the 
BGS Bright sample.}

%% file: svda.tex
%%%%%%%%%%%%%%%%%%%%%%%%%%%%%%%%%%%%%%%%%%%%%
\section{BGS Early Data Release} \label{sec:survey}
In the previous section, we demonstrated with DESI Survey Validation data that BGS will 
meet its requirements on completeness and redshift efficiency.  
In the following, we showcase the key advantages of the BGS galaxy 
samples based on the first public dataset, the Early Data Release (EDR).

\begin{figure}
\begin{center}
    \includegraphics[width=0.8\textwidth]{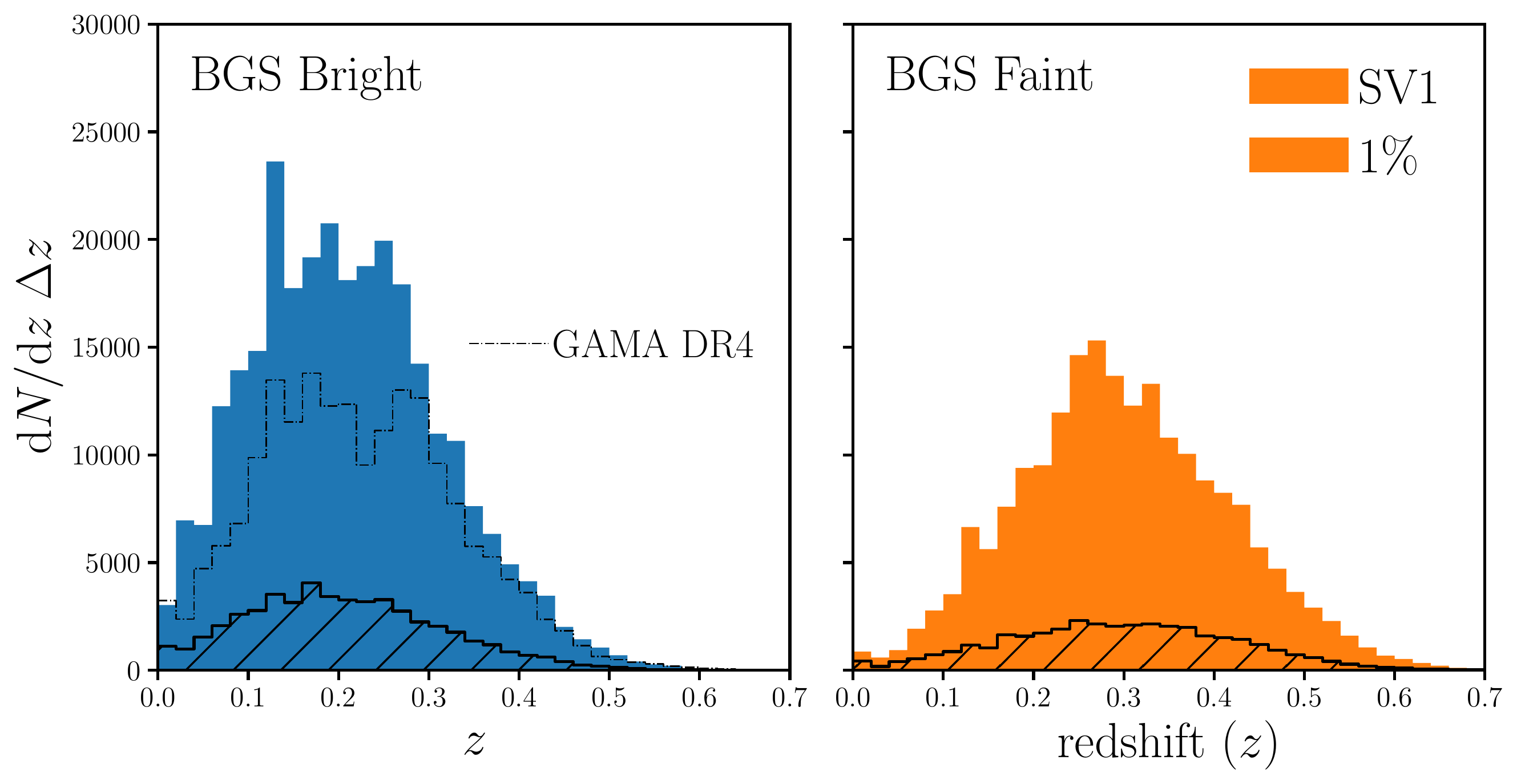}
    \caption{
    Redshift distribution of BGS galaxies from \svi~(hashed) and the \sviii~(solid).
    We present the distributions of the BGS Bright and Faint galaxies in
    the left and right panels, respectively. 
    The success of BGS targeting delivers galaxies in the desired redshift range,
    $0.0 < z < 0.6$. 
    The Bright and Faint samples have median redshifts of $z\sim0.2$ and 0.3,
    respectively.  
    During $\sim$5 months of SV observations, DESI amassed 286,934 BGS Bright 
    and 202,590 Faint BGS redshifts at an average rate of $\simeq 4,200$ per 
    20 minute exposure. 
    The BGS Bright sample from SV already exceeds the number of redshifts in
    GAMA DR4 (dot dashed). 
    \emph{Over the course of the next 5 years, BGS will expand to an unprecedented 
    $>10$ million spectra over a third of the sky.}
    }\label{fig:dndz}
\end{center}
\end{figure}

In total, DESI amassed redshifts of 285,335 of BGS Bright galaxies and 201,532 
BGS Faint galaxies in $\sim$5 months of operations. 
DESI acquired $\simeq 4,200$ redshifts per exposure at a rate of one exposure 
per 20 minutes on average. % --- with a reconfiguration time of \todo{XXX} minutes.
%\todo{which state-of-the-art to date?}
%Thanks to the success of the instrument design, a dataset larger than the state-of-the-art to date was delivered in only approximately 5 months.  
Over the next five years, this will expand to an unprecedented
>10 million galaxies spanning a third of the sky at two
magnitudes deeper than the SDSS MGS. 
In Figure~\ref{fig:dndz}, we highlight the progress of BGS and present the absolute 
number of redshifts in $\Delta z=0.02$ bins for BGS Bright (left; blue) and
Faint (right; orange) galaxies observed during the SV programs.  
BGS targeting successfully delivers galaxies spanning the desired redshift 
range, $0.0 < z < 0.6$.  
The Bright and Faint samples have median redshifts of $z=0.2$ and $0.3$ 
respectively, which is more than double that of the SDSS MGS. 
We include the redshift distribution of GAMA DR4 for comparison.
Even with the EDR alone, BGS exceeds the total number of spectroscopic
redshift of GAMA. 

\begin{figure}
\begin{center}
    \includegraphics[width=0.5\textwidth]{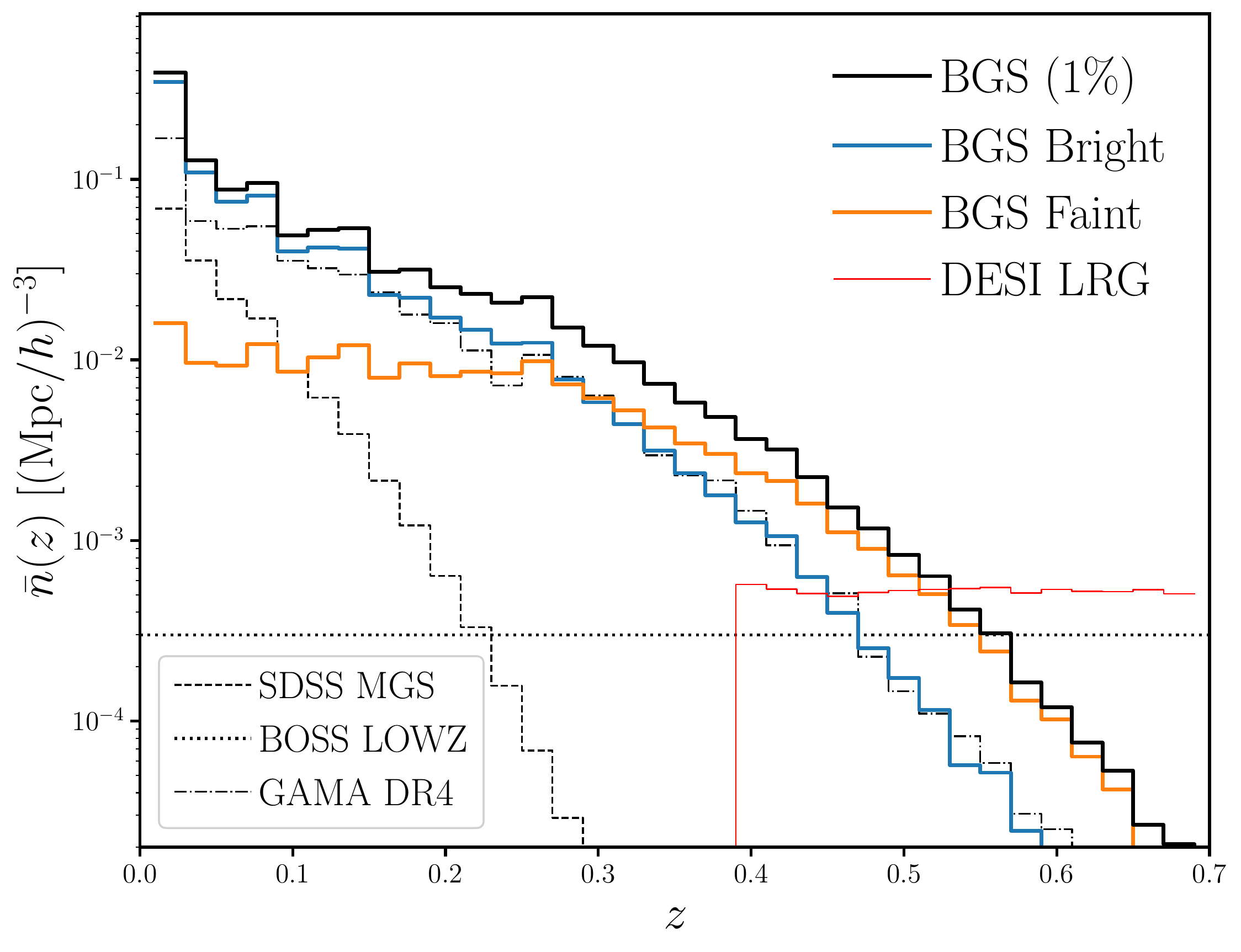}
    \caption{
    Comoving number density distribution, $\bar{n}(z)$, of BGS galaxies from 
    the EDR (black; \svi~and the \sviii). 
    We plot $\bar{n}(z)$ for BGS Bright (blue) and Faint (orange).  
    We show external datasets of interest, SDSS (dashed), BOSS 
    (dotted), and GAMA DR4 (dash dotted), and DESI LRGs (red) in
    the overlapping range.  
    BGS will provide the highest density cosmological galaxy sample to date at 
    low redshifts.
    It will provide maximum leverage against higher redshift measurements and 
    CMB constraints for dark energy constraints. 
    Furthermore, its high density and broad galaxy selection will enable a wide 
    range of new approaches to galaxy clustering and galaxy evolution studies.
    }\label{fig:nz}
\end{center}
\end{figure}

In Figure~\ref{fig:nz}, we further highlight the advantages of BGS through its
comoving number density distribution, $\bar{n}(z)$.
BGS will have significantly higher number density than any previous  survey in 
this redshift range.  
We include $\bar{n}(z)$ for SDSS (dashed), BOSS (dotted), and GAMA DR4 (dash dotted) for
comparison. 
We also include $\bar{n}(z)$ for the DESI dark-time LRGs above  $z\gtrsim 0.4$, 
the redshift limit that will be imposed for LRG galaxy clustering analyses (red). 
At $z = 0.2$, the BGS number density is more than an order-of-magnitude larger 
than SDSS.
%Based on the validated performance of BGS in Sections~\ref{sec:survey} and~\ref{sec:fibeff}, we expect \todo{X,Y,Z} redshifts over \todo{X,Y,Z} deg$^2$  for the Y1, Y3 and Y5 data releases, respectively. 

%% file: summary.tex
\section{Summary} \label{sec:summary}
Over the next five years, DESI will conduct dark and bright time spectroscopic
galaxy surveys on the 4m Mayall Telescope at Kitt Peak National Observatory,
using robotically-actuated fiber-fed spectrographs that can collect 5000
spectra  simultaneously.  
As the first Stage-IV dark energy experiment to be realized, it promises to make unprecedented 
measurements of cosmic acceleration and advance our understanding of the nature of dark energy. 
During dark conditions, DESI will measure the redshifts of 20 million LRGs, 
ELGs, and quasars from $z\approx 0.4$ to $3.0$.
During bright conditions, DESI will survey the low redshift 
Universe,  $0 < z < 0.6$, when dark energy is most dominant, with 
>10 million BGS galaxies.
By targeting brighter galaxies at closer distances, BGS proceeds effectively 
in slower conditions and makes optimal use of bright time. 

BGS will enable a broad range of science goals from probing dark energy to
studying dwarf galaxies. 
To achieve this, we require BGS to sample a wide range of galaxy types, have a
high and well-characterized completeness, and be at least an order of magnitude 
larger than SDSS MGS. 
These requirements translate, in practice, to the following: the primary galaxy 
sample (BGS Bright) will be selected using a magnitude limit that yields a target 
density >800 targets/deg$^2$; BGS targets will have a stellar contamination 
rate of <1\%; >80\% of BGS Bright targets will be assigned to fibers; and redshifts
will be successfully measured for >95\% of those assigned fibers.  
BGS will also  cover a footprint of 14,000 deg$^2$ and complete its 5 year operations 
with 20\% margins (in $\sim$4 years) in simulated operations.  

% target selection summary
In this work, we present the finalized target selection and survey design for BGS.  
BGS targets are selected from the LS DR9 imaging surveys (DECaLS, BASS, and M$z$LS),
with supplementary data from external catalogs (including \gaia~DR2, Tycho-2, and SGA). 
We apply spatial masking together with fiber magnitude, quality, and bright-end cuts
to remove spurious sources and contaminants. 
We further impose a $(G_{Gaia} - r_{\rm raw}) > 0.6$ cut to remove stars while retaining a 
high galaxy completeness.  
From the target set, we select the following samples designed to achieve well defined 
goals:  
\begin{itemize}
    \item \textbf{BGS Bright} is the primary sample assigned with highest priority. 
    As a magnitude-limited sample ($r < 19.5$) BGS Bright meets the stated 
    requirements on completeness and density, with $\sim$860 targets/deg$^2$ . 
    \item \textbf{BGS Faint} selects fainter galaxies at $19.5 < r < 20.175$ with 
    an additional $r_{\rm fiber}$--~color cut to ensure a high redshift efficiency 
    and boost the comoving density. 
    We find the color, $(z - W1) - 1.2 (g - r) + 1.2$, to be an accurate proxy for 
    emission line flux. 
    Hence, the $r_{\rm fiber}$--~color cut identifies galaxies faint in $r$ that either 
    have relatively bright fiber magnitudes or have strong emission lines. 
    \item \textbf{BGS AGN} is a supplementary sample of AGN host galaxies that 
    are otherwise rejected by the BGS star-galaxy separation and raise the 
    completeness of the dark time DESI quasar selection.  
    The primary selection criteria are based on optical and WISE infrared colors 
    that trace the signatures of hot, AGN-heated dust. 
\end{itemize}
We note that the selection criteria above are significantly updated from preliminary 
versions presented in \cite{ruiz-macias2021} and \cite{zarrouk2021}, particularly for BGS Faint. 
See Section~\ref{sec:ts} for further information.

%survey design summary
After target selection, BGS targets are assigned fibers based on a strategy optimized 
to ensure the highest completeness for BGS Bright and otherwise maximize the number 
of successful assignments.  
We assign in the first priority tier all BGS Bright and 20\% of BGS Faint targets.  
The remaining 80\% of BGS Faint and BGS AGN targets are assigned to a lower priority tier. 
We promote a random subsample of BGS Faint targets to facilitate later  corrections 
for fiber assignment incompleteness. 
Conflicts caused by cases where multiple targets of the same priority are  available 
to a fiber are resolved by a random sub-priority given to each target.

DESI will observe BGS during `slower' conditions, according to a predetermined 
threshold on `survey speed' --- a dynamically calculated metric for the impact 
of observing conditions derived from seeing, transparency, airmass, and sky
brightness.  
BGS will observe a footprint of 14,000 deg$^2$ as closely matched to the dark time 
program as possible.
Each point on the footprint will be visited three times on average, using a four 
pass strategy, to ensure a high (>80\%) fiber assignment completeness. 
Exposure times are dynamically scaled based on the measured observing conditions 
to yield uniform redshift efficiency and a close-to-homogeneous survey.  
We set the anchoring exposure time to $t_{\rm nom}$ = 180s based on spectral 
simulations. 
$t_{\rm nom}$ is defined as the exposure time required to achieve >95\% redshift efficiency
for the BGS Bright sample under nominal dark conditions.  
Based on forecasts using simulations of survey operations, we confirm that we 
complete the BGS survey as detailed above in 5 years with a margin of 20\%
(Section~\ref{sec:strat}).  

A primary goal of this work is to demonstrate that these BGS design choices 
achieve its stated requirements --- specifically, those on target density, 
redshift efficiency, and fiber  assignment efficiency.  
To do so, we utilize spectroscopic observations from the SV program conducted by DESI 
before the start of the main survey.
SV was divided into the \svi~and \sviii, observed over $\sim$110 nights (Section~\ref{sec:sv_obs}). 
\svi~aimed to characterize the redshift performance under different observing 
conditions and optimize the target selection. 
Meanwhile, the \sviii~aimed to validate the main survey design choices and 
provide clustering samples of particularly high completeness for additional tests.
Using LS imaging data and these SV observations, we have demonstrated in
Section~\ref{sec:sv} that:  
\begin{itemize}
    \item Our stellar mask together with our fiber magnitude, quality, and bright-end 
    cuts successfully remove spurious sources  without significantly impacting the 
    target completeness. 
    Furthermore, we find <1\% stellar contamination for BGS targets. 
    Overall, we find <5\% variation in the target densities of BGS Bright and Faint samples
    and no strong dependence on imaging properties (Figures~\ref{fig:stargalaxy} and~\ref{fig:image_sys}). 
    \item We achieve the required 95\% redshift success rate for the BGS Bright sample 
    with an exposure time of $t_{\rm nom} = 180$s under nominal conditions 
    (Figures~\ref{fig:zsuccess}). 
    This redshift efficiency does depend markedly on $r_{\rm fiber}$ and optical colors, 
    as expected, but our target selection achieves a high redshift efficiency for a broad 
    range of galaxies 
    (Figures~\ref{fig:zsuccess_rmag_rfib} and~\ref{fig:zsucc_color}). 
    \item We achieve >95\% redshift success rate for the BGS Faint sample
    using the  $r_{\rm fiber}$--~color based selection,  
    which effectively identifies targets with relatively bright $r$ fiber magnitudes 
    or strong emission lines (Figures~\ref{fig:zsuccess_rmag_rfib} and~\ref{fig:zsucc_faint}). 
    \item We achieve >95\% redshift success rate for both BGS Bright and Faint over the  
    range of survey speeds expected to be available to BGS (Figure~\ref{fig:zsucc_speed}). 
    This demonstrates that the impact of observing conditions is well understood and that
    the spectroscopic pipeline is robust to systematic error in challenging conditions. 
    \item Lastly, we achieve >80\% fiber assignment efficiency for the BGS Bright sample 
    with a four pass strategy (Figure~\ref{fig:fa_eff}). 
    This estimate is based on the focal plane status as of October 2021, which has 
    $\simeq 4,200$ positioners available to science targets. 
    Efforts to fix malfunctioning positioners are ongoing and should ultimately 
    result in a higher fiber assignment efficiency for the same four pass strategy. 
\end{itemize}

% paragraph reemphasizing the many applications of the BGS samples 
Overall, we have demonstrated that BGS will successfully deliver a >10 million galaxy 
sample within $0 < z < 0.6$ at a high completeness over a wide range of galaxy properties.
As such, BGS will be an order of magnitude larger than the SDSS MGS and will provide the 
densest galaxy sample out to $z\approx0.45$ to date. 
Clustering analyses of BGS galaxies will produce the most precise low-$z$ BAO and RSD 
measurements, thereby maximizing the leverage against higher redshift measurements, 
such as the CMB. 
BGS presents a unique discovery space for testing the predictions of dark energy and 
modified gravity models.  
Furthermore, its high sampling density makes BGS ideal for state-of-the-art analyses 
using galaxy-galaxy lensing, higher-order statistics, small-scale clustering, and 
multi-tracer techniques. 
In addition, BGS will be an unprecedented sample for studying galaxy formation and
decoding the relation between galaxies and dark matter.

%% file: acknowledgements.tex
%%%%%%%%%%%%%%%%%%%%%%%%%%%%%%%%%%%%%%%%%%%%%
\section*{acknowledgements}
% first-tier author acknowledgements
We thank Marla Geha for leading the Bright Galaxy Survey design in its early stages.
CH is supported by the AI Accelerator program of the Schmidt Futures Foundation. 
MJW, OR-M, SMC, CSF and PN  acknowledge STFC support (grant ST/T000244/1).
DHW is supported by NSF grant AST-2009735. 
%ADM was supported by the U.S. Department of Energy, Office of Science, Office of High Energy Physics, under Award Number DE-SC0019022. 
%HZ acknowledges the support of the National Natural Science Foundation of China  (NSFC) under grants 12120101003 and 11890691.

This research is supported by the Director, Office of Science, Office of High Energy Physics 
of the U.S. Department of Energy under Contract No. DE-AC02-05CH11231, and by the National 
Energy Research Scientific Computing Center, a DOE Office of Science User Facility under the 
same contract; additional support for DESI is provided by the U.S. National Science Foundation, 
Division of Astronomical Sciences under Contract No. AST-0950945 to the NSF's National 
Optical-Infrared Astronomy Research Laboratory; the Science and Technologies Facilities 
Council of the United Kingdom; the Gordon and Betty Moore Foundation; the Heising-Simons 
Foundation; the French Alternative Energies and Atomic Energy Commission (CEA); the National 
Council of Science and Technology of Mexico (CONACYT); the Ministry of Science and Innovation 
of Spain (MICINN), 
and by the DESI Member Institutions: 
\url{https://www.desi.lbl.gov/collaborating-institutions}.

The DESI Legacy Imaging Surveys consist of three individual and complementary projects: the Dark Energy Camera Legacy Survey (DECaLS), the Beijing-Arizona Sky Survey (BASS), and the Mayall z-band Legacy Survey (MzLS). DECaLS, BASS and MzLS together include data obtained, respectively, at the Blanco telescope, Cerro Tololo Inter-American Observatory, NSF’s NOIRLab; the Bok telescope, Steward Observatory, University of Arizona; and the Mayall telescope, Kitt Peak National Observatory, NOIRLab. NOIRLab is operated by the Association of Universities for Research in Astronomy (AURA) under a cooperative agreement with the National Science Foundation. Pipeline processing and analyses of the data were supported by NOIRLab and the Lawrence Berkeley National Laboratory. Legacy Surveys also uses data products from the Near-Earth Object Wide-field Infrared Survey Explorer (NEOWISE), a project of the Jet Propulsion Laboratory/California Institute of Technology, funded by the National Aeronautics and Space Administration. Legacy Surveys was supported by: the Director, Office of Science, Office of High Energy Physics of the U.S. Department of Energy; the National Energy Research Scientific Computing Center, a DOE Office of Science User Facility; the U.S. National Science Foundation, Division of Astronomical Sciences; the National Astronomical Observatories of China, the Chinese Academy of Sciences and the Chinese National Natural Science Foundation. LBNL is managed by the Regents of the University of California under contract to the U.S. Department of Energy. The complete acknowledgments can be found at \url{https://www.legacysurvey.org/}.

The authors are honored to be permitted to conduct scientific research on Iolkam Du’ag (Kitt Peak), a mountain with particular significance to the Tohono O’odham Nation.

%% file: sv1.tex
\section{\svi~Selection} \label{sec:sv1}
Survey Validation for DESI was divided into two main programs: \svi~and the \sviii. 
We primarily focus on \sviii~in this work, since it used the same target selection 
as the main survey (Section~\ref{sec:select}) as well as a similar observing 
strategy (Section~\ref{sec:sv_obs}).
\svi~provided the prior observations that we used to devise and optimize the 
\sviii~and final target selection and observing strategy. 
In this appendix, we describe the target selection that was used for \svi. 
The observing strategy for the \svi~program is described in Section~\ref{sec:sv_obs}
and also in~\citep{sv}.

Preliminary BGS target selection established that observing a magnitude-limited BGS Bright
sample and a fainter BGS Faint sample would enable a survey that can achieve a broad 
range of science goals~\citep{ruiz-macias2021, zarrouk2021}. 
With \svi~observations, we aimed to test more specific choices in the star-galaxy 
separation, BGS Faint sample selection, and quality cuts. 
First, for the star-galaxy separation, we wanted to confirm whether the final criteria
of excluding \emph{Gaia} objects with $(G_{\rm Gaia} - r_{\rm raw}) \leq 0.6$ (Section~\ref{sec:ts}) 
sufficiently removes stellar contaminants. 
Therefore, for all \svi~targets, we used a relaxed star-galaxy separation that only excludes
\emph{Gaia} objects that have $(G_{\rm Gaia} - r_{\rm raw}) \leq 0.6$ and is also best fit
by a PSF model in {\sc Tractor}. 
We also did not impose the FMC (Eq.~\ref{eq:fmc}) or the bright limit (Eq.~\ref{eq:bright_limit}).

\begin{figure}
    \centering
    \includegraphics[width=0.5\textwidth]{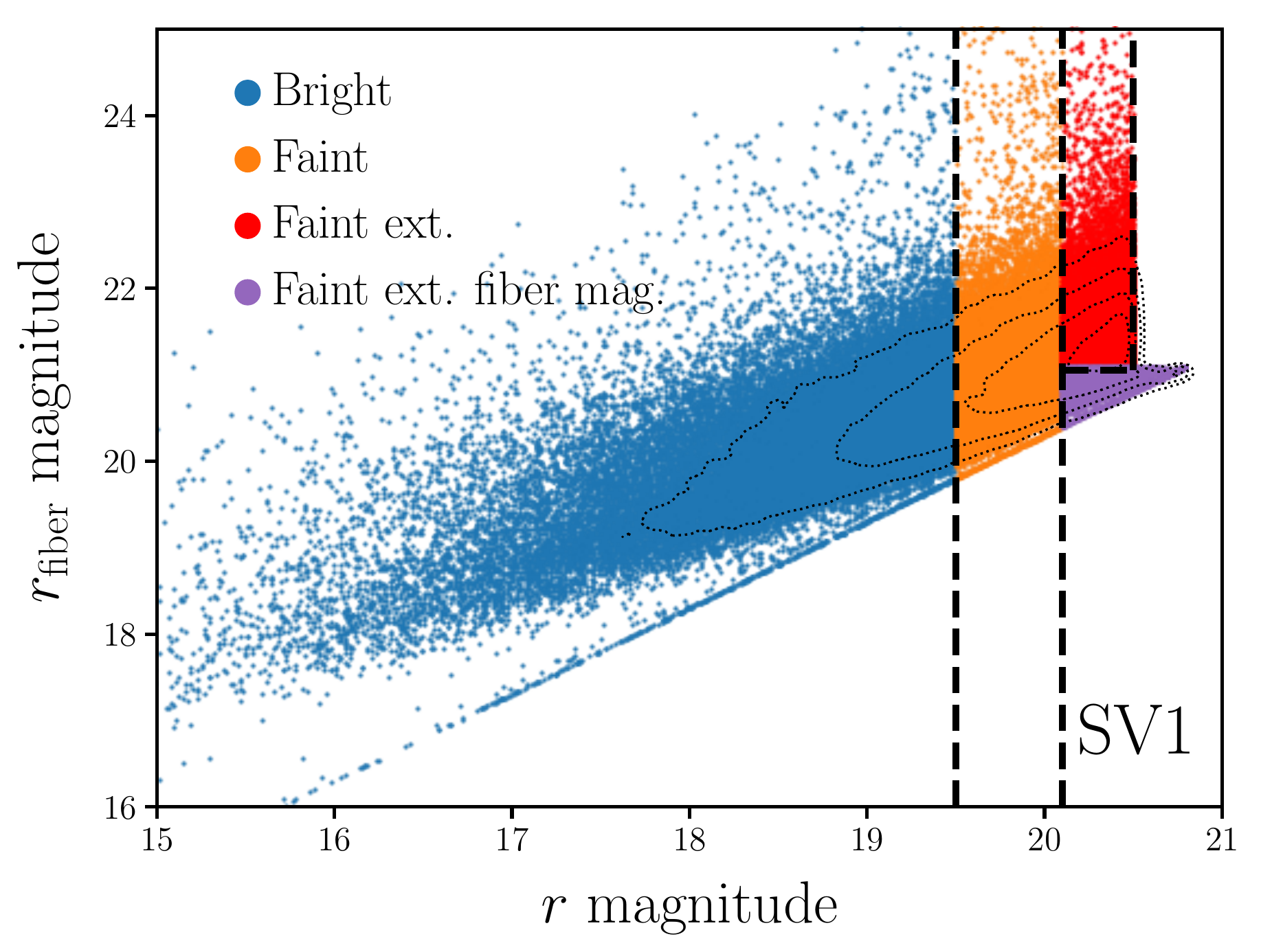}
    \caption{\label{fig:sv1}
    $r_{\rm fib}$ versus $r$ distribution of \svi~target in the `Bright' (blue),
    `Faint' (orange), `Faint extended' (red), and `Faint extended fiber magnitude' 
    (purple) target classes. 
    We mark the $r$ and $r_{\rm fib}$ limits of the target classes in black dashed.
    Targets selection for \svi~was designed to test the star-galaxy separation, 
    BGS Faint sample selection, and quality cuts. 
    Therefore, the \svi~selection, which we present in Appendix~\ref{sec:sv1}, is 
    overall a more relaxed version of the final BGS target selection (Section~\ref{sec:ts}). 
    }
    \label{fig:sv1selection}
\end{figure}
Next, we designed \svi~to explore the BGS Faint sample selection and whether 
additional cuts could significantly increase its redshift success rate.  
In addition to a `Bright' target class that corresponds the $r < 19.5$ BGS Bright 
sample, \svi~also included three additional fainter target classes.  
(1) A `Faint' sample with $19.5 < r < 20.1$ targets that extends 0.1 magnitudes
fainter than the preliminary BGS Faint selection. 
(2) A `Faint extended' sample with even fainter targets $20.1 < r < 20.5$ that
also have $r_{\rm fib} > 21.051$.
(3) A `Faint extended fiber magnitude' sample also with fainter targets, $r > 20.1$,
but with brighter fiber magnitudes, $r_{\rm fib} < 21.051$.
The goal of (1) was to explore the relationship between magnitude and redshift 
success rate in further detail for the BGS Faint sample. 
The goal of (2) and (3) was to probe whether there are any subsets of fainter 
galaxies that would produce higher redshift success rates. 
In Figure~\ref{fig:sv1}, we present $r$ versus $r_{\rm fib}$ relation of \svi~target 
classes. 
The final selection of the BGS Faint sample ($19.5 < r < 20.175$ and 
$r_{\rm fib}$-color cut), was determined using observations of the fainter \svi~targets. 

Lastly, we used \svi~observations to test the preliminary quality cuts 
from~\citep{ruiz-macias2021}. 
In addition to the quality cuts in the final selection (Eqs.~\ref{eq:nobs} 
and \ref{eq:quality_color}), the preliminary cuts also include cuts based on 
{\sc Tractor} photometric quality flags: ($\mathtt{FRACMASK}_i$ < 0.4) and 
($\mathtt{FRACIN}_i$ > 0.3) and ($\mathtt{FRACFLUX}_i$ < 5).
To test whether each of the preliminary quality cuts minimize spurious targets 
from photometric artifacts  without sacrificing completeness, we include a 
`Low Quality' target class in \svi.  
This target class included $r < 20.1$ objects (131 objects deg$^{-2}$) without 
any quality cuts. 
Later, through visually inspection, we confirm that the {\sc Tractor}  based 
quality cuts removes a significant number of real galaxies so we exclude 
them in the final quality cut (Section~\ref{sec:ts}). 
When we construct the target catalog for \svi, we randomly sample the targets 
to meet densities of 300, 50, 50, and 20 targets/deg$^{2}$ for the `Faint', 
`Faint extended',  `Faint extended fiber magnitude', and `Low Quality' classes, 
respectively. 

%% file: sky.tex
\section{Bright time Sky Brightness Model} \label{sec:sky}
Survey simulations of DESI operations were a key component of determining 
the final design and observing strategy of BGS.
They simulate the detailed operations of DESI observations and account for 
expected configuration and dead times of the instrument as well as historical 
weather, seasonal, and environmental factors. 
Nightly operations are simulated as a sequence of tile exposures, where the
exposure time is scaled according to the predicted sky brightness at the 
time of observation. 
For the dark program, the simulations include a simplified model of sky 
brightness that is sufficiently accurate for dark time~\citep{ops}. 
However, accurately simulating BGS operations requires a more accurate model
of sky brightness during bright time. 
In this appendix, we describe how we construct this bright time sky brightness 
model using an empirical data-driven approach.

\begin{figure}
    \centering
    \includegraphics[width=0.85\textwidth]{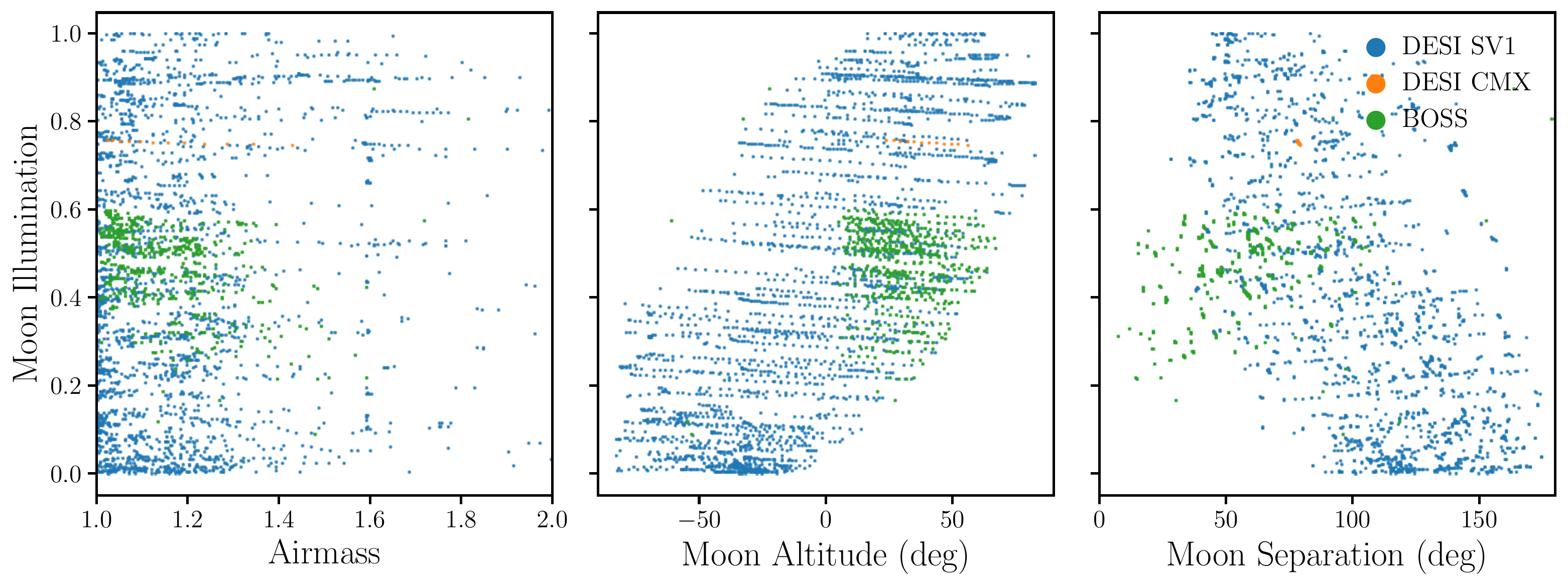}
    \caption{\label{fig:sky_obscond}
    The distribution of airmass and lunar conditions (illumination, altitude, separation)
    of DESI \svi~(blue), DESI commissioning (CMX; orange), and BOSS (green) sky spectra 
    used to construct the empirical bright time sky brightness model. 
    The exposures enable us to construct a model that predicts sky brightness given 
    airmass, moon illumination, altitude, and separation as input. 
    }
\end{figure}

To construct our sky brightness model, we use sky spectra primarily observed 
during \svi. 
A large number of sky spectra were observed by DESI because a significant number 
of fibers are dedicated to measuring the sky flux in every exposure to
perform sky subtraction in the spectroscopic pipeline~\citep{spec2022}. 
We only use sky spectra observed by DESI exposures during high transparency 
and exclude any sky spectra observed during twilight, when the altitude of the 
sun is above -18 deg. 
We supplement the \svi~data with sky spectra measured during the DESI commissioning 
(CMX) campaign as well as sky spectra from SDSS-III BOSS (also used in~\citealt{fagrelius2018})
under similar high transparency and non-twilight conditions. 
In total, we use sky spectra from 2,331 \svi~exposures, 12 CMX exposures,
and 990 BOSS exposures. 
In Figure~\ref{fig:sky_obscond}, we present the airmass and lunar conditions of these
exposures: DESI \svi~(blue), CMX (orange), and BOSS (green). 
The exposures span a broad range of observing conditions that fully encompasses 
typical BGS bright time conditions.

For each exposure, we compute the median sky spectrum of all the observed sky spectra
in the exposure. 
We then convert the sky spectrum flux to sky surface brightness using the area of the DESI
and BOSS fiber apertures. 
Afterwards, we smooth the median sky brightness and estimate its amplitude at 5000\AA, 
$I_{\rm sky}^{5000A}$. 
We use $I_{\rm sky}^{5000A}$ because we find good agreement between the redshift 
success rates predicted by the spectral simulations (Section~\ref{sec:texp}) and of
SV observations when we scale exposure time of the spectral simulations based on 
this amplitude. 

\begin{figure}
    \centering
    \includegraphics[width=0.45\textwidth]{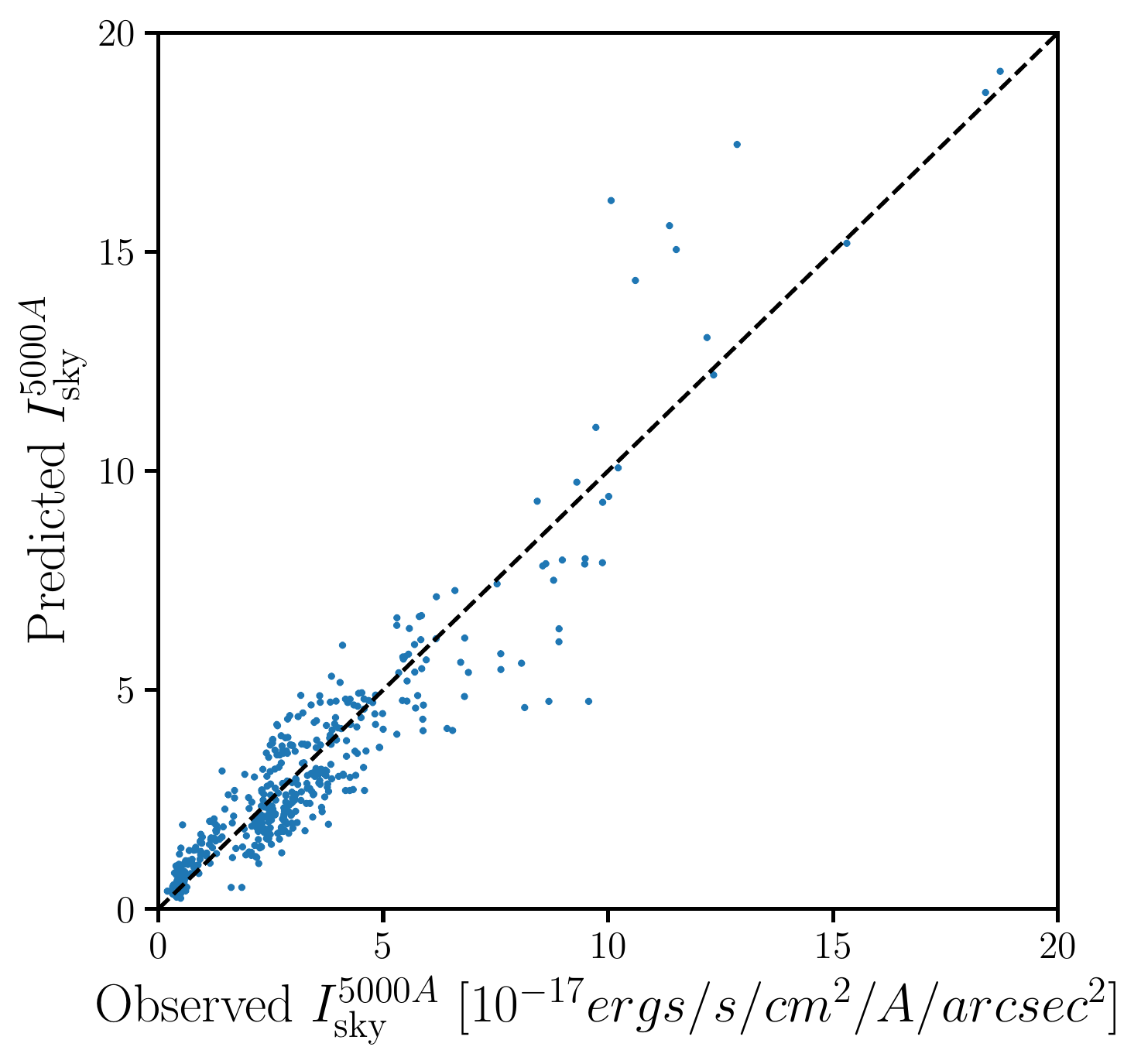}
    \caption{\label{fig:sky_pred}
    Comparison of predicted sky brightness at 5000\AA, $I_{\rm sky}^{5000A}$, of the 
    bright time sky brightness model versus the observed $I_{\rm sky}^{5000A}$ for 
    a test set of observations. 
    The bright time sky brightness model is trained using ridge regression on 
    $I_{\rm sky}^{5000A}$ compiled from the exposures in Figure~\ref{fig:sky_obscond}. 
    The bright time sky brightness model shows overall good agreement with observations
    and, therefore, accurately estimates the sky brightness during bright time.
    }
\end{figure}
Next, we use the $I_{\rm sky}^{5000A}$, airmass, moon illumination, altitude, and 
separation values of all the exposures to construct our sky brightness model. 
We use ridge regression to train our model with 80\% of the exposures. 
The other 20\% is reserved for testing the model.
The L2 regularization in ridge regression helps reduce model complexity and 
multicollinearity. 
We train polynomials of different orders (up to 8) and choose one with the minimum 
cross validation score. 
For our sky brightness model, we use a 5$^{th}$-order polynomial model that takes 
airmass, moon illumination, altitude, and separation as inputs and predicts 
$I_{\rm sky}^{5000A}$. 
In Figure~\ref{fig:sky_pred}, we compare the sky brightness, $I_{\rm sky}^{5000A}$,
of observations versus our model for the test set of exposures.
Overall, we find good agreement between the bright time sky brightness model and
observations. 
Moreover, survey simulations for BGS using this sky brightness model are in good
agreement with the \sviii~and early main survey operations. 